\documentclass[preprint]{elsarticle}

\usepackage{lineno,hyperref}
\usepackage{amsmath}
\usepackage{amssymb}
\usepackage{amsfonts}
\usepackage{graphicx,subcaption,float}
\usepackage{epstopdf}
\usepackage{tikz}
\usepackage{wasysym}

\usepackage{slashed}

\allowdisplaybreaks

\modulolinenumbers[1]

\DeclareMathOperator{\Tr}{Tr}

\usepackage{geometry}
\makeatletter
\def\ps@pprintTitle{%
	\let\@oddhead\@empty
	\let\@evenhead\@empty
	\def\@oddfoot{\reset@font\hfil\thepage\hfil}
	\let\@evenfoot\@oddfoot
}
\makeatother

\biboptions{sort&compress}

\begin{document}

\begin{frontmatter}

\title{One-Loop Renormalization of the Higgs Sector of the Electroweak Chiral Lagrangian extended by N Scalar Singlets}

\author[LMU]{Andreas Lindner}
\ead{and.lindner@physik.uni-muenchen.de}
\author[WWU]{Khoirul Faiq Muzakka}
\ead{khoirul.muzakka@uni-muenster.de}

\address[LMU]{Arnold Sommerfeld Center for Theoretical Physics, Ludwig-Maximilians-Universit\"at M\"unchen\\Theresienstr. 37, D-80333 M\"unchen, Germany}
\address[WWU]{Institut f{ü}r Theoretische Physik, Westf{ä}lische Wilhelms-Universit{ä}t M{ü}nster, Wilhelm-Klemm-Stra{ß}e 9, D-48149 M{ü}nster, Germany}

\begin{abstract}
The framework of the electroweak chiral Lagrangian with a light Higgs is extended by an additional scalar and then generalized to N scalars in the Higgs sector.
Divergences from scalar fluctuations are renormalized up to one loop using the background field method.
The results are crosschecked against the case of one scalar.
A subset of the divergences is demonstrated and crosschecked diagrammatically.
Together with the complete one-loop renormalization of the electroweak chiral theory with one light Higgs conducted previously, this constitutes a renormalization framework of any pure scalar extension to the electroweak chiral theory.

\end{abstract}

\begin{keyword}
renormalization, background field method, Higgs effective theory, chiral power counting
\end{keyword}

\end{frontmatter}

\tableofcontents{}
%\linenumbers

\section{Motivation\label{Sec:Introduction}}

%Since the Standard Model (SM) \textit{is} an effective theory, working only within the SM when comparing to measurements will increase the accuracy of its parameters, but not give any hint on what lies behind.
With the widely held conviction about the incompleteness of the Standard Model (SM) of particle physics, many extensions to the SM exist in the literature.
Effective field theory (EFT) has become an essential tool to study the impact of the ultraviolet (UV) completion of the SM to low energy observables. 
Assuming that the local gauge symmetry of the SM, $SU(3)_c \otimes SU(2)_L \otimes U(1)_Y$, is a remnant of a larger UV symmetry, the low energy EFT that generalizes the SM must preserve the SM gauge symmetry.
Based on how the terms in an effective Lagrangian are ordered in the EFT expansion, a bottom-up construction can be divided into two kinds of EFTs, the Standard Model effective field theory (SMEFT) and the electroweak chiral Lagrangian (EWCL).
The SMEFT is renormalizable at the leading order (LO) and the effective Lagrangian is ordered based on canonical dimensions.
The EWCL on the other hand is nonrenormalizable in the classical sense at leading order and expanded accordingly in loop orders, or equivalently, chiral dimensions \cite{chiralnlo}.

The electroweak chiral Lagrangian and the extension to one Higgs scalar have been treated to a large extend  \cite{longhitano1980,longhitano1981,herrero1,herrero2,appelquist1980,Appelquist1993,feruglio,Alonso1,Alonso2,Alonsoetal2013,Alonsoetal2013b,Alonsoetal2014,Brivioetal2014,sanz,gavela,Gavelaetal2016,Delgade2014,ewnlo,silh,Buchallaetal2014,chiralnlo,power,higgscoupling,fithiggs,addsinglet,completerenorm,guo}.
An excellent historical review can be found in \cite{BrivioTrott2019}.

The focus of this work is on the EWCL.
It is particularly suited to parameterize beyond the Standard Model physics admitting dynamical symmetry breakings that require no fundamental scalar, just like in the chiral symmetry breaking of quantum chromodynamics (QCD).
These dynamic models are interesting because they are free from the naturalness problem.
Examples of such models are the composite Higgs models (CHMs) \cite{kaplan,kaplan1,custodialsu2,dugan}, first proposed in unified composite model of all fundamental particles and forces \cite{Terazawa1977,Terazawa1980}.
Considering one scalar Higgs, we have the electroweak chiral Lagrangian with one light Higgs (EWCLh), which we also denote as $\mathcal{L} _{ew\chi}$.
It can parameterize, for example, the SM and the minimal CHM based on the coset $SO(5)/SO(4)$ \cite{Krause}.

%The Higgs sector is the most mysterious in the Standard Model.
%One could rephrase the statement and state that it is the sector with the most opportunities.
The situation of the highly fine-tuned SM can be considered unsatisfying.
There exist more natural solutions to electroweak symmetry breaking (EWSB), like the CHMs.
If true, they would induce some small deviations from the SM.
The discovery of CP violation in the Kaon system by Cronin and Fitch who went beyond the 1 \% level in their measurements, where others have long discarded the CP violation thesis \cite{futurepp}, reminds on the duty to be as precise in the Higgs sector as well.

The EWCLh can accommodate for various scenarios of electroweak symmetry breaking because of its general nature.
It is an effective framework at the electroweak scale with all the present particle content, where any modifications of the Higgs couplings, in comparison to the Standard Model, can be captured.
It does not need a specific UV-completion for itself to be of use, but serves as a broad effective framework that can be used to particularize to specific models \cite{chiralnlo}.
Any resulting Lagrangian of models beyond the Standard Model can be matched to $\mathcal{L} _{ew\chi}$ at the electroweak scale $ v $ \cite{addsinglet, silh}.
The matching makes the modified phenomenology in the Lagrangian compared to the Standard Model Lagrangian, $ \mathcal{L}_{SM} $ directly visible.
Matching the different models and experimental data to the $\mathcal{L} _{ew\chi} $ is therefore maybe the most general way to investigate the nature of the Higgs.\footnote{See also the discussions in \cite{higgscoupling, fithiggs}.}

A natural extension of the EWCLh would be an electroweak effective theory with more than one Higgs.
As can be seen in Table \ref{tb}, several CHMs, such as the ones based on $SO(6)/SO(5)$ and $SO(7)/SO(6)$ have more than one scalar singlet.

\begin{table}[ht]
	\centering
	\begin{tabular}{c c c c}
		$G$ & $H$ & $N_G$ & NGBs rep.[$H$] =rep.[$SU(2)_L\times SU(2)_R$]\\
		\hline
		$SO(5)$ & $SO(4)$ &4&\textbf{4}=\textbf{(2,2)}\\
		$SO(6)$ & $SO(5)$ &5& \textbf{5}=\textbf{(1,1)+(2,2)}\\
		$SO(6)$ & $SO(4)\times SO(2)$ & 8 & $\mathbf{{4_{2}+\bar{4}_{-2}}}$=2$\times\textbf{(2,2)}$\\		
		$SO(7)$ &$SO(6)$ &6& $\mathbf{6}=2\times \mathbf{(1,1)+(2,2)}$\\
		$SO(7)$ & $G_2$ & 7& $\mathbf{7= (1,3)+(2,2)}$\\
		$SO(7)$ & $SO(5)\times SO(2)$ & 10 & $\mathbf{10_0=(3,1)+(1,3)+(2,2)}$\\
		$SO(7)$ & $[SO(3)]^3$ & 12& $\mathbf{(2,2,3)}=3\times\mathbf{(2,2)}$\\
		$SU(5)$ & $SU(4)\times U(1)$ & 8 & $\mathbf{(4_{5},\bar{4}_{-5})}=2\times \mathbf{(2,2)}$\\
		$SU(5)$ & $SO(5)$ & 14 & $\mathbf{14} = (3,3)+(2,2)+(1,1)$\\
		\hline
	\end{tabular}
	\caption{Coset structures of the global breaking $G \rightarrow H$ for various CHMs and the number of Nambu-Goldstone Bosons (NGBs) for each case along with their transformations under the custodial symmetry $SU(2)_L\times SU(2)_R$.}
	\label{tb}
\end{table}

The additional singlets are for instance considered as dark matter candidates \cite{dm} or providers of CP violation for baryogenesis \cite{baryogenesis}.
With the addition of more scalar singlets it is therefore imperative to generalize the EWCLh to an electroweak chiral Lagrangian with an undefined number of N singlet scalars, we abbreviate it as EWCLNh.
This generalizaton is done in the present work.
Furthermore, the scalar sector of the arising Lagrangian is renormalized to one loop.

As mentioned above, the effective Lagrangian for theories with dynamical symmetry breaking is nonrenormalizable even at leading order.
Non-renormalizability of the effective theory in the classical sense is not obstructive.
It indicates non-decoupling new physics and hence the influence of the latter on the measurable regime.
And therefore, the indirect discovery of new physics.
Loops, however, induce divergences that must be absorbed by next-to-leading order (NLO) operators.
To this end, each operator gets a loop order assigned, which tells the number of loops for which UV divergences can be canceled.
The whole effective Lagrangian is thus organized by a loop expansion and a finite set of counterterms is provided at each order.
Since the loops come with derivative interactions in this strongly coupled theory, it as at the same time an energy, or momentum, expansion of the low-energy effective theory \cite{weinberg}.
In order to account for mass terms or couplings to gauge fields that obstruct the derivative or loop counting, and to keep things consistent, an assignment of \textit{chiral dimensions} was developed \cite{Leutwyler,chiraldynamics,power}.
Furthermore, because of the way higher order operators are generated, the evolution of the effective couplings (Wilson coefficients) is determined by the leading order Lagrangian alone \cite{chiralnlo}.

We investigate the idea of the observed Higgs-like particle deviating a little from the SM behavior.
The path we want to pursue is first to extend the electroweak chiral Lagrangian by two scalar fields.
An additional scalar brings a lot of new physics opportunities and widens the narrow corset of parameter space which constrains the electroweak chiral Lagrangian with only one scalar that necessarily has to be identified with the observed one.
A second scalar particle in addition to the observed one has been extensively discussed, e.g., in \cite{qi2021effective,chala,Redi}.
One aspect is to consider it as a dark matter candidate \cite{dm} which leads to new CP violating processes that we need to explain the matter-antimatter asymmetry \cite{chala,EWPhT}.
Ultimately, the framework of the electroweak chiral Lagrangian is extended by an arbitrary number of scalar fields at leading chiral order and all divergences arising through scalar loops to one-loop order are extracted via the background field method.

The present work is based on our master's theses \cite{faiq,andi} and directly expands on \cite{completerenorm} and the ideas outlined in \cite{chiralnlo}.
A basically equivalent, but formally different approach to renormalize a general scalar sector, based on geometric considerations, was taken in \cite{rgehiggseft}.
We believe that the formulation which relies on a clearly defined operator basis \cite{Krause} is less abstract and better apt for phenomenology.

\subsection{Outline}

We begin in Section \ref{sec:ewclh} with a recap of the electroweak chiral Lagrangian with one light Higgs and its renormalization that has been worked out in \cite{completerenorm,guo}.
The procedure to extract the divergences is extended to two scalar singlet fields in Section \ref{generaltwoscalars} and to N singlets in Section \ref{NSinglets}.
The result is tested against the limit of one scalar singlet and a subset of the results which are obtained via the background field method are checked diagrammatically in Section \ref{sec:crosschecks}.
A short discussion of the counterterms to cancel the divergences concludes that section.

The impact of the new framework is summarized in the conclusions.
The next-to-minimal composite Higgs model with $ SO(6)/SO(5) $ coset structure is addressed as an application in a follow-up work \cite{LindnerMuzakka2022b}.

\section{Recap: Electroweak Chiral Lagrangian with a light Higgs}\label{sec:ewclh}
We want to tackle explicitly the nonrenormalizable sector of the electroweak chiral Lagrangian including a light Higgs -- the scalar, the Higgs, sector.
The nonrenormalizable sector generates NLO counterterms that are not present in the renormalizable Standard Model effective field theory \cite{completerenorm}.
We are going to renormalize the divergences arising from scalar loops, which are the only ones leading to NLO divergences scaling as $ \mathcal{O}(p^4) $ due to their derivative interactions, and reproduce the results of \cite{guo,completerenorm}.

Instead of summing up all the contributing Feynman diagrams, we use functional integral methods, resulting in an enormous reduction of work in this use case (and in many other scenarios).
The complete one-loop renormalization of the electroweak chiral Lagrangian with a Higgs via the background field method has been performed in \cite{completerenorm}.
The method provides a means of calculating the effective action while keeping gauge invariance in the counterterms.
The procedure was used in the similar scenario of resonance chiral theory in \cite{Rosell1,Rosell2}, where the loop effects of the scalar Goldstone bosons were considered as well as of some scalar resonances included in that theory.

\subsection{The Compact EWCLh}
For computational convenience we compactify the Lagrangian.
At leading order, the EWCLh can be written in a compact form \cite{completerenorm}:

\begin{equation}
\begin{aligned}
\mathcal{L} _{ew\chi}= & -\frac{1}{4}G_{\mu\nu}^{i2}-\frac{1}{4}B_{\mu\nu}^2-\frac{1}{4}W_{\mu\nu}^{a2}+\frac{1}{2}\partial_\mu h\,\partial^\mu h-V(h)  +\frac{v^2}{4}\langle (D_\mu U)^\dagger D^\mu U \rangle F(h)+\bar{\psi}i \slashed{D}\psi-\bar{\psi}m(h,U)\psi \ ,
\end{aligned}
\label{eqn:EWCLhcompact}
\end{equation}
where $G^i_{\mu\nu}$, $W_{\mu\nu}$, and $B_{\mu\nu}$ are the gauge field strength tensors of the strong and electroweak interactions and the covariant derivative is given by

\begin{equation}
D_\mu U=\partial_\mu U+igW_\mu U-ig'B_\mu UT^3
\end{equation}
and

\begin{equation}
U=\exp\left(2i\varphi/v\right) \ , \quad \varphi = \varphi^a \ T^a \ ,
\end{equation}
is the Goldstone matrix and the $ \varphi^a $ are the Goldstone fields.
The $ SU(2) $ generators $ T^a $ are normalized such that

\begin{equation}\label{eq:su2gen}
\langle T^a T^b \rangle = \frac{1}{2} \delta^{ab} \  \Rightarrow T^a = \frac{\sigma^a}{2} \ .
\end{equation} 
The $ \sigma^a $ are the Pauli sigma-matrices.
The Yukawa term $ m(h,U) $ is short for

\begin{equation}
m(h,U) \doteq UM(h) P_R + M^\dagger(h,s) U^\dagger P_L \ ,
\end{equation}
with the mass matrix $ M $ and the fermion field $ \psi $ defined as (generation indices suppressed)

\begin{equation}
M(h) = \text{diag} (M_u(h),M_d(h),M_\nu(h),M_e(h)) \ ,\quad \psi= (u,d,\nu,e)^T \ .
\label{eqn:MandPsi}
\end{equation}
Note that $U$ is a 2 x 2 matrix and $M(h)_{ij}^\alpha $ as defined in (\ref{eqn:MandPsi}) a 4 x 4 matrix. 
The action is thus split into two diagonal blocks, $ U $ acting on the two doublets each.
Since we are working with the SM particle content we have $ M_\nu = 0 $.
Further, we have the expansions of the Higgs-dependent functions

\begin{equation}\label{eq:polynomial_funcs}
\begin{aligned}
& F(h) = 1+ \sum_{n=1}^\infty F_n \left( \frac{h}{v} \right)^n  , \, V(h) = v^4 \sum_{n=2}^\infty V_n \left( \frac{h}{v} \right)^n  , \,  M(h) = \sum_{n=0}^\infty M_n \left( \frac{h}{v} \right)^n  .
\end{aligned}
\end{equation}
Custodial symmetry breaking terms are deferred to NLO to avoid large corrections to the $ \rho $ parameter, $\rho \doteq M_W^2/(c_W^2 M_Z^2)$. 
Singlet interactions with the fermion kinetic terms can be compensated by field redefinitions \cite{chiralnlo}.

Further simplification of the upcoming computations is obtained by means of a \textit{Stueckelberg Transformation}.

\subsection{Unitary Gauge for Background Goldstones}
\label{UnitaryGauge}
In the framework of the background field method we decompose the fields under consideration, $B_\mu$, $W^a_\mu$, and $h$ into their classical background and quantum fields,

\begin{eqnarray}
B_\mu=B_{c \mu}+\tilde{B}_\mu,\quad\quad W^a_\mu=W^a_{c \mu}+\tilde{W}^a_\mu, \quad\quad h=h_c+\tilde{h} \ ,
\end{eqnarray}
while we split the Goldstone field $ U $ in a multiplicative nonlinear fashion,

\begin{eqnarray}
U=U_c\tilde{U}  \, ,
\end{eqnarray}
for a reason which becomes clear in a moment.
The $U_c$ and $\tilde{U}$ are the matrices for the background and quantum Goldstone fields $\varphi_c$ and $\varphi$. 
The full fields transform under $SU(2)_L\times U(1)_Y$ gauge transformations,

\begin{equation}
S_L=\exp\left(iT^a\alpha^a\right) \, , \quad S_Y=\exp \left(iT^3\beta\right) \, ,
\end{equation}
as

\begin{equation}
\begin{aligned}
& B_\mu T^3 \longrightarrow B_\mu T^3 -\frac{i}{g'} S_Y \partial_\mu S_Y^\dagger  \, , \quad W_\mu\longrightarrow S_LW_\mu S_L^\dagger-\frac{i}{g} S_L \partial_\mu S_L^\dagger  \, , \\
& h\longrightarrow h \ , \quad U \longrightarrow S_L U S_Y^\dagger \ , \quad  D_\mu U\longrightarrow S_L( D_\mu U)S_Y^\dagger \ .
\end{aligned}
\end{equation}
In the background field formalism, however, we retain the full gauge symmetry of the classical background gauge fields while spoiling the gauge symmetry of quantum gauge fields \cite{schwartz},

\begin{equation}
\begin{aligned}
& B_{c \mu} T^3 \longrightarrow  B_{c \mu} T^3 - \frac{i}{g'} S_Y \partial_\mu S_Y^\dagger  \, ,\quad\tilde{B}_\mu\longrightarrow \tilde{B}_\mu  \, ,\\
& W_{c \mu}\longrightarrow S_L W_{c \mu} S_L^\dagger-\frac{i}{g} S_L \partial_\mu S_L^\dagger  \, ,\quad \tilde{ W}_\mu\longrightarrow S_L\tilde{ W}_\mu S_L^\dagger\,.
\end{aligned}
\end{equation}
With this choice of gauge transformation, provided the gauge fixing function respects the background gauge symmetry, the resulting effective action exhibits gauge symmetry explicitly. 

For convenience in the calculations we choose unitary gauge for the background field, $U_c=1$, such that the background Goldstone fields disappear completely from the effective action.
In the end all observables are gauge independent and this convenient gauge simplifies the computations the most. 
The unitary gauge is achieved by requiring \cite{Dittmaier1, Dittmaier2}

\begin{equation}
\langle (D_\mu U)^\dagger D^\mu U \rangle \longrightarrow \langle (D_\mu \tilde{ U})^\dagger D^\mu\tilde{ U} \rangle.
\end{equation}
To obtain the explicit transformation we first write

\begin{equation}
D_\mu U=D_\mu (U_c \tilde{ U})=\left(\partial_\mu U_c\right)\tilde{ U}+U_c\partial_\mu \tilde{ U}+igW_\mu U_c \tilde{ U}-ig'B_\mu U_c\tilde{ U}T^3 \, .
\end{equation}
Under $SU(2)_L\times U(1)_Y$ gauge transformations we find

\begin{equation}
\begin{aligned}
D_\mu U\longrightarrow & \left(\partial_\mu U_c' \right)\tilde{ U'}+U_c'\partial_\mu \tilde{ U'} +ig\left(S_L W_{c \mu} S_L^\dagger-\frac{i}{g} S_L \partial_\mu S_L^\dagger+S_L\tilde{W}_\mu  S_L^\dagger\right) U_c' \tilde{ U'}\\
&-ig'
U_c'\tilde{ U'}\left(B_{c \mu}T^3-\frac{i}{g'} S_Y \partial_\mu S_Y^\dagger+\tilde{ B}_\mu T^3 \right)\\
&=\left(\partial_\mu U_c' \right)\tilde{ U'}+ S_L\partial_\mu S_L^\dagger U_c' \tilde{ U'} - S_Y \partial_\mu S_Y^\dagger U_c' \tilde{ U'}\\
&+\left[U_c' \partial_\mu \tilde{ U'}+igS_L\left(W_{c \mu}+\tilde{ W}_\mu\right)S_L^\dagger U_c' \tilde{ U'}-ig'(B_{c \mu}+\tilde{ B}_\mu) U_c'\tilde{ U'}T^3 \right].
\end{aligned}
\end{equation}
It is now obvious that unitary gauge for the background Goldstones can be obtained by choosing

\begin{equation}
S_L=U_c \, , \quad S_Y=1  \, ,
\end{equation}
which renders

\begin{equation}
	D_\mu U\longrightarrow U_c D_\mu \tilde{ U} \, .
\end{equation}
The resulting transformation is called \textit{Stueckelberg transformation}:

\begin{equation}
\begin{aligned}
& B_{c \mu}\longrightarrow  B_{c \mu} \, ,\quad\tilde{B}_\mu\longrightarrow \tilde{B}_\mu \, ,\\
& W_{c \mu}\longrightarrow U_cW_{c \mu}  U_c^\dagger-\frac{i}{g}  U_c \partial_\mu  U_c^\dagger  \, ,\quad \tilde{ W}_\mu\longrightarrow  U_c\tilde{ W}_\mu U_c^\dagger \, .
\label{eqn:StueckW}
\end{aligned}
\end{equation}
In this gauge the covariant derivative is, as demanded,

\begin{equation}
D_\mu U\longrightarrow U_c D_\mu \tilde{U}  \, .
\end{equation}
Unitary gauge also removes the background Goldstone field from the Yukawa mass term since it implies the gauge transformation

\begin{equation}
\psi_L \rightarrow U_c \psi_L  \, .
\label{eqn:StueckPsi}
\end{equation}
This guarantees the Lagrangian to be $U_c$ free.
From now on we also drop the subscript "c" in $U_c$, which denotes the classical background fields, for notational convenience.

In the background field method the gauge fixings for the background fields and the quantum fields are unrelated, so we are free to choose different gauges \cite{Dittmaier3}.
Since the background Goldstones do not appear in the loop calculations that we are going to perform, it is a matter of convenience to remove only these by the gauge choice.
Unitary gauge for the quantum fields would complicate loop calculations on the other hand.
We regain the Goldstones by inverting the Stueckelberg transformation in the end of our computations.
This is essentially a replacement of $ U $ by $ U^\dagger $ in (\ref{eqn:StueckW}) and (\ref{eqn:StueckPsi}).
The inversion of lengthier expressions is more involved.

\subsection{t'Hooft Master Formula}

The background field method provides a convenient way of computing the one-loop effective action \cite{thooft,Abbott,Ramond,dewittquantum}.
By separating the scalar field configurations into a classical background and quantum fluctuations around it,

\begin{equation}
\phi = \phi_c + \phi_q \ ,
\end{equation}
it is obtained

\begin{equation}\label{eq:Seff}
S_{\text{eff}}^{\text{1loop}} = \frac{i}{2} \textrm{tr} \ \int d^4x \int \frac{d^4p}{(2\pi)^d} \ \ln \ \Delta(x,\partial_x+ip) \ ,
\end{equation}
where tr is the trace only over operator space and the differential operator $\Delta(x,\partial_x+ip)$ is given by

\begin{equation}
\dfrac{ \delta^2 \mathcal{L}}{\delta \phi(x) \delta \phi(y)} \bigg|_{\phi_c} \doteq  \Delta(x,\partial_x) \ \delta^4(x-y) \ .
\label{eqn:diffop}
\end{equation}
In full generality the differential operator can be written as

\begin{equation}
\Delta(x,\partial_x) = - \mathcal{D}_\mu \mathcal{D}^\mu - Y \ , \quad \mathcal{D}_\mu = \partial_\mu + X_\mu \ .
\label{eqn:diffopx}
\end{equation}
In these terms it is found

\begin{equation}
\Delta(x,\partial_x+ip) = p^2 \left(1- \dfrac{2ip \mathcal{D} + \mathcal{D}^2 +Y}{p^2} \right) \ .
\end{equation}
The logarithm in \eqref{eq:Seff} can be expanded and the integrals evaluated using dimensional regularization \cite{dimreg1,dimreg2}.
The t'Hooft master formula for the divergent terms of the effective Lagrangian (which he found using symmetry arguments \cite{thooft}) is then given by

\begin{equation}
\mathcal{L}_{\text{eff,div}}^{\text{1loop}} = \dfrac{1}{32\pi^2 \epsilon} \text{tr} \left[ \frac{1}{12} X_{\mu\nu}X^{\mu\nu} + \frac{1}{2} Y^2 \right] \, ,
\label{eqn:tHooft}
\end{equation}
with

\begin{equation}\label{eq:Xmunu}
X_{\mu\nu} = [\mathcal{D}_\mu,\mathcal{D}_\nu] \, .
\end{equation}

\subsection{One-Scalar-Loop Divergence}

%We restrict ourselves to the Higgs sector, the nonrenormalizable part of the EWCL with scalars.
We integrate out the quadratic singlet and Goldstone quantum fluctuations to extract the divergences arising from scalar loops. 
The procedure is therefore to expand the EWCLh to second order in the scalar fluctuations to find the differential operator (\ref{eqn:diffop}) which yields the effective one-loop Lagrangian.
Writing the differential operator in the form of (\ref{eqn:diffopx}) one obtains the divergence via the t'Hooft master formula (\ref{eqn:tHooft}).
This calculation was demonstrated in \cite{guo,completerenorm}.
It has proven convenient to define $L_\mu = iU (D_\mu U)^\dagger$ and $\tau_L = U\sigma_3 U^\dagger$.
%Both transform as $L_{\mu}\longrightarrow S_LL_{\mu}S_L^\dagger$.
With primes denoting differentiation with respect to $ h $, the additional shorthands $ \kappa = vF'/2\sqrt{F} $ and $ \mathcal{B}= v^2F'^2/4F^2 - v^2F''/2F $ are introduced.
The result of the one-scalar-loop divergence, for later reference, is given by \cite{completerenorm}

\begin{equation}
\begin{aligned}
\mathcal{L}_{ew\chi,\text{eff,div}}^{\text{1loop}}& = \frac{1}{32\pi^2 \epsilon} \Bigg\{ - \frac{\kappa^2+1}{24} \left( 2g^2 \langle W_{\mu\nu} W^{\mu\nu} \rangle +  g'^2 B_{\mu\nu} B^{\mu\nu} \right)  + \frac{(\kappa^2-1)^2}{6}\langle L_\mu L_\nu \rangle ^2 \\
&+ \left( \frac{(\kappa^2-1)^2}{12} + \frac{F^2 \mathcal{B}^2}{8} \right) \langle L_\mu L^\mu \rangle ^2  + \frac{\kappa^2-1}{6} gg' \langle \tau_L W^{\mu\nu} \rangle B_{\mu\nu} - \frac{\kappa^2-1}{12} \bigg( ig\langle W_{\mu\nu} [L^\mu,L^\nu] \rangle\\
& + ig' B_{\mu\nu} \langle \tau_L [L^\mu , L^\nu ] \rangle \bigg) -\left( (\kappa^2-1) \frac{\mathcal{B}}{v^2} +\frac{\kappa'^2 }{6} \right) (\partial h)^2 \langle L_\mu L^\mu \rangle + \frac{2}{3} \kappa'^2 (\partial^\mu h \partial^\nu h) \langle L_\mu L_\nu \rangle \\
& - \frac{1}{3} \kappa \kappa' (\partial_\mu h) \left(  g\langle W^{\mu\nu} L_\nu \rangle - g' B^{\mu\nu} \langle \tau_L L_\nu \rangle  \right)  - \left( \frac{F'V'}{2F} (\kappa^2-1) - \frac{V'' F}{2} \mathcal{B} \right)  \langle L_\mu L^\mu \rangle \\
&- \frac{(\kappa^2-1)}{F} \left[ \bar{\psi}_L U \left( \frac{F'}{2} M' - \frac{M}{v^2} \right) \psi_R + \textrm{h.c.}\right] \langle L_\mu L^\mu \rangle  \\
&  + \frac{3}{8} \left( \frac{F'V'}{F} \right)^2 + \frac{3}{2} \frac{F'V'}{F^2} \left[ \bar{\psi}_L U \left( \frac{F'}{2} M' - \frac{M}{v^2} \right) \psi_R + \textrm{h.c.}\right] + \frac{3}{2} \frac{F'V'}{F} \frac{\mathcal{B}}{v^2} (\partial h)^2 \\
& + \frac{3}{2F^2}  \left[ \bar{\psi}_L U \left( \frac{F'}{2} M' - \frac{M}{v^2} \right) \psi_R + \textrm{h.c.}\right]^2 + \frac{3 \mathcal{B}}{v^2F} \left[ \bar{\psi}_L U \left( \frac{F'}{2} M' - \frac{M}{v^2} \right) \psi_R + \textrm{h.c.}\right] (\partial h)^2 \\
&+ \frac{3\mathcal{B}^2}{2v^4} (\partial h)^2 (\partial h)^2  + \frac{4}{v^2} \left[i \bar{\psi}_L U T^a \left( \frac{M}{\sqrt{F}} \right)'  \psi_R+ \textrm{h.c.}\right]^2  + \frac{2\kappa'}{v^2} (\partial^\mu h)\left[i \bar{\psi}_L L_\mu U \left( \frac{M}{\sqrt{F}} \right)'  \psi_R+ \textrm{h.c.}\right]  \\
& + \frac{V''^2}{2} +  V'' \left[ \bar{\psi}_L U M'' \psi_R+\textrm{h.c.}\,\right]+ \frac{1}{2} \left( \bar{\psi}_L U M'' \psi_R+\textrm{h.c.}\,\right)^2   + \frac{F \mathcal{B}}{2} \left[ \bar{\psi}_L U M'' \psi_R+\textrm{h.c.}\,\right] \langle L_\mu L^\mu \rangle \Bigg\} .
\label{eqn:finalewclh}
\end{aligned}
\end{equation}

\section{Renormalization of Kinetic Terms with two Singlets via Matrix Field Redefinitions}
\label{generaltwoscalars}

In this chapter we proceed to the general kinetic Lagrangian for two scalars to introduce the renormalization procedure that we generalize in the next section.
First, we set up the most general form of a kinetic term for two scalar fields:

\begin{equation}
\mathcal{L}_{\text{kin,2}}=\frac{1}{2}\, (\partial h)^2(1+F_{11}(h,s))+\frac{1}{2}\,(\partial s)^2(1+F_{22}(h,s))+\partial h\, \partial s \,F_{12}(h,s) \ .
\label{eqn:generalkinetictwoscalar}
\end{equation}
However, when renormalizing this form via the background field method, it leads to kinetic mixing of quantum fluctuations.
This yields a differential operator $ \Delta $ with $ \partial^2 $ operations in off-diagonal elements in contrast to the desired form \eqref{eqn:diffopx}.
We need matrix field redefinitions to obtain a properly defined differential operator.

Defining $ h_i = (h,s)^T $, \eqref{eqn:generalkinetictwoscalar} is compactly written as

\begin{equation}
\mathcal{L}_{\text{kin,2}}=\frac{1}{2}\left(\delta_{ij}+F_{ij}(h)\right) \partial_\mu h_i \,\partial^\mu h_j \, ,
\end{equation} 
where $F_{ij}$ is a symmetric matrix consisting of polynomial expansions of the form of \ref{eq:polynomial_funcs}.
We split the fields into classical background plus quantum fluctuations $h_i \rightarrow h_i +\tilde{h}_i $.
The Lagrangian is expanded to second order in the quantum fluctuations:

\begin{equation}
\begin{aligned}
\mathcal{L}_{\text{kin,2}} & \overset{\mathcal{O}(\tilde{h}^2)}{=} \frac{1}{2}\left(\delta_{ij}+F_{ij}\right)\partial_\mu \tilde{h}_i \,\partial^\mu \tilde{h}_j + F'_{ij,k}(\partial_\mu h_i)\, \tilde{h}_k \partial^\mu \tilde{h}_j  + \frac{1}{4}F''_{ij,kl} (\partial_\mu h_i)(\partial^\mu h_j) \tilde{h}_k \tilde{h}_l \\
& \ \ = -\frac{1}{2} \tilde{h}_i \Big[\left(\delta_{ij}+F_{ij}\right)\partial^2+\left(F'_{ij,k}-2F'_{jk,i}\right)(\partial_\mu h_k)\partial^\mu -\frac{1}{2}F''_{kl,ij}(\partial_\mu h_k)(\partial^\mu h_l) \Big] \tilde{h}_j \ .
\end{aligned}
\end{equation}
For any symmetric matrix $S^\mu$ and some vector $\eta$ we have

\begin{equation}
\eta^T S^\mu \partial_\mu \eta= -\frac{1}{2}\eta^T (\partial_\mu S^\mu)\eta +\text{total derivative} \ .
\end{equation}
With this we find

\begin{equation}
\begin{aligned}
\mathcal{L}_{\text{kin,2}} &\overset{\mathcal{O}(\tilde{h}^2)}{=}-\frac{1}{2} \tilde{h}_i\Big[\left(\delta_{ij}+F_{ij}\right)\partial^2+\left(F'_{ik,j}-F'_{jk,i}\right)(\partial_\mu h_k)\partial^\mu \\
&\qquad -\partial_\mu\left[(F'_{ij,k}-F'_{jk,i}-F'_{ik,j})\partial^\mu h_k\right] -\frac{1}{2}F''_{kl,ij}(\partial_\mu h_k)(\partial^\mu h_l) \Big]\tilde{h}_j\\
& = -\frac{1}{2} \tilde{h}^T \left[A(h)\partial^2+2B^\mu(h)\partial_\mu+C \right] \tilde{h} \ ,
\end{aligned}
\end{equation}
where 

\begin{equation}
\begin{aligned}
A_{ij} (h)=\delta_{ij}+F_{ij}(h)\,,\quad B^\mu_{ij}(h)=\frac{1}{2}\left(F'_{ik,j}-F'_{jk,i}\right)(\partial_\mu h_k)\,,\\
C_{ij}(h) = -\partial_\mu\left[(F'_{ij,k}-F'_{jk,i}-F'_{ik,j})\partial^\mu h_k\right] -\frac{1}{2}F''_{kl,ij}(\partial_\mu h_k)(\partial^\mu h_l) \ .
\end{aligned}
\end{equation}
As required for the method, $A$ and $C$ are symmetric, while $B^\mu$ is antisymmetric.
We express the components of $ B $ and $ C $ in terms of the background scalars $ h $ and $ s $.
Derivatives with respect to $ h $ are denoted with a prime and those with respect to $ s $ with a dot.
We deduce the components

\begin{equation}
\begin{aligned}
B^\mu_{12} & = - B^\mu(h)_{21}= \frac{1}{2} (\dot{F}_{11} - F_{12}')\, \partial^\mu h+ (\dot{F}_{12} - F_{22}') \, \partial^\mu s \ , \\
C_{11} & = \frac{1}{2} F_{11}'' (\partial h)^2 + F_{12}'' (\partial h)(\partial s) + (-\ddot{F}_{11} + 2 \dot{F}_{12}' - \frac{1}{2} F_{22}'') (\partial s)^2  + F_{11}' (\partial^2 h) + (-\dot{F}_{11} + 2F_{12}') (\partial^2 s) \ , \\
C_{22} & = \frac{1}{2} \ddot{F}_{22} (\partial s)^2 + \ddot{F}_{12} (\partial h)(\partial s) + (-F_{22}'' + 2 \dot{F}_{12}' - \frac{1}{2} \ddot{F}_{11}) (\partial h)^2  + \dot{F}_{22} (\partial^2 s) + (-F'_{22} + 2\dot{F}_{12}) (\partial^2 h) \ , \\
C_{12} & = C_{21} = \frac{1}{2} \dot{F}_{11}' (\partial h)^2 +\frac{1}{2} \dot{F}_{22}' (\partial s)^2 + (F_{22}''+ \ddot{F}_{11}-\dot{F}_{12}') (\partial h)(\partial s) + F_{22}' (\partial^2 s) + \dot{F}_{11} (\partial^2 h) \ .
\end{aligned}
\end{equation}
After this operation, the fields are not canonical anymore.
To obtain a differential operator in the standard form of Equation (\ref{eqn:diffopx}) we now perform a matrix field rescaling:

\begin{equation}
\tilde{h} \rightarrow A^{-1/2} \, \tilde{h} \ .
\end{equation}
Hence, we have 

\begin{equation}
\begin{aligned}
&\partial_\mu \tilde{h} \longrightarrow (\partial_\mu A^{-1/2}) \tilde{h}+ A^{-1/2}\partial_\mu \tilde{h} \ , \\
& \partial^2 \tilde{h} \longrightarrow (\partial^2 A^{-1/2})\tilde{h} + 2(\partial_\mu A^{-1/2})\partial^\mu \tilde{h} + A^{-1/2} \partial^2\tilde{h} \ .
\end{aligned}
\end{equation}
With this, we obtain  

\begin{equation}
\begin{aligned}
\mathcal{L}_{\text{kin,2}} &\overset{\mathcal{O}(\tilde{h}^2)}{=} -\frac{1}{2}\tilde{h}^T \Big[\partial^2+2\left(A^{1/2} (\partial^\mu A^{-1/2})+A^{-1/2} B^\mu A^{-1/2}  \right)\partial_\mu\\
& \qquad+ A^{-1/2} CA^{-1/2}+A^{1/2} (\partial^2 A^{-1/2})+2A^{-1/2} B^\mu (\partial_\mu A^{-1/2}) \Big]\tilde{h} \ .
\end{aligned}
\end{equation}
Next, we antisymmetrize the coefficient of $\partial_\mu$ using

\begin{equation}
A^{1/2} (\partial^\mu A^{-1/2}) = - (\partial^\mu A^{1/2})A^{-1/2} \, ,
\end{equation}
and symmetrize the constant terms again using partial integration.
We obtain (remember that $B_\mu$ is antisymmetric and $C$ is symmetric)

\begin{equation}
\begin{aligned}
\mathcal{L}_{\text{kin,2}} &\overset{\mathcal{O}(\tilde{h}^2)}{=} -\frac{1}{2}\tilde{h}^T \Big[ \ \partial^2 \\
&\qquad +2\Big(A^{-1/2} B^\mu A^{-1/2} +\frac{1}{2}A^{1/2} (\partial^\mu A^{-1/2})- \frac{1}{2}(\partial^\mu A^{-1/2}) A^{1/2}  \Big)\partial_\mu\\
&\qquad+ A^{-1/2} CA^{-1/2} +A^{-1/2} B^\mu (\partial_\mu A^{-1/2}) -   (\partial_\mu A^{-1/2}) B^\mu  A^{-1/2} \\
&\qquad-\frac{1}{2}  \left(\partial_\mu A^{-1/2} \,\partial^\mu A^{1/2}+  \partial_\mu A^{1/2} \,\partial^\mu A^{-1/2}\right)\Big]\tilde{h}\\
& \quad \doteq -\frac{1}{2}\tilde{h}^T \left[(\partial_\mu+X_\mu)^2+Y \right]\tilde{h} \ .
\end{aligned}
\end{equation}
From this result, we extract the components of the t'Hooft master formula (\ref{eqn:tHooft})

\begin{equation}
\begin{aligned}
X^\mu & = A^{-1/2} B^\mu A^{-1/2} +\frac{1}{2}A^{1/2} (\partial^\mu A^{-1/2})- \frac{1}{2}(\partial^\mu A^{-1/2}) A^{1/2} \ , \\
Y &=A^{-1/2} CA^{-1/2} +A^{-1/2} B^\mu (\partial_\mu A^{-1/2}) -   (\partial_\mu A^{-1/2}) B^\mu  A^{-1/2} \\
&-\frac{1}{2}  \left(\partial_\mu A^{-1/2} \,\partial^\mu A^{1/2}+  \partial_\mu A^{1/2} \,\partial^\mu A^{-1/2}\right)-X_\mu X^\mu \ .
\end{aligned}
\label{eqn:XandY}
\end{equation}
These are well-defined objects for the formula of the differential operator (\ref{eqn:diffopx}), from which the divergence is obtained.
We leave the explicit formulation of the divergence of the theory to the general case discussed in the following section.

\section{Generalization to N Scalar Singlet Fields}
\label{NSinglets}

The matrix field redefinitions make the computations hard, but the great benefit of the formalism in the previous section is its generality and that it is easily extendable to N scalar fields.
We can directly write down the most general leading order electroweak chiral Lagrangian with N scalar singlet fields:

\begin{equation}
\begin{aligned}
\mathcal{L}_{ew\chi ,N}= & -\frac{1}{4}G_{\mu\nu}^{i2}-\frac{1}{4}B_{\mu\nu}^2-\frac{1}{4}W_{\mu\nu}^{a2} + \frac{1}{2}\left(\delta_{ij}+F_{ij}(h)\right)\partial_\mu h_i\,\partial^\mu h_j -V(h) \\
& +\frac{v^2}{4}\langle (D_\mu U)^\dagger D^\mu U \rangle \, F_U(h)+\bar{\psi}i \slashed{D}\psi-\bar{\psi}m(h,U)\psi \ ,
\end{aligned}
\end{equation}
where $ i= 1,...,N $.
Clearly, $ F_{ij}=F_{ji} $.

\subsection{Quadratic Scalar Fluctuations}

We proceed analogously to the scenario of two singlets, expanding the scalar fields up to quadratic order.
The scalar fields $ h_{i} $ are expanded again as $ h_{i} \rightarrow h_{i} + \tilde{h}_i $ into classical background and quantum fluctuations.
Keep in mind that unitary gauge is chosen for background Goldstones.
The one-loop effective Lagrangian takes the form

\begin{equation*}
\begin{aligned}
\mathcal{L}_{ew\chi ,N,\text{eff}}^{\text{1loop}}=& - \frac{1}{2} \varphi^a \Delta^{ab} \, \varphi^b - \frac{1}{2} \varphi^a \Delta^{ai} \, \tilde{h}^i- \frac{1}{2} \tilde{h}^i \Delta^{ia} \, \varphi^a - \frac{1}{2} \tilde{h}^i \Delta^{ij} \, \tilde{h}^j \ ,
\end{aligned}
\end{equation*}
where $\varphi^a $ are quantum fluctuations of the Goldstone fields.
The components of the $ \Delta $ operator are now computed in turn.

\begin{itemize}
	\item First, we compute the quadratic expansions of
	
	\begin{equation}
	\frac{v^2}{4}\Tr \left(D_\mu U^\dagger D^\mu U\right)F_U(h_k)
	\end{equation}
	to second order in both Goldstone and Higgs fields.
	From
	
	\begin{equation}
		\begin{split}
			\left(D_\mu U\right)^\dagger D^\mu U &= \partial_\mu U^\dagger \, \partial^\mu U+\left[ ig\left(\partial_\mu U^\dagger\right) W^\mu U+\textrm{h.c.}\right]   +\left[ -ig'\left(\partial_\mu U^\dagger\right)UB^\mu T^3+\textrm{h.c.}\right] \\
			&\quad +g^2\, U^\dagger W_\mu W^\mu U + g'^2 B_\mu B^\mu T^3 T^3+\left[-gg' U^\dagger W_\mu U B^\mu T+\textrm{h.c.}\right] \, ,
		\end{split}
	\end{equation}
	we have up to quadratic order in $\varphi$:
	
	\begin{equation}
		\begin{split}
			&\frac{v^2}{4}\left\langle\partial_\mu U^\dagger \, \partial^\mu U\right\rangle = \frac{1}{2}\left(\partial_\mu \varphi^a\right)\partial^\mu\varphi^a \, , \\
			&\frac{v^2}{4}\left\langle ig\left(\partial_\mu U^\dagger\right) W^\mu U+\textrm{h.c.}\right\rangle =\frac{gv}{2}\left(\partial_\mu\varphi^a\right)W^{\mu a}-\frac{g}{2}\epsilon^{abc}W_\mu^a \varphi^b\partial^\mu \varphi^c \, , \\
			&\frac{v^2}{4}\left\langle-ig'\left(\partial_\mu U^\dagger\right)UB^\mu T^3+\textrm{h.c.}\right\rangle =-\frac{g'v}{2}B^\mu\partial_\mu\varphi^3+\frac{g'}{2}B^\mu\left((\partial_\mu \varphi^1)\varphi^2-(\partial_\mu \varphi^2)\varphi^1\right)\, , \\
			&\frac{v^2}{4}\left\langle g^2\, U^\dagger W_\mu W^\mu U
			+ g'^2 B_\mu B^\mu T^3 T^3\right\rangle =\frac{1}{8}g^2v^2 W_\mu^a W^{\mu a}+\frac{1}{8}g'^2v^2B_\mu B^\mu \, , \\
			&\frac{v^2}{4} \left\langle-gg' U^\dagger W_\mu U B^\mu T^3+\textrm{h.c.}\right\rangle =-\frac{v^2gg'}{4}B^\mu W_\mu^3+\frac{gg'v}{2}B^\mu\left(W_\mu^1\varphi^2-W_\mu^2\varphi^1\right)  \\
			&\qquad\qquad\qquad\qquad\qquad\qquad\qquad\qquad-\frac{gg'}{2}B^\mu \left((\varphi^a W^a_\mu )\varphi^3-(\varphi^a \varphi^a)W^3_\mu\right) \, .
		\end{split}
	\end{equation}
	By expanding the function $F_U(h_k)$ around the background with $h_k\longrightarrow h_k+\tilde{h}_k$, we obtain
	
	\begin{eqnarray}
		F_U(h_k)\longrightarrow F_U(h_k)+ F^{(i)}_U(h_k)\tilde{h}_i+\frac{1}{2}F^{(ij)}_U(h_k)\tilde{h}_i\tilde{h}_j+... \, ,
	\end{eqnarray}
	where $F^{(ij...)}_U=\partial F_U/(\partial h_i\, \partial h_j...)$.
	Therefore, we have
	
	\begin{equation}
		\begin{split}
			\frac{v^2}{4}\Tr\left(\left(D_\mu U\right)^\dagger D^\mu U\right)F_U(h_k) &\overset{\mathcal{O}(\varphi^2)}{=} \varphi^a \ F_U(h) \Bigg[-\frac{1}{2}\partial^2 \delta^{ab}-\frac{g}{2}\epsilon^{cab}W^c_\mu \partial^\mu-\frac{g'}{2}B^\mu(\delta^{a1}\delta^{b2}-\delta^{a2}\delta^{b1})\partial_\mu  \\
			&\quad\quad-\frac{gg'}{2}B^\mu\left(W^a_\mu \delta^{b3}-W^3_\mu \delta^{ab}\right)\Bigg]\varphi^b \, , \\
			\frac{v^2}{4}\Tr\left(\left(D_\mu U\right)^\dagger D^\mu U\right)F_U(h_k) &\overset{\mathcal{O}(\tilde{h}\varphi)}{=}\tilde{h}_i vF^{i}_U(h) \Bigg[ \frac{g}{2}W^{\mu a}\partial_\mu-\frac{g'}{2}B^\mu\delta^{a3}\partial_\mu + \frac{gg'}{2}B^\mu\left(W^1_\mu\delta^{a2}-W^2_\mu \delta^{a1}\right)\Bigg]\varphi^a \ , \\
			\frac{v^2}{4}\Tr\left(\left(D_\mu U\right)^\dagger D^\mu U\right)F_U(h) &\overset{\mathcal{O}(\tilde{h}^2)}{=}\tilde{h}_iv^2\Bigg[  \frac{1}{8}g^2W_\mu^a W^{\mu a}+\frac{1}{8}g'^2B_\mu B^\mu -\frac{gg'}{4}B^\mu W_\mu^3\Bigg]\frac{1}{2}F^{(ij)}_U(h_k)\, \tilde{h}_j \ .
		\end{split}
		\label{qu1}
	\end{equation}
	
	\item Second, we consider the light Higgs terms:
	
	\begin{equation}\label{qu2}
		\begin{split}
			\frac{1}{2}\left(\delta_{ij}+f_{ij}(h_k)\right)\partial_\mu h_i\,\partial^\mu h_j&-V(h_k)\overset{\mathcal{O}(\tilde{h}^2)}{=}-\frac{1}{2}\tilde{h}_i\Bigg[ (\delta_{ij}+f_{ij}(h))\partial^2+\left(f^{(j)}_{ik}-f^{(i)}_{jk}\right)(\partial_\mu h_k)\,\partial^\mu \\
			& + V^{(ij)}-\frac{1}{2}f^{(ij)}_{kl}(\partial_\mu h_k)(\partial^\mu h_l) -\partial_\mu\left[(f^{(k)}_{ij}-f^{(i)}_{jk}-f^{(j)}_{ik})\partial^\mu h_k\right] \Bigg] \tilde{h}_j  \ .
		\end{split}
	\end{equation}
	
	\item  Third, we expand the Yukawa terms:
	
	\begin{equation}
		\begin{split}
			\bar{\psi}_L &UM(h) \psi_R  =\bar{\psi}_L \Bigg[\frac{1}{2}M^{(ij)}(h)\tilde{h}_i\tilde{h}_j+\frac{2i}{v}\varphi M^{(i)}(h)\tilde{h}_i -\frac{2}{v^2}\varphi\varphi M(h)\Bigg]\psi_R\\
			&= \tilde{h}_i \Big[ \frac{1}{2}\bar{\psi}_L M^{(ij)}(h) \psi_R\Big]\tilde{h}_j+\varphi^a \Big[ \frac{2i}{v}\bar{\psi}_L T^aM^{(i)}(h) \psi_R\Big]\tilde{h}_i+\varphi^a \Big[ \frac{-2}{v^2}\bar{\psi}_L T^a T^bM(h) \psi_R\Big]\varphi^b\,.
		\end{split}
	\end{equation}
	Therefore,
	
	\begin{equation}\label{qu3}
		\begin{split}
			-\bar{\psi}m(h,U)\psi&\overset{\mathcal{O}(\tilde{h}^2)}{=} -\tilde{h} \Big[ \frac{1}{2}\bar{\psi}_L M''(h) \psi_R+\textrm{h.c.}\,\Big]\tilde{h}\,,\\
			-\bar{\psi}m(h,U)\psi&\overset{\mathcal{O}(\tilde{h}\varphi)}{=}-\varphi^a \Big[ \frac{2i}{v}\bar{\psi}_L T^aM'(h) \psi_R+ \textrm{h.c.}\,\Big]\tilde{h}\,,\\
			-\bar{\psi}m(h,U)\psi&\overset{\mathcal{O}(\varphi^2)}{=}- \varphi^a \Big[ \frac{-2}{v^2}\bar{\psi}_L T^a T^bM(h) \psi_R+\textrm{h.c.}\,\Big]\varphi^b\,.
		\end{split}
	\end{equation}
\end{itemize}
From these results ((\ref{qu1}), (\ref{qu2}), and (\ref{qu3})) the differential operator for the quadratic scalar fluctuations, $\Delta$, can be read off easily.
Its components are:

\begin{equation}
\begin{split}
\Delta^{ab} &= F_U \Bigg[  \delta^{ab}\left( \partial^2 + \frac{1}{F_U}(\partial^\mu F_U)\partial_\mu\right) 
+ \left( g\epsilon^{cab}W^{\mu c} +g'B^\mu \epsilon^{3ab} \right) \partial_\mu +gg'B^\mu\left(W_\mu^a \delta^{b3}-W^3_\mu \delta^{ab}\right) \Bigg]\\
& \quad +\left[\frac{-4}{v^2}\bar{\psi}_L T^a T^b M \psi_R+\textrm{h.c.}\right]\\
&\overset{PI}{=} F_U \delta^{ab} \partial^2 
+ F_U\left( g\epsilon^{cab}W^{\mu c} +g'B^\mu \epsilon^{3ab} \right) \partial_\mu +F_U gg'B^\mu\left(\frac{1}{2}W_\mu^a \delta^{b3}+\frac{1}{2}W_\mu^b \delta^{a3}-W^3_\mu \delta^{ab}\right) \\
&\quad\quad +\delta^{ab}\left[ - \frac{1}{2}(\partial^2 F_U)+\frac{-1}{v^2}\bar{\psi}_L M \psi_R+\textrm{h.c.}\right] \, ,
\end{split}
\end{equation}
\begin{equation*}
\begin{split}
\Delta^{ai} &= \left[  \frac{4i}{v}\bar{\psi}_L T^aM^{(i)}\psi_R+ \textrm{h.c.} \right]  -  F_U^{(i)} vgg'B^\mu\left(W^1_\mu\delta^{a2}-W^2_\mu \delta^{a1}\right)\\
&\overset{PI}{=} \left[  \frac{2i}{v}\bar{\psi}_L T^aM^{(i)}\psi_R+ \textrm{h.c.} \right]  -  \frac{F_U^{(i)}}{2} vgg'B^\mu\left(W^1_\mu\delta^{a2}-W^2_\mu \delta^{a1}\right)+ \frac{1}{4} \partial_\mu \left[F^{(i)}_U v\left( gW^{\mu a}-g'B^\mu\delta^{a3} \right)\right]\\
&\quad\quad+   \frac{F^{(i)}_U}{2} v\left( gW^{\mu a}-g'B^\mu\delta^{a3} \right)\partial_\mu \, , \\
\Delta^{ia} &= - F^{(i)}_U v\left( gW^{\mu a}-g'B^\mu\delta^{a3} \right)\partial_\mu\\
&\overset{PI}{=} \left[  \frac{2i}{v}\bar{\psi}_L T^aM^{(i)}\psi_R+ \textrm{h.c.} \right]  -  \frac{F_U^{(i)}}{2} vgg'B^\mu\left(W^1_\mu\delta^{a2}-W^2_\mu \delta^{a1}\right)+ \frac{1}{4} \partial_\mu \left[F^{(i)}_U v\left( gW^{\mu a}-g'B^\mu\delta^{a3} \right)\right]\\
&\quad\quad- \frac{F^{(i)}_U}{2} v\left( gW^{\mu a}-g'B^\mu\delta^{a3} \right)\partial_\mu\,,\\
\Delta^{ij} &=(\delta_{ij}+f_{ij}(h))\partial^2+\left(f^{(j)}_{ik}-f^{(i)}_{jk}\right)(\partial_\mu h_k)\,\partial^\mu +\Bigg[ \left[ \bar{\psi}_L M^{(ij)} \psi_R+\textrm{h.c.}\,\right] + V^{(ij)}-\frac{1}{2}f^{(ij)}_{kl}(\partial_\mu h_k)(\partial^\mu h_l) \\
&\quad -\partial_\mu\left[(f^{(k)}_{ij}-f^{(i)}_{jk}-f^{(j)}_{ik})\partial^\mu h_k\right]   
-\left(  \frac{1}{8}g^2W_\mu^a W^{\mu a}+\frac{1}{8}g'^2B_\mu B^\mu -\frac{gg'}{4}B^\mu W_\mu^3\right)v^2 F_U^{(ij)}\Bigg] \\
&\doteq A^{ij}\partial^2+2B^\mu_{ij}\partial_\mu +C_{ij}\, .
\end{split}
\end{equation*}
It is useful to introduce the shorthand

\begin{equation}
C^a_\mu = gW^a_\mu -g' B^a_\mu \, ,
\end{equation}
where $B^a_\mu = B_\mu \delta^{a3}$.
It has advantageous properties under the inverse Stueckelberg transformation as is shown in the following subsection.
Defining the new index $\alpha=\{a,i\}$ the above differential operator for quadratic fluctuations can be recast into the form

\begin{equation}\label{eq:deltaalpha}
\Delta^{\alpha\beta}=\mathbf{A}^{\alpha\beta}\partial^2+2\mathbf{B}^{\alpha\beta}\partial_\mu+\mathbf{C}^{\alpha\beta}\,,
\end{equation}
where now

\begin{equation}\label{abc}
\begin{split}
\mathbf{A}^{ab} &= F_U \delta^{ab},\qquad \mathbf{A}^{ij} = A^{ij}, \qquad \mathbf{A}^{ai} = 0 \, , \\
\mathbf{B}_\mu^{ab}& = \frac{F_U}{2}\epsilon^{abc}\left(g W^c_\mu + g' B_\mu^c\right),\qquad \mathbf{B}_\mu^{ai} =\frac{vF_U^{(i)}}{4}{C^a_\mu}= -\mathbf{B}_\mu^{ia},\qquad \mathbf{B}_\mu^{ij} = B_\mu^{ij} \, , \\
\mathbf{C}^{ab} &=F_U gg'B^\mu\left(W_\mu^a \delta^{b3}-W^3_\mu \delta^{ab}\right) +\delta^{ab}\left[ - \frac{1}{2}(\partial^2 F_U)+\frac{-1}{v^2}\bar{\psi}_L M \psi_R+\textrm{h.c.}\right] \, , \\
\mathbf{C}^{ai} &=\left[  \frac{2i}{v}\bar{\psi}_L T^aM^{(i)}\psi_R+ \textrm{h.c.} \right]  -  \frac{F_U^{(i)}}{2} vgg'B^\mu\left(W^1_\mu\delta^{a2}-W^2_\mu \delta^{a1}\right)+ \frac{1}{4} \partial^\mu \left[F^{(i)}_U vC^a_\mu\right]=\mathbf{C}^{ia} \, , \\
\mathbf{C}^{ij} &= C^{ji} \, .
\end{split}
\end{equation}
Note that the matrices $\mathbf{A}$ and $\mathbf{C}$ are symmetric, while $\mathbf{B}$ is antisymmetric.

\subsection{Extracting the Divergences}
As the operator $\Delta$ as given in \eqref{eq:deltaalpha} does not take the standard form for use in the t'Hooft master formula \eqref{eqn:tHooft}, we can not directly extract the divergences.
Field redefinitions can cure this again.
To see this in detail, by introducing $\phi^\alpha=(\varphi^a, \tilde{h}^i)$ the LO Lagrangian that contains quadratic fluctuations of scalars can be rewritten as

\begin{equation}
\begin{split}\label{ns}
\mathcal{L}_{ew\chi ,N,\text{eff}}^{\text{1loop}} &\overset{\mathcal{O}(\phi^2)}{=} -\frac{1}{2}\phi^T \left[\mathbf{A}(h)\partial^2+2\mathbf{B}^\mu(h)\partial_\mu+\mathbf{C} \right]\phi \, .
\end{split}
\end{equation}
Since $\mathbf{A}$ is symmetric it can be written as $\mathbf{A}=\left(\mathbf{A}^{1/2}\right)^T\mathbf{A}^{1/2 }$.
We assume that $\mathbf{A}^{1/2 }$ is invertible and denote its inverse as $\mathbf{A}^{-1/2 }$.
We define a new field $\eta$ as

\begin{equation}
\phi = \mathbf{A}^{-1/2 } \eta \ .
\end{equation}
Thus we have 

\begin{equation}
\begin{split}
&\partial_\mu \phi = (\partial_\mu \mathbf{A}^{-1/2 }) \eta+ \mathbf{A}^{-1/2 }\partial_\mu \eta \ , \\
& \partial^2 \phi = (\partial^2 \mathbf{A}^{-1/2 })\eta +2(\partial_\mu \mathbf{A}^{-1/2 })\partial^\mu \eta + \mathbf{A}^{-1/2 } \partial^2\eta \ .
\end{split}
\end{equation}
The quadratic fluctuations become

\begin{equation}
\begin{split}
\mathcal{L}_{ew\chi ,N,\text{eff}}^{\text{1loop}} &\overset{\mathcal{O}(\eta^2)}{=} -\frac{1}{2}\eta^T \Bigg[\partial^2+2\left(\mathbf{A}^{1/2 } (\partial^\mu \mathbf{A}^{-1/2 })+\mathbf{A}^{-1/2 } B^\mu \mathbf{A}^{-1/2 }  \right)\partial_\mu\\
&\quad \quad\quad+ \mathbf{A}^{-1/2 } C\mathbf{A}^{-1/2 }+\mathbf{A}^{1/2 } (\partial^2 \mathbf{A}^{-1/2 })+2\mathbf{A}^{-1/2 } B^\mu (\partial_\mu \mathbf{A}^{-1/2 }) \Bigg]\eta\,.
\end{split}
\end{equation}
Now, we perform a partial integration to antisymmetrize the coefficient of $\partial_\mu$ and symmetrize the constant terms:

\begin{equation}
\begin{split}
\mathcal{L}_{ew\chi ,N,\text{eff}}^{\text{1loop}} &\overset{\mathcal{O}(\eta^2)}{=} -\frac{1}{2}\eta^T \Bigg[\partial^2+2\left(\mathbf{A}^{-1/2 } B^\mu \mathbf{A}^{-1/2 } +\frac{1}{2}\mathbf{A}^{1/2 } (\partial^\mu \mathbf{A}^{-1/2 })- \frac{1}{2}(\partial^\mu \mathbf{A}^{-1/2 }) \mathbf{A}^{1/2 }  \right)\partial_\mu+ \mathbf{A}^{-1/2 } C\mathbf{A}^{-1/2 } \\
&\quad +\mathbf{A}^{-1/2 } B^\mu (\partial_\mu \mathbf{A}^{-1/2 }) -   (\partial_\mu \mathbf{A}^{-1/2 }) B^\mu  \mathbf{A}^{-1/2 } -\frac{1}{2}  \left(\partial_\mu \mathbf{A}^{-1/2 } \,\partial^\mu \mathbf{A}^{1/2 }+  \partial_\mu \mathbf{A}^{1/2 } \,\partial^\mu \mathbf{A}^{-1/2 }\right)\Bigg]\eta \\
&  \doteq -\frac{1}{2}\eta^T \left[(\partial_\mu+X_\mu)^2+Y \right]\eta \, ,
\end{split}
\end{equation}
where we identify

\begin{equation}\label{XY}
\begin{split}
&X^\mu = \mathbf{A}^{-1/2 } B^\mu \mathbf{A}^{-1/2 } +\frac{1}{2}\mathbf{A}^{1/2 } (\partial^\mu \mathbf{A}^{-1/2 })- \frac{1}{2}(\partial^\mu \mathbf{A}^{-1/2 }) \mathbf{A}^{1/2 } \ , \\
&Y =\mathbf{A}^{-1/2 } C\mathbf{A}^{-1/2 } +\mathbf{A}^{-1/2 } B^\mu (\partial_\mu \mathbf{A}^{-1/2 }) -   (\partial_\mu \mathbf{A}^{-1/2 }) B^\mu  \mathbf{A}^{-1/2 }\\
&\qquad -\frac{1}{2}  \left(\partial_\mu \mathbf{A}^{-1/2 } \,\partial^\mu \mathbf{A}^{1/2 }+  \partial_\mu \mathbf{A}^{1/2 } \,\partial^\mu \mathbf{A}^{-1/2 }\right)-X_\mu X^\mu\doteq Z-X_\mu X^\mu \ .
\end{split}
\end{equation}
With this choice of $X_\mu$ and $Y$, one can bring the Lagrangian (\ref{ns}) into the desired canonical form, with $\phi$ replaced by $\eta$. 
Next we define $A^{1/2}$ and its inverse $A^{-1/2}$ such that $A=\left(A^{1/2}\right)^T A^{1/2}$.
In terms of these newly defined matrices, $\mathbf{A}^{1/2}$ and its inverse can be obtained right away from (\ref{abc}) as

\begin{equation}
\begin{split}
\mathbf{A}^{1/2}_{ab} = \sqrt{F_U}\delta^{ab},\qquad  \mathbf{A}^{1/2}_{ij} = A^{1/2}_{ij}, \qquad \mathbf{A}^{1/2}_{ai}=0 \ , \\
\mathbf{A}^{-1/2}_{ab} = \frac{1}{\sqrt{F_U}}\delta^{ab},\qquad  \mathbf{A}^{-1/2}_{ij} = A^{-1/2}_{ij}, \qquad \mathbf{A}^{-1/2}_{ai}=0 \ .
\end{split}
\end{equation}
Using (\ref{XY}) we obtain

\begin{equation}
\begin{split}
X_\mu^{ab} &= \frac{ 1}{2}\epsilon^{ab}_c\left( gW^c_\mu +g'B^c_\mu\right) = \frac{i}{2}\tilde{T}^{ab}_c \left( gW^c_\mu +g'B^c_\mu\right)  \ ,\\
X_\mu^{ai} &= \frac{vF^{(j)}}{4\sqrt{F_U}}C^a_\mu A_{ji}^{-1/2} \doteq \frac{1}{2}\kappa^i C^a_\mu \ ,\\
X_\mu^{ij} &= A_{ik}^{-1/2}B^\mu_{kl} A_{lj}^{-1/2}+\frac{1}{2} A_{ik}^{1/2} \partial_\mu A_{k j}^{-1/2}-\frac{1}{2}\partial_\mu A_{ik}^{-1/2} A_{k j}^{1/2}\doteq \frac{1}{2}\mathcal{X}_\mu^{ij} \ ,
\end{split}
\end{equation}	
where

\begin{equation}
\kappa_i= \frac{vF^{(j)}_U}{2\sqrt{F_U}} A^{-1/2}_{ji} \, .
\end{equation}
Employing these, we can calculate the field strength tensor $X_{\mu\nu}$ \eqref{eq:Xmunu} of the t'Hooft master formula.
The result is:

\begin{equation}
\begin{split}
X_{\mu\nu}^{ab} &= \frac{1}{2}i\tilde{T}^{ab}_c \left(gW^c_{\mu\nu}+g'B^c_{\mu\nu}\right)-\frac{1}{4}\left(\kappa^2-1\right)\left(C^a_\mu C^b_\nu-C^a_\nu C^b_\mu\right) \, , \\
X_{\mu\nu}^{ai} &= \frac{1}{2} (\partial_\mu \kappa^i)C^a_\nu-\frac{1}{2} (\partial_\nu \kappa^i)C^a_\mu+\frac{\kappa^i}{2}\left(gW^a_{\mu\nu}-g'B^a_{\mu\nu}\right) -\frac{1}{4} \kappa_j\left(C^a_\mu\mathcal{X}^{ji}_{\nu}-C^a_\nu\mathcal{X}^{ji}_{\mu}\right) \, , \\
X_{\mu\nu}^{ij} &= \frac{1}{2}\partial_\mu \mathcal{X}^{ij}_{\nu}- \frac{1}{2}\partial_\nu \mathcal{X}^{ij}_{\mu}+\frac{1}{4}\left[\mathcal{X}^{i}_{\mu k} \mathcal{X}^{kj}_\nu-\mathcal{X}^{i}_{\nu k}  \mathcal{X}^{kj}_\mu   \right] \, ,
\end{split}
\end{equation}
with

\begin{equation}\label{eq:Ttilde}
\kappa^2=\sum_i \kappa_i^2 \textrm{ and } i\tilde{T}^c_{\ ab}=\epsilon_{\ ab}^{ c} \, .
\end{equation}

Having obtained these results, we are ready to calculate the divergences from $\text{tr} X_{\mu\nu} X^{\mu\nu}$.
However, it is useful to discuss first how the inverse Stueckelberg transformation (IST) can be performed to restore the dependency of the effective action on the background $\hat{U}$.
Recall the Stueckelberg transformation

\begin{equation}
W_\mu \longrightarrow U W_\mu U^\dagger - \frac{i}{g} U\partial_\mu U^\dagger \, .
\end{equation}
Thus, the IST reads 

\begin{equation}
W_\mu \overset{\text{IST}}{\longrightarrow} U^\dagger W_\mu U +\frac{i}{g}(\partial_\mu U^\dagger) U \, .
\end{equation}
In the computation of the one loop effective action it is instructive to use notations which make the gauge symmetry manifest.
We define

\begin{equation}
C_\mu = T^a C^a_\mu \, ,
\end{equation}
where $T^a=\sigma^a/2$ are the usual $SU(2)$ generators from \eqref{eq:su2gen}.
We summarize some useful relations:

\begin{equation}
\left\langle T^aT^b\right\rangle = \frac{1}{2}\delta^{ab}\,,\quad\quad \left\langle T^a T^b T^c\right \rangle = -\frac{1}{4i} \epsilon^{abc} \,,\quad\quad \left\langle \tilde{T}^a\tilde{T}^b\right\rangle= 2\delta^{ab} \, ,
\end{equation}
with the generators of the adjoint representation $\tilde{T}^a$ defined in \eqref{eq:Ttilde}.
The nice thing about $C_\mu$ is that under the IST it transforms into $U^\dagger L_\mu U$ with $L_\mu = i U(D_\mu U)^\dagger$ as defined in Section \ref{sec:ewclh}.
The proof is straightforward:

\begin{equation}
C_\mu \overset{IST}{\longrightarrow} gU^\dagger W_\mu U+i(\partial_\mu U^\dagger) U-g'B_\mu T^3 =i\left(D_\mu U\right)^\dagger U=U^\dagger L_\mu U \, .
\end{equation}
Since $L_\mu$ transforms under the electroweak symmetry as $L_\mu \longrightarrow S_L L_\mu S_L^\dagger$, $C_\mu$ transforms as $C_\mu\longrightarrow S_Y C_\mu S_Y^\dagger$, i.e, it transforms in the adjoint representation of $U(1)_Y$.
Thus, it is useful to define the covariant derivative acting on $C_\mu$ as 

\begin{equation}
\left(D_\mu C_\nu\right)^a = \left(\partial_\mu+i\tilde{T}_3 B_\mu\right)^a_{\ b} C^b_\nu=\partial_\mu C^a_\nu +g'\epsilon^{3a}_{\ \ b} B_\mu C^b_\nu = \partial_\mu C^a_\nu +gg'\epsilon^{3a}_{\ \ b} B_\mu W^b_\nu \ .
\end{equation}
This equation is going to be useful in the computation of $ \text{tr}\, Y^2$. 
Employing the fact that $X_\mu$ is an antisymmetric matrix, we can make use of the following equation:

\begin{equation}\label{rel}
\text{tr}\, X_{\mu\nu} X^{\mu\nu} = X_{ab} X^{ba}-2X_{ai}X^{ai}-X_{ij}X^{ij} \ .
\end{equation}
Note that the Greek indices are spacetime components and the Roman indices the $SU(2)_L$ gauge group components.
We compute the divergences as follows:

\begin{equation}
\begin{split}
X^{ab} X^{ba} &=-\frac{1}{4}(gW^c_{\mu\nu}+g'B^c_{\mu\nu})(gW^{\mu\nu d}+g'B^{\mu\nu d})\left\langle \tilde{T}_c \tilde{T}_d\right\rangle +\frac{\kappa^2-1}{2}(gW^c_{\mu\nu}+g'B^c_{\mu\nu})\epsilon_{cab} C^{\mu a} C^{\nu b}\\
&\qquad-\frac{(\kappa^2-1)^2}{8} C^{\mu a} C^{\nu b} \left(C^a_\mu C^b_\nu -C^a_\nu C^b_\mu\right)\\
&\overset{IST}{\longrightarrow} - g^2 \left \langle W_{\mu\nu} W^{\mu\nu}\right\rangle -\frac{1}{2}g^2 B_{\mu\nu} B^{\mu\nu} -2gg'\left\langle W_{\mu\nu} \tau_L\right\rangle B^{\mu\nu}-i\left(\kappa^2-1\right)\Big( g\left \langle W_{\mu\nu} \left[L^\mu, L^\nu \right]\right\rangle\\
&\qquad+g'B_{\mu\nu}\left\langle \tau_L \left[L^\mu, L^\nu \right]\right\rangle\Big)-\frac{(\kappa^2-1)^2}{2}\Big(\left\langle L_\mu L^\mu\right\rangle^2-\left\langle L_\mu L_\nu\right\rangle^2\Big) \, ,
\end{split}
\end{equation}

\begin{equation*}
\begin{aligned}
-2X^{ai} X^{ai} &= -2 \Bigg[ \frac{1}{2}\left((\partial_\mu \kappa^i) C^a_\nu-(\partial_\nu \kappa^i) C^a_\mu\right)(\partial^\mu \kappa^i)C^{\nu a}+\frac{\kappa^2}{4}(gW^a_{\mu\nu}-g'B^a_{\mu\nu})^2\\
&\qquad +\frac{\kappa_j \kappa_k}{16} (C^a_\mu \mathcal{X}^{ji}_{\nu}-C^a_\nu \mathcal{X}^{ji}_{ \mu})(C^{\mu a} \mathcal{X}^{\nu ki}-C^{\nu a} \mathcal{X}^{\mu ki}) +\kappa^i(\partial^\mu \kappa^i) C^{\nu a} (gW^a_{\mu\nu}-g'B^a_{\mu\nu}) \\
&\qquad-\frac{\kappa_j}{2}(\partial^\mu \kappa^i) C^{\nu a} (C^a_\mu \mathcal{X}^{ji}_{\nu}-C^a_\nu \mathcal{X}^{ji}_{\mu})-\frac{\kappa_i\kappa_j}{2}(gW^a_{\mu\nu}-g'B^a_{\mu\nu})C^{\mu a} \mathcal{X}^{\nu ji} \Bigg]
\end{aligned}
\end{equation*}

\begin{equation*}
\begin{aligned}
&\overset{IST}{\longrightarrow} -2(\partial_\mu\kappa^i)^2\left\langle L_\mu L^\mu\right\rangle +2 (\partial_\mu \kappa^i)(\partial_\nu \kappa^i)\left\langle L^\mu L^\nu\right\rangle -\kappa^2g^2 \left\langle W_{\mu\nu} W^{\mu\nu}\right\rangle\\
&\qquad-\frac{\kappa^2}{2}g'^2B_{\mu\nu}B^{\mu\nu}+2\kappa^2 gg' \left\langle W_{\mu\nu}\tau_L\right\rangle B^{\mu\nu}-\frac{\kappa_j\kappa_k}{2}\Big(\left\langle L_\mu L^\mu\right\rangle \mathcal{X}^{ji}_{\mu}\mathcal{X}^{\mu ki}-\left\langle L^\mu L^\nu\right\rangle \mathcal{X}^{ji}_{\mu}\mathcal{X}^{ki}_\nu\Big) \\
&\qquad -4\kappa^i(\partial_\mu \kappa^i)\Big( g\left\langle W^{\mu\nu}L_\nu\right\rangle -g'B^{\mu\nu}\left\langle \tau_L L_\nu\right\rangle\Big)-2\kappa_j \mathcal{X}^{ji}_\mu (\partial^\mu \kappa^i)\left\langle L_\nu L^\nu\right\rangle\\
&\qquad+\kappa_j\left(\mathcal{X}^{ji}_\mu (\partial_\nu\kappa^i)+\mathcal{X}^{ji}_\nu (\partial_\mu\kappa^i)\right) \left\langle L^\mu L^\nu \right\rangle+2\kappa^i\kappa_j  \mathcal{X}^{ji}_{\nu}\Big(g\left\langle L_\mu W^{\mu\nu} \right\rangle-\left\langle L_\mu \tau_L \right\rangle g'B^{\mu\nu}\Big) \, , \\
-X^{ij} X^{ij} & =-\Bigg[ \frac{1}{2}\partial_\mu \mathcal{X}^{ij}_{\nu}- \frac{1}{2}\partial_\nu \mathcal{X}^{ij}_{\mu}+\frac{1}{4}\left[\mathcal{X}^{ik}_\mu  \mathcal{X}^{kj}_\nu-\mathcal{X}^{ik}_\nu  \mathcal{X}^{kj}_\mu   \right]\Bigg]^2 \, .
\end{aligned}
\end{equation*}
Using (\ref{rel}), we therefore obtain:\footnote{Note that $L_\mu$ is real, namely $L_\mu^\dagger =L_\mu$. So $\langle L_\mu L^\mu \rangle = \langle L_\mu^\dagger L^\mu \rangle$.}

\begin{equation}\label{x}
\begin{aligned}
\frac{1}{12}\text{tr}\, X_{\mu\nu} X^{\mu\nu}& =  -\frac{1+\kappa^2}{12}g^2\left\langle W_{\mu\nu} W^{\mu\nu}\right\rangle -\frac{1+\kappa^2}{24}g'^2 B_{\mu\nu} B^{\mu\nu} +\frac{ \kappa^2-1}{6}gg'\left\langle W_{\mu\nu} \tau_L\right\rangle B^{\mu\nu}\\
&-i\frac{\kappa^2-1 }{12}\Big(g\left\langle W_{\mu\nu} \left[L^\mu, L^\nu \right]\right\rangle+g'B_{\mu\nu}\left\langle \tau_L \left[L^\mu, L^\nu \right]\right\rangle\Big) -\frac{(\kappa^2-1)^2}{24}\Big(\left\langle L_\mu L^\mu\right\rangle^2-\left\langle L_\mu L_\nu\right\rangle^2\Big)\\
&+\frac{1}{12}\left(-2(\partial_\mu\kappa_i)^2 -\frac{\kappa_i\kappa_j}{2} \mathcal{X}^{ji}_{\nu}\mathcal{X}^{ji}_{\nu}-2\kappa_j \mathcal{X}^{ji}_{\mu} (\partial_\mu \kappa_i)  \right) \left\langle L_\mu L^\mu\right\rangle +\frac{1}{12}\Big(2 (\partial_\mu \kappa_i)(\partial_\nu \kappa^i) \\
& +\frac{\kappa_j\kappa_k}{2}\mathcal{X}^{ji}_{\mu}\mathcal{X}^{ki}_\nu +\kappa_j\left(\mathcal{X}^{ji}_\mu (\partial_\nu\kappa_i)+\mathcal{X}^{ji}_\nu (\partial_\mu\kappa_i)\right)    \Big)\left\langle L^\mu L^\nu\right\rangle-\frac{1}{3}\kappa_i(\partial_\mu \kappa^i) \Big( g\left\langle W^{\mu\nu}L_\nu\right\rangle\\
& -g'B^{\mu\nu}\left\langle \tau_L L_\nu\right\rangle\Big)  - \frac{1}{12}\Bigg[ \frac{1}{2}\partial_\mu \mathcal{X}^{ij}_{\nu}- \frac{1}{2}\partial_\nu \mathcal{X}^{ij}_{\mu}+\frac{1}{4}\left[\mathcal{X}^{ik}_\mu  \mathcal{X}^{kj}_\nu-\mathcal{X}^{ik}_\nu  \mathcal{X}^{kj}_\mu   \right]\Bigg]^2 \ .
\end{aligned}
\end{equation}

Next, we compute divergences coming from $\text{tr} Y^2$.
First, we calculate $X_\mu X^\mu$:

\begin{equation}
\begin{split}
(X^\mu X_\mu)^{ab} &= -\frac{1}{4} (gW+g'B)_\mu^c (gW+g'B)^{\mu d} \left(\tilde{T}_c \tilde{T}_d\right)^{ab}-\frac{1}{4}\kappa^2C^a_\mu C^{\mu b} \ , \\
(X^\mu X_\mu)^{ai} &= -\frac{\kappa_i}{4}\epsilon^{a}_{\ cd} (gW+g'B)^c_\mu C^{\mu d}-\frac{1}{4}\kappa_j C^{\mu a} \mathcal{X}^{ji}_\mu \ , \\
(X^\mu X_\mu)^{ij} &= -\frac{1}{4} \kappa^i\kappa^j C^a_\mu C^{\mu a} -\frac{1}{4}\mathcal{X}^{ik}_\mu \mathcal{X}^{\mu jk}\ .
\end{split}
\end{equation}
We then calculate $Z$ using (\ref{XY}):

\begin{equation}
\begin{split}
Z^{ab} &= gg'B^\mu\left(\frac{1}{2}W_\mu^a \delta^{b3}+\frac{1}{2}W_\mu^b \delta^{a3}-W^3_\mu \delta^{ab}\right) +\delta^{ab}\left[\frac{(\partial_\mu F_U)^2}{4F_U^2} - \frac{1}{2}\frac{\partial^2 F_U}{F_U}-\frac{1}{v^2F_U}\bar{\psi}_L M \psi_R+\textrm{h.c.}\right] \, ,
\end{split}
\end{equation}

\begin{equation}
\begin{split}
Z^{ai} &= \frac{1}{\sqrt{F_U}}\Bigg[\left[  \frac{2i}{v}\bar{\psi}_L T^aM^{(j)}\psi_R+ \textrm{h.c.} \right]  -  \frac{F_U^{(j)}}{2} vgg'B^\mu\left(W^1_\mu\delta^{a2}-W^2_\mu \delta^{a1}\right)\Bigg] A_{j i}^{-1/2}\\
&+\partial^\mu\left(\frac{vF^{(j)}_U}{4\sqrt{F_U}} C^a_\mu   A_{j i}^{-1/2}  \right) +\frac{vF^{(j)}_U}{2\sqrt{F_U}} C^a_\mu    A_{j i}^{-1/2}\frac{\partial^\mu F_U}{2F_U} \, , \\
Z^{ij} &=  A_{im}^{-1/2} \Bigg[ \left[ \bar{\psi}_L M^{(mn)} \psi_R+\textrm{h.c.}\,\right]+V^{(mn)}-\partial_\mu\left[(f^{(k)}_{mn}-f^{(m)}_{nk}-f^{(n)}_{mk})\partial^\mu h^k\right] \\
&-\frac{1}{2}f^{(mn)}_{kl}(\partial_\mu h^k)(\partial^\mu h^l) \Bigg]  A_{nj}^{-1/2} \\
& - A_{im}^{-1/2}\left(  \frac{1}{8}g^2W_\mu^a W^{\mu}_a +\frac{1}{8}g'^2B_\mu B^\mu -\frac{gg'}{4}B^\mu W_\mu^3\right)v^2 F_U^{(mn)}  A_{nj}^{-1/2}\\
&+A^{-1/2}_{ik} B^\mu_{kl} \partial_\mu A^{-1/2}_{lj}-(\partial_\mu A^{-1/2}_{ik})B^\mu_{kl} A^{-1/2}_{lj}-\frac{1}{2}\left[\partial_\mu A^{-1/2}_{ik}\,\partial^\mu A^{1/2}_{kj}+\partial_\mu A^{1/2}_{ik}\,\partial^\mu A^{-1/2}_{kj}\right] \, .
\end{split}
\end{equation}

For further computations, we make use of the equations of motion of the scalar fields.
%set the background fields to coincide with their equation of motion solution.
The equations of motion for the Higgses and Goldstones are given by

\begin{equation}
\begin{split}
0&=- \partial^2 h_i + (\partial h_m)(\partial h_n) A^{-1}_{ij} \left(f^{(j)}_{mn}-\frac{1}{2}f^{(n)}_{jm}-\frac{1}{2}f^{(m)}_{jn}\right)+\frac{v^2}{8} A^{-1}_{ij} F^{(j)}_U C_\mu^a C^\mu_a-A^{-1}_{ij} V^{(j)}-A^{-1}_{ij} \bar{\psi} \tilde{m}^{(j)}\psi  \, , \\
0&= D^\mu (C_\mu^a F_U) + \frac{4}{v^2} \left[ i\bar{\psi}_L T^aM\psi_R+ \textrm{h.c.} \right] \, ,
\end{split}
\end{equation}
where

\begin{equation}
\tilde{m} \doteq M(h_k)P_R+ M^\dagger(h_k)P_L \, .
\end{equation}
Using the relation $Y= Z-X_\mu X^\mu$ and the above equations gives us after some straightforward calculations

\begin{equation}
\begin{split}
Y^{ab} &=\delta^{ab}\Big[ (\partial h_i)(\partial h_j) D^{ij}+\mathcal{Y} -\frac{1}{4}\left(\kappa^2-1\right) C_\mu^c C^\mu_c \Big] + \frac{1}{4} (\kappa^2-1) C_\mu^a C^\mu_b \, , \\
Y^{ai} &=  \frac{2A^{-1/2}_{ji} }{v\sqrt{F_U}} \left[  i\bar{\psi}_L T^aM^{(j)}\psi_R+ \textrm{h.c.} \right]  +\frac{1}{4}\kappa_j C^{\mu a} \mathcal{X}^{ji}_\mu +\frac{1}{2} (\partial^\mu  \kappa^i ) C^a_\mu + \frac{\kappa^i}{2F_U}  D^\mu (C_\mu^a F_U) \, , \\
Y^{ij} &= \mathcal{S}^{ij}+\tilde{\mathcal{B}}^{ij}  C_\mu^a C^\mu_a +\tilde{\mathcal{Y}}^{ij} \ ,
\end{split}
\end{equation}
where the lenghty expressions for $\mathcal{X}^{ji}_\mu, \, \mathcal{S}^{ij}, \, \tilde{\mathcal{B}}^{ij}, \ \textrm{and}\ \tilde{\mathcal{Y}}^{ij}$ after the inverse Stueckelberg transformation are explicitly given below in (\ref{1}) and (\ref{3}).
The computation of $\text{tr} \, Y^2$ is very similar to what was done before.
Using the relation

\begin{equation}
\text{tr} \, Y^2= Y^{ab}Y^{ab} + 2 Y^{ai}Y^{ai} + Y^{ij}Y^{ij} \, ,
\end{equation}
we compute it term by term:

\begin{equation}\label{456}
\begin{split}
\frac{1}{2}Y^{ab}Y^{ab} &= \frac{1}{2}\Bigg\{3\left[ (\partial h_i)(\partial h_j) D^{ij}+\mathcal{Y} -\frac{1}{4}\left(\kappa^2-1\right) C_\mu^c C^\mu_c \right]^2 + \frac{1}{2} (\kappa^2-1) \Big[ (\partial h_i)(\partial h_j) D^{ij}+\mathcal{Y} \\
&\qquad-\frac{1}{4}\left(\kappa^2-1\right) C_\mu^c C^\mu_c \Big] C_\mu^a C^\mu_a +\frac{1}{16} (\kappa^2-1)^2 C^a_\mu C^{b\mu}  C^a_\nu C^{b\nu} 
\Bigg\}\\
&\overset{IST}{\longrightarrow}\frac{1}{8}\left(\kappa^2-1\right)^2\left[ \langle L_\mu L^\mu\rangle^2+ \langle L_\mu L_\nu\rangle^2 \right]-\left(\kappa^2-1\right)\left[ (\partial h_i)(\partial h_j) D^{ij}+\mathcal{Y}' \right]\langle L_\mu L^\mu \rangle\\
&\qquad + \frac{3}{2}\left[(\partial h_i)(\partial h_j) D^{ij}\right]^2+\frac{3}{2}\mathcal{Y}'^2+3\mathcal{Y}'(\partial h_i)(\partial h_j) D^{ij} \ ,
\end{split}
\end{equation}

\begin{equation}
\begin{aligned}
Y^{ai}Y^{ai} &= \frac{4}{v^2} \left( {A^{-1/2}_{ji}}\left[  i\bar{\psi}_L T^a \left( \frac{M}{\sqrt{F_U}} \right)^{(j)}  \psi_R+ \textrm{h.c.} \right]\right) ^2 + \frac{1}{16}\kappa^n \kappa^m (C^a_\mu \mathcal{X}_{ni}^\mu )(C^a_\mu \mathcal{X}_{mi}^\mu ) \\
& + \frac{1}{4} \partial^\mu \kappa_i \partial^\nu \kappa_i C_\mu^a C_\nu^a +  \frac{2 A^{-1/2}_{ji}}{v}\left[  i\bar{\psi}_L T^a \left( \frac{M}{\sqrt{F_U}} \right)^{(j)}  \psi_R+ \textrm{h.c.} \right](\partial_\mu \kappa^i) C_\mu^a\\
& + \frac{\kappa^n A^{-1/2}_{ji} }{v}\left[  i\bar{\psi}_L T^a \left( \frac{M}{\sqrt{F_U}} \right)^{(j)}  \psi_R+ \textrm{h.c.} \right]\mathcal{X}_{ni}^\mu C_\mu^a+ \frac{1}{4} \kappa^j (\partial_\mu \kappa^i)C^{\mu a} (\mathcal{X}_{ji}^\nu C_\nu^a) \\
\overset{IST}{\longrightarrow} \ & \frac{4}{v^2}\left( {A^{-1/2}_{ji}}\left[  i\bar{\psi}_L U \left( \frac{M}{\sqrt{F_U}} \right)^{(j)}  \psi_R+ \textrm{h.c.} \right]\right) ^2 + \frac{1}{8}\kappa^n \kappa^m \mathcal{X}_{ni}^\mu \mathcal{X}_{mi}^\nu \langle L_\mu L_\nu\rangle + \frac{1}{2} \partial^\mu \kappa^i \partial^\nu \kappa_i \langle L_\mu L_\nu \rangle \\
&+ \frac{2A^{-1/2}_{ji}}{v}\left[  i\bar{\psi}_L L_\mu U \left( \frac{M}{\sqrt{F_U}} \right)^{(j)}  \psi_R+ \textrm{h.c.} \right](\partial^\mu \kappa^i) + \frac{\kappa_n A^{-1/2}_{ji}}{v}\Big[  i\bar{\psi}_L L_\mu U \left( \frac{M}{\sqrt{F_U}} \right)^{(j)}  \psi_R\\
&+ \textrm{h.c.} \Big] \mathcal{X}_{ni}^\mu+ \frac{1}{4} \kappa^j \left( (\partial^\mu \kappa^i)\mathcal{X}_{ji}^\nu + (\partial^\nu \kappa^i) \mathcal{X}^\mu_{ji}\right)  \langle L_\mu L_\nu \rangle \ ,
\end{aligned}
\end{equation}

\begin{equation}
\begin{split}
\frac{1}{2} Y^{ij}Y^{ij} &=  \frac{1}{2} \mathcal{S}^{ij}\mathcal{S}^{ij}+\frac{1}{2} \tilde{\mathcal{B}}^{ij} \tilde{\mathcal{B}}^{ij}  C^a_\mu C^{a\mu} C^b_\nu C^{b\nu} +\frac{1}{2} \tilde{\mathcal{Y}}^{ij}\tilde{\mathcal{Y}}^{ij}+\mathcal{S}^{ij}\tilde{\mathcal{B}}^{ij}  C_\mu^a C^{\mu a} +\mathcal{S}^{ij}\tilde{\mathcal{Y}}^{ij}+\tilde{\mathcal{B}}^{ij} C_\mu^a C^{\mu a} \tilde{\mathcal{Y}}^{ij}\\
& \overset{IST}{\longrightarrow} \frac{1}{2} \mathcal{S}^{ij}\mathcal{S}^{ij}+2\tilde{\mathcal{B}}^{ij} \tilde{\mathcal{B}}^{ij} \langle L_\mu L^\mu\rangle^2+\frac{1}{2} \tilde{\mathcal{Y}}^{ij}\tilde{\mathcal{Y}}^{ij}+2\left[\mathcal{S}^{ij}\tilde{\mathcal{B}}^{ij}+\tilde{\mathcal{Y}  }^{ij} \tilde{\mathcal{B}}^{ij}\right] \langle L_\mu L^\mu\rangle+ \mathcal{S}^{ij}\tilde{\mathcal{Y}}^{ij} \ ,
\end{split}
\end{equation}
Therefore, the divergences that come from $\text{tr} \, Y^2$ read

\begin{equation}\label{y}
\begin{aligned}
\frac{1}{2}\text{tr}\, Y^2& =\Bigg[ \frac{1}{8}\left(\kappa^2-1\right)^2  +2\tilde{\mathcal{B}}^{ij} \tilde{\mathcal{B}}^{ij}  \Bigg]\langle L_\mu L^\mu\rangle^2+\frac{1}{8}\left(\kappa^2-1\right)^2 \langle L_\mu L_\nu\rangle^2 \\
&+\Bigg[-\left(\kappa^2-1\right)\left[ (\partial h_i)(\partial h_j) D^{ij}+\mathcal{Y}' \right]  +2\left[\mathcal{S}^{ij}\tilde{\mathcal{B}}^{ij}+\tilde{\mathcal{Y}  }^{ij} \tilde{\mathcal{B}}^{ij}\right]\Bigg]\langle L_\mu L^\mu \rangle\\
&+\Bigg[\frac{1}{8}\kappa_n \kappa_m \mathcal{X}_{ni}^\mu \mathcal{X}_{mi}^\nu  + \frac{1}{2} \partial^\mu \kappa_i \partial^\nu \kappa_i + \frac{1}{4} \kappa_j \left( (\partial^\mu \kappa_i)\mathcal{X}_{ji}^\nu + (\partial^\nu \kappa_i) \mathcal{X}^\mu_{ji}\right)  \Bigg]\langle L_\mu L_\nu \rangle  \\
& +\frac{3}{2}\mathcal{Y}'^2+3\mathcal{Y}'(\partial h_i)(\partial h_j) D^{ij} +\frac{4}{v^2}\left( A^{-1/2}_{ji}\left[  i\bar{\psi}_L U \left( \frac{M}{\sqrt{F_U}} \right)^{(j)}  \psi_R+ \textrm{h.c.} \right]\right) ^2 \\
&+ \frac{2A^{-1/2}_{ji}}{v}\left[  i\bar{\psi}_L L_\mu U \left( \frac{M}{\sqrt{F_U}} \right)^{(j)}  \psi_R+ \textrm{h.c.} \right](\partial^\mu \kappa_i) + \frac{\kappa_n A^{-1/2}_{ji} }{v}\Big[  i\bar{\psi}_L L_\mu U \left( \frac{M}{\sqrt{F_U}} \right)^{(j)}  \psi_R\\
&+ \textrm{h.c.} \Big] \mathcal{X}_{ni}^\mu +\frac{1}{2} \tilde{\mathcal{Y}}^{ij}\tilde{\mathcal{Y}}^{ij}+ \mathcal{S}^{ij}\tilde{\mathcal{Y}}^{ij} + \frac{1}{2} \mathcal{S}^{ij}\mathcal{S}^{ij}+ \frac{3}{2}\left[(\partial h_i)(\partial h_j) D^{ij}\right]^2 \ .
\end{aligned}
\end{equation}
%\newpage
Finally, we end up with the divergence formula for the EWCLNh;
the main result of the present work.
It is obtained by summing up (\ref{x}) and (\ref{y}):

\begin{equation}
\begin{aligned}
\mathcal{L}_{ew\chi ,N,\text{eff,div}}^{\text{1loop}}& = \frac{1}{32\pi^2 \epsilon}\Bigg\{
-\frac{1+\kappa^2}{12}g^2\left\langle W_{\mu\nu} W^{\mu\nu}\right\rangle -\frac{1+\kappa^2}{24}g'^2 B_{\mu\nu} B^{\mu\nu} +\frac{ \kappa^2-1}{6}gg'\left\langle W_{\mu\nu} \tau_L\right\rangle B^{\mu\nu} \\
&-i\frac{\kappa^2-1 }{12}\Big(g\left\langle W_{\mu\nu} \left[L^\mu, L^\nu \right]\right\rangle+g'B_{\mu\nu}\left\langle \tau_L \left[L^\mu, L^\nu \right]\right\rangle\Big) -\frac{1}{3}\kappa_i \kappa_i^{(n)} (\partial_\mu h_n) \Big( g\left\langle W^{\mu\nu}L_\nu\right\rangle \\
&-g'B^{\mu\nu}\left\langle \tau_L L_\nu\right\rangle\Big)+\Bigg[ \frac{1}{12}\left(\kappa^2-1\right)^2  +2\tilde{\mathcal{B}}^{ij} \tilde{\mathcal{B}}^{ij}  \Bigg]\langle L_\mu L^\mu\rangle^2+\frac{1}{6}\left(\kappa^2-1\right)^2 \langle L_\mu L_\nu\rangle^2 \\
&+\Bigg[\frac{1}{12}\left(-2(\kappa_i^{(n)} \partial_\mu h_n)^2 -\frac{\kappa_i\kappa_j}{2} \mathcal{X}^{ji}_{\nu}\mathcal{X}^{ji}_{\nu}-2\kappa_j \mathcal{X}^{ji}_{\mu} \kappa_i^{(n)} (\partial_\mu h_n)  \right) -\left(\kappa^2-1\right)\Big( (\partial h_i)(\partial h_j) D^{ij}\\
&+ \frac{1}{2 F_U} F^{(i)}_U A^{-1}_{ij} \left(V^{(j)}+[ \bar{\psi}_L U {M}^{(j)}\psi_R+\textrm{h.c.}]\right)  -\frac{1}{v^2F_U}[\bar{\psi}_L U M \psi_R+\textrm{h.c.}] \Big)\\
&  +2\left[\mathcal{S}^{ij}\tilde{\mathcal{B}}^{ij}+\tilde{\mathcal{Y}  }^{ij} \tilde{\mathcal{B}}^{ij}\right]\Bigg]\langle L_\mu L^\mu \rangle +\Bigg[ \frac{2}{3} \kappa_i^{(n)} \kappa_i^{(m)} (\partial^\mu h_n)(\partial^\nu h_m) +\frac{\kappa_j\kappa_k}{6}\mathcal{X}^{ji}_{\mu}\mathcal{X}^{ki}_\nu \\
&+\frac{1}{3}\kappa_j\left(\mathcal{X}^{ji}_\mu \kappa_i^{(n)} (\partial_\nu h_n)+\mathcal{X}^{ji}_\nu \kappa_i^{(m)} (\partial_\mu h_m)\right)   \Bigg]\langle L_\mu L_\nu \rangle  \\
& +\frac{3}{2}\left( \frac{1}{2 F_U} F^{(i)}_U A^{-1}_{ij} \left(V^{(j)}+\left[ \bar{\psi}_L U {M}^{(j)}\psi_R+\textrm{h.c.}\right]\right)  -\frac{1}{v^2F_U}[\bar{\psi}_L U M \psi_R+\textrm{h.c.}]\right)^2\\
&+3\left(\frac{1}{2 F_U} F^{(i)}_U A^{-1}_{ij} \left(V^{(j)}+\left[ \bar{\psi}_L U {M}^{(j)}\psi_R+\textrm{h.c.}\right]\right)  -\frac{1}{v^2F_U}[\bar{\psi}_L U M \psi_R+\textrm{h.c.}]\right)(\partial h_i)(\partial h_j) D^{ij} \\
&+\frac{4}{v^2}\left( A^{-1/2}_{ji}\left[  i\bar{\psi}_L U \left( \frac{M}{\sqrt{F_U}} \right)^{(j)}  \psi_R+ \textrm{h.c.} \right]\right) ^2 + \frac{2A^{-1/2}_{ji}}{v}\left[  i\bar{\psi}_L L_\mu U \left( \frac{M}{\sqrt{F_U}} \right)^{(j)}  \psi_R+ \textrm{h.c.} \right]\kappa_i^{(n)} (\partial^\mu h_n) \\
& + \frac{\kappa_n A^{-1/2}_{ji}}{v}\left[  i\bar{\psi}_L L_\mu U \left( \frac{M}{\sqrt{F_U}} \right)^{(j)}  \psi_R+ \textrm{h.c.} \right] \mathcal{X}_{ni}^\mu +\frac{1}{2} \tilde{\mathcal{Y}}^{ij}\tilde{\mathcal{Y}}^{ij}+ \mathcal{S}^{ij}\tilde{\mathcal{Y}}^{ij} \\
&+ \frac{1}{2} \mathcal{S}^{ij}\mathcal{S}^{ij}+ \frac{3}{2}\left[(\partial h_i)(\partial h_j) D^{ij}\right]^2  - \frac{1}{12}\Bigg[ \frac{1}{2}\partial_\mu \mathcal{X}^{ij}_{\nu}- \frac{1}{2}\partial_\nu \mathcal{X}^{ij}_{\mu}+\frac{1}{4}\left[\mathcal{X}^{ik}_\mu  \mathcal{X}^{kj}_\nu-\mathcal{X}^{ik}_\nu  \mathcal{X}^{kj}_\mu   \right]\Bigg]^2 
\Bigg\} \ ,
\end{aligned}
\label{finalN}
\end{equation}		
where 

\begin{equation}\label{1}
\begin{split}
\mathcal{X}_\mu^{ij}& = A^{-1/2}_{ik}\left(f^{(l)}_{km}-f^{(k)}_{lm}\right)(\partial_\mu h_m) A^{-1/2}_{lj}+A^{1/2}_{ik}  A^{-1/2(n)}_{kj} (\partial_\mu h_n)- A^{-1/2(n)}_{ik}A^{1/2}_{kj}(\partial_\mu h_n) \ , \\
\kappa_i &= \frac{vF^{(j)}_U}{2\sqrt{F_U}} A^{-1/2}_{ji}\,,\quad\quad \kappa^2=\sum_i \kappa_i^2 \ , 
\end{split}
\end{equation}

\begin{equation*}
\begin{split}
\tilde{\mathcal{B}}^{ij} &=\frac{1}{4} \kappa^i\kappa^j - \frac{1}{8}v^2A^{-1/2}_{im}F_U^{(mn)}A^{-1/2}_{nj}  - \frac{v^2}{8} A^{-1/2}_{im}(f^{(k)}_{mn}-f^{(m)}_{nk}-f^{(n)}_{mk})A^{-1/2}_{nj}  A^{-1}_{kp} F^{(p)}_{U}\,,\quad\quad \\
\tilde{\mathcal{Y}}^{ij} &=A^{-1/2}_{im}\left[ \bar{\psi}_LU M^{(mn)} \psi_R+\textrm{h.c.}\,\right] A^{-1/2}_{nj}	+ A^{-1/2}_{im}(f^{(k)}_{mn}-f^{(m)}_{nk}-f^{(n)}_{mk})A^{-1/2}_{nj}  A^{-1}_{kp} \left[\bar{\psi}_L U M^{(p)} \psi_R+\textrm{h.c.}\right] \, , \\
D^{ij} &= \left(\frac{F^{(i)}_UF^{(j)}_U}{4F_U^2} -\frac{F_U^{(ij)}}{2F_U}\right) -\frac{F_U^{(m)}}{2F_U} A^{-1}_{mn}\left(f^{(n)}_{ij}-\frac{1}{2}f^{(j)}_{ni}-\frac{1}{2}f^{(i)}_{nj}\right) \ ,
\end{split}
\end{equation*}

\begin{equation}\label{3}
\begin{split}
\mathcal{S}^{ij} &=A^{-1/2}_{im}\Bigg[ V^{(mn)}-\left(f^{(kl)}_{mn}-f^{(ml)}_{nk}-f^{(nl)}_{mk}  -\frac{1}{2}f^{(mn)}_{kl} 			\right)(\partial_\mu h_k)(\partial^\mu h_l) \Bigg] A^{-1/2}_{nj}\\
&+\frac{1}{2}\left[A^{-1/2}_{ik}\left(f^{(l)}_{km}-f^{(k)}_{lm}\right)(\partial_\mu h_m) A^{-1/2(n)}_{lj}(\partial_\mu h_n)- A^{-1/2(n)}_{ik}(\partial_\mu h_n)\left(f^{(l)}_{km}-f^{(k)}_{lm}\right)(\partial_\mu h_m) A^{-1/2}_{lj}   \right]\\
&-\frac{1}{2}\left[ A^{-1/2(n)}_{ik}(\partial_\mu h_n)\, A^{1/2(m)}_{kj}(\partial^\mu h_m)+ A^{1/2(n)}_{ik}(\partial_\mu h_n)\, A^{-1/2(m)}_{kj}(\partial^\mu h_m)\right] \\
&+\frac{1}{4}\mathcal{X}^{ik}_\mu \mathcal{X}^{jk}_\mu-  A^{-1/2}_{im}(f^{(k)}_{mn}-f^{(m)}_{nk}-f^{(n)}_{mk})A^{-1/2}_{nj} A^{-1}_{kr}\left(f^{(r)}_{pq}-f^{(p)}_{rq}- f^{(q)}_{rp} \right)(\partial h_p)(\partial h_q)\\
&+    A^{-1/2}_{im}(f^{(k)}_{mn}-f^{(m)}_{nk}-f^{(n)}_{mk})A^{-1/2}_{nj}  A^{-1}_{kp}V^{(p)} \ .\\
\end{split}
\end{equation}

\section{Crosschecks}\label{sec:crosschecks}

\subsection{Matching to the EWCLh}
Since the EWCLNh reduces to the EWCLh for one singlet field, it is instructive to check our final result by comparing it with the existing one loop divergence formula \eqref{eqn:finalewclh}.
The matching is given by the correspondences:

\begin{equation}
\begin{split}
&\mathcal{X}_\mu^{ij}\rightarrow 0, \quad\kappa_i\rightarrow\kappa= \frac{vF'}{2\sqrt{F}}, \quad S^{ij}\rightarrow V''\,,\quad v^2D^{ij}\rightarrow  \mathcal{B}= \frac{v^2 {F'}^2}{4F^2}-\frac{v^2 F''}{2F} \ , \\
&\tilde{\mathcal{B}}^{ij}\rightarrow \frac{F_U \mathcal{B}}{4} \,,\quad \tilde{\mathcal{Y}}^{ij} \rightarrow \bar{\psi}_LU M'' \psi_R+\textrm{h.c.}\, .
\end{split}
\end{equation}
Inserting these into (\ref{finalN}), we obtain the one loop divergence formula \ref{eqn:finalewclh}.
Furthermore, we checked that for two singlet fields there exists the limit of the renormalizable SM extended by a scalar, as discussed, e.g., in \cite{addsinglet}, in which all NLO operator structures in the corresponding one-loop divergence formula vanish.

\subsection{Some Diagrammatic Checks}
We compute some divergences diagrammatically to understand how the factors in \ref{finalN} arise.
First, we consider $ \varphi^+ \varphi^0 \rightarrow \varphi^+ \varphi^0 $ Goldstone scattering.
We expand the kinetic Goldstone term that gives us the interaction Lagrangian

\begin{equation}
\frac{v^2}{4} \langle \partial_\mu U^\dagger \partial^\mu U \rangle \overset{\mathcal{O}(\varphi^4)}{=} \frac{1}{6v^2} \left[ (\vec{\varphi} \cdot \partial_\mu \vec{\varphi})^2 - \vec{\varphi}^2 (\partial_\mu \vec{\varphi} \cdot \partial^\mu \vec{\varphi} )  \right] =\mathcal{L}_{GI} \ .
\end{equation}
We rewrite it in terms of the \textit{physical} Goldstones

\begin{equation}
\varphi^\pm = \frac{1}{\sqrt{2}} ( \varphi^1 \mp i\varphi^2) \, , \ \varphi^0 = \varphi^3 \ ,
\label{eqn:physGold}
\end{equation}
with

\begin{equation}
\sigma^\pm = \frac{1}{\sqrt{2}} ( \sigma^1 \pm i \sigma^2) \, , \ \text{with} \ \langle \sigma^+ \sigma^- \rangle = 2 \, , \ \langle \sigma^\pm \sigma^3 \rangle = 0 \ .
\end{equation}
The Goldstone interaction Lagrangian becomes

\begin{equation}
\begin{aligned}
\mathcal{L}_{\text{G,I}}=\frac{1}{6v^2} &\Big[ (\varphi^+ \partial_\mu \varphi^- + \varphi^- \partial_\mu \varphi^+ + \varphi^0 \partial_\mu \varphi^0 )^2 - (2 \varphi^+ \varphi^- + \varphi^0 \varphi^0) (2 \partial_\mu \varphi^+ \partial^\mu \varphi^- + \partial_\mu \varphi^0 \partial^\mu \varphi^0) \Big] \\
= \frac{1}{6v^2} & \Big[ (\varphi^+)^2 (\partial_\mu \varphi^-)^2 + (\varphi^-)^2 (\partial_\mu \varphi^+)^2  - 2 \varphi^+ \varphi^- \partial_\mu \varphi^+ \partial^\mu \varphi^- +2 (\varphi^+ \partial_\mu \varphi^- + \varphi^- \partial_\mu \varphi^+) \varphi^0 \partial^\mu \varphi^0 \\
& - 2 \varphi^+ \varphi^- (\partial_\mu \varphi^0) -2 (\varphi^0)^2 \partial_\mu \varphi^+ \partial^\mu \varphi^- \Big] \ .
\end{aligned}
\end{equation}
This gives us all the vertex factors for Goldstone-Goldstone scattering (Figure \ref{fig:gg_tree}):

\if false
\begin{center}
	\begin{tikzpicture}
	\begin{feynman}
	\vertex (a);
	\vertex [above left=of a] (p1) {\( \varphi_{p_1} \)};
	\vertex [below left=of a] (p2) {\( \varphi_{p_2} \)};
	\vertex [above right=of a] (f1) {\( \varphi_{p_3} \)};
	\vertex [below right=of a] (f2) {\( \varphi_{p_4} \)};

	\diagram*{
		(a) -- (p1);
		(a) -- (p2);
		(a) --  (f1);
		(a) --  (f2);
		
	};
	
	\end{feynman}
	\end{tikzpicture}
    \label{fig:gg_tree}
\end{center}
\fi
\begin{figure}[ht]
	\centering
	\includegraphics[width=0.25\linewidth]{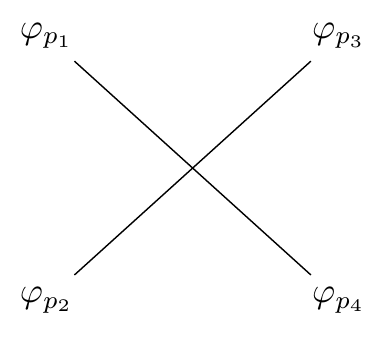}
	\caption{Goldstone-Goldstone-Scattering at tree level.}
	\label{fig:gg_tree}
\end{figure}

\begin{equation}
\begin{aligned}
& i \mathcal{M}_{\varphi^+ \varphi^- \rightarrow \varphi^+ \varphi^-} = \frac{i}{3v^2} (s+t-2u) \ , \quad i \mathcal{M}_{\varphi^+ \varphi^- \rightarrow \varphi^- \varphi^+} = \frac{i}{3v^2} (s+u-2t) \ , \\
& i \mathcal{M}_{\varphi^+ \varphi^0 \rightarrow \varphi^+ \varphi^0} = \frac{i}{3v^2} (2t-s-u) \ , \quad i \mathcal{M}_{\varphi^+ \varphi^0 \rightarrow \varphi^0 \varphi^+} = \frac{i}{3v^2} (2u-s-t) \ ,  \\
& i \mathcal{M}_{\varphi^+ \varphi^- \rightarrow \varphi^0 \varphi^0} = \frac{i}{3v^2} (2s-t-u) \ .
\end{aligned}
\label{eqn:vertices}
\end{equation}
The NLO operators $ \langle L_\mu L^\mu \rangle^2 $ and $ \langle L_\mu L_\nu \rangle^2 $ give local contributions to this process.
We find

\begin{equation}
\langle L_\mu L^\mu \rangle^2 = \frac{16}{v^4} \, \partial_\mu \varphi^+ \partial^\mu \varphi^- \partial_\nu \varphi^0 \partial^\nu \varphi^0 + ... \ ,
\end{equation}

\begin{equation}
\langle L_\mu L_\nu \rangle^2 = \frac{16}{v^4} \, \partial_\mu \varphi^+ \partial_\nu \varphi^- \partial^\mu \varphi^0 \partial_\nu \varphi^0 + \ ... \, .
\end{equation}
We recall the Mandelstam variables

\begin{equation}
s = (p_1+p_2)^2 \, , \ t = (p_1-p_3)^2 \, , \ u = (p_1-p_4)^2 \ ,
\end{equation}
where for massless on-shell fields we make use of the relations

\begin{equation}
s+t+u = 0 \, , \ st = \frac{1}{2}(u^2-s^2-t^2) \, , \ su = \frac{1}{2}(t^2-s^2-u^2) \, , \ tu = \frac{1}{2}(s^2-t^2-u^2) \, .
\end{equation}
The contribution of these operators to the scattering process can thus be evaluated to be

\begin{equation}
\begin{aligned}
& \frac{16i}{v^4} \int d^4x \langle 0 | T \lbrace a_{+3} a_{04} \, ( \partial_\mu \varphi^+ \partial^\mu \varphi^- \, \partial_\nu \varphi^0 \partial^\nu \varphi^0) \, a^\dagger_{+1} a^\dagger_{02} \rbrace | 0 \rangle \doteq (2\pi)^4 \delta^4(\sum_i p_i) i \mathcal{M} \\
& \Rightarrow i \mathcal{M} = i \frac{16}{v^4} \left(  2( p_1 \cdot p_3) (p_2 \cdot p_4) \right)  = \frac{8i}{v^4} t^2 \ ,
\end{aligned}
\end{equation}
where $ T $ here denotes the time-ordering operator,
and 

\begin{equation}
\begin{aligned}
& \frac{16i}{v^4} \int d^4x \langle 0 | T \lbrace a_{+3} a_{04} \, ( \partial_\mu \varphi^+ \partial_\nu \varphi^- \partial^\mu \varphi^0 \partial_\nu \varphi^0 ) \, a^\dagger_{+1} a^\dagger_{02} \rbrace | 0 \rangle \doteq (2\pi)^4 \delta^4(\sum_i p_i) i \mathcal{M} \\
& \Rightarrow i \mathcal{M} = i \frac{16}{v^4} \left(  2( p_1 \cdot p_2) (p_3 \cdot p_4) + (p_1 \cdot p_4) (p_2 \cdot p_3) \right)  = \frac{8i}{v^4} (s^2+u^2) \ .
\end{aligned}
\end{equation}
In (\ref{finalN}) we therefore find a divergent contribution which arises from diagrams containing only Goldstones, also in the loops.
These are the divergences

\begin{equation}
\frac{1}{32\pi^2 \epsilon}\left[ \frac{1}{12} \langle L_\mu L^\mu \rangle^2 + \frac{1}{6} \langle L_\mu L_\nu \rangle^2 \right]  {\Rightarrow}\ i\mathcal{M} = \frac{i}{96\pi^2 \epsilon v^4}  2(s^2+t^2+u^2)   \ .
\label{eqn:agreement}
\end{equation}
There are three diagrams contributing to this process, see Figure \ref{fig:ggg},

\if false
\begin{figure}[ht]
	\centering
	\begin{tikzpicture}
	\begin{feynman}
	\vertex (a);
	\vertex (b);
	\vertex [above left=of a] (p1) {\( \varphi^+_{p_1} \)};
	\vertex [below left=of a] (p2) {\( \varphi^0_{p_2} \)};
	\vertex [right =1.5cm of a] (b) ;
	\vertex [above right=of b] (f1) {\( \varphi^+_{p_3} \)};
	\vertex [below right=of b] (f2) {\( \varphi^0_{p_4} \)};

	\diagram*{
		(a) -- (p1);
		(a) -- (p2);
		(b) --  (f1);
		(b) --  (f2);
		(a) -- [half right, edge label = \( \varphi^0 \)] (b) ;
		(a) -- [half left, edge label = \( \varphi^+ \)] (b) ;
	};
	
	\end{feynman}
	\end{tikzpicture}
	\begin{tikzpicture}
	\begin{feynman}
	\vertex (a);
	\vertex (b);
	\vertex [above left=of a] (p1) {\( \varphi^+_{p_1} \)};
	\vertex [above right=of a] (p2) {\( \varphi^+_{p_3} \)};
	\vertex [below =1.5cm of a] (b) ;
	\vertex [below left=of b] (f1) {\( \varphi^0_{p_2} \)};
	\vertex [below right=of b] (f2) {\( \varphi^0_{p_4} \)};

	\diagram*{
		(a) -- (p1);
		(a) -- (p2);
		(b) --  (f1);
		(b) --  (f2);
		(a) -- [half right, edge label = \( \varphi^- \)] (b) ;
		(a) -- [half left, edge label = \( \varphi^- \)] (b) ;
	};
	
	\end{feynman}
	\end{tikzpicture}
	\begin{tikzpicture}     
	\begin{feynman} 
	\vertex (a);
	\vertex (b);
	\vertex [above left=of a] (p1) {\( \varphi^+_{p_1} \)};
	\vertex [above right=of a] (p2) {\( \varphi^+_{p_3} \)};
	\vertex [below =1.5cm of a] (b) ;
	\vertex [below left=of b] (f1) {\( \varphi^0_{p_2} \)};
	\vertex [below right=of b] (f2) {\( \varphi^0_{p_4} \)};            
	
	\diagram*{
		(a) -- (p1);
		(a) -- [half left] (f2);
		(b) --  (f1);
		(b) -- [half right](p2);
		(a) -- [half right, edge label = \( \varphi^0 \)] (b) ;
		(a) -- [half left, edge label = \( \varphi^- \)] (b) ;
		
	};
	
	\end{feynman}
	\draw node at (2.24,-0.74) [circle,fill=white, inner sep=2pt] (a) {} ;  
	\end{tikzpicture}
    \caption{Goldstone-Goldstone-Scattering via Goldstone loops}
	\label{fig:ggg}
\end{figure}
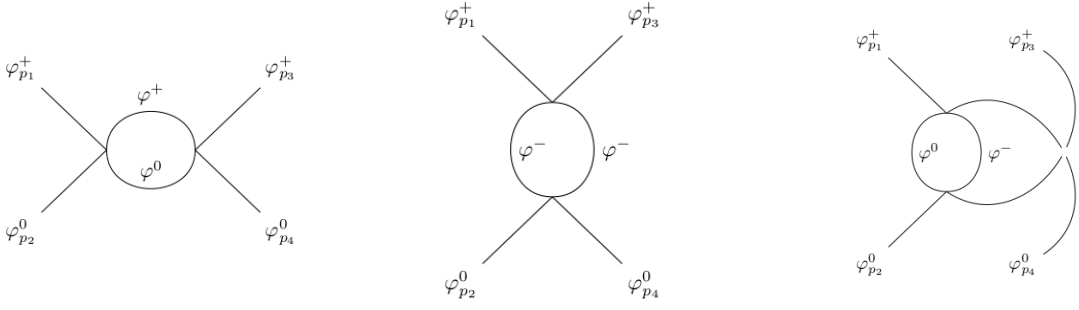
\fi
%\if false
\begin{figure}[ht]
	\centering
	\begin{subfigure}{.3\textwidth}
		\centering
		\includegraphics[width=.8\linewidth]{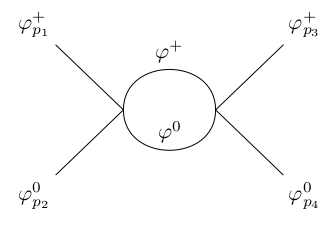}
	\end{subfigure}
	\begin{subfigure}{.3\textwidth}
		\centering
		\includegraphics[width=0.6\linewidth]{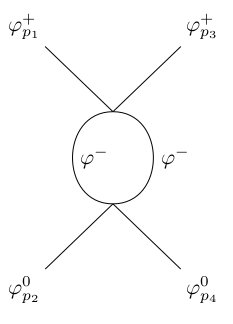}
	\end{subfigure}
	\begin{subfigure}{.3\textwidth}
		\centering
		\includegraphics[width=0.6\linewidth]{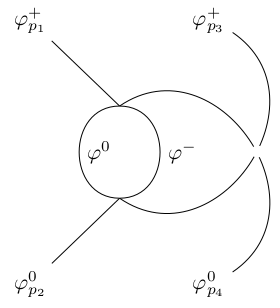}
	\end{subfigure}
	\caption{Goldstone-Goldstone-Scattering via Goldstone loops}
	\label{fig:ggg}
\end{figure}
%\fi
which we compute in turn to obtain

\begin{equation}
\begin{aligned}
& \int \frac{d^4k}{(2\pi)^4} \frac{i}{3v^2} \left( 2(p_2-k)^2-s-(p_1-k)^2 \right)  \frac{i}{(p_1+p_2-k)^2} \frac{i}{k^2} \left( 2(p_4-k)^2-s-(p_3-k)^2 \right) \\
& \overset{\mathcal{O}(1/\epsilon)}{=}  \frac{i}{(4\pi)^2\epsilon} \frac{1}{9v^4}\left[ \frac{7}{3} s^2+ \frac{5}{6} st - \frac{2}{3} su \right] ,
\end{aligned}
\end{equation}
\begin{equation}
\begin{aligned}
& \int \frac{d^4k}{(2\pi)^4} \frac{i}{3v^2} \left( (p_3+k)^2+t-2(p_1-k)^2 \right)  \frac{i}{(p_1-p_3-k)^2} \frac{i}{k^2} \left( 2t-(p_4-k)^2-(p_2+k)^2 \right) \\
& \overset{\mathcal{O}(1/\epsilon)}{=}  \frac{i}{(4\pi)^2\epsilon} \frac{1}{9v^4}\left[ \frac{9}{2}t^2\right] ,
\end{aligned}
\end{equation}
\begin{equation}
\begin{aligned}
& \int \frac{d^4k}{(2\pi)^4} \frac{i}{3v^2} \left( 2(p_1-k)^2-2(p_4+k)^2-u \right) \times \frac{i}{(p_1-p_3-k)^2} \frac{i}{k^2} \left( 2(p_3-k)^2-u-(p_2+k)^2 \right) \\
& \overset{\mathcal{O}(1/\epsilon)}{=}  \frac{i}{(4\pi)^2\epsilon} \frac{1}{9v^4}\left[ \frac{7}{3} u^2+ \frac{5}{6} ut - \frac{2}{3} su \right] .
\end{aligned}
\end{equation}
Combining the results gives

\begin{equation}
\begin{aligned}
& \frac{i}{96\pi^2\epsilon v^4} \frac{2}{3} \left[ \frac{7}{3} (s^2+u^2) + \frac{5}{6} t(s+u) - \frac{4}{3} su + \frac{9}{2} t^2 \right] =\frac{i}{96\pi^2 \epsilon v^4} 2(s^2+t^2+u^2) \ ,
\end{aligned}
\end{equation}
in agreement with \eqref{eqn:agreement}.

For a convenient computation of the above integrals in dimensional regularization \cite{dimreg1,dimreg2}, we calculate the set of integrals

\begin{equation}
\begin{aligned}
& \int \frac{d^4k}{(2\pi)^4} \frac{1}{(p-k)^2 k^2}  \overset{\mathcal{O}(1/\epsilon)}{=} \frac{i}{(4\pi)^2 \epsilon} \ ,  \\
& \int \frac{d^4k}{(2\pi)^4} \frac{k^\mu}{(p-k)^2 k^2} \overset{\mathcal{O}(1/\epsilon)}{=} \frac{i}{(4\pi)^2 \epsilon} \frac{p^\mu}{2} \ , \\
& \int \frac{d^4k}{(2\pi)^4} \frac{k^2}{(p-k)^2 k^2} \overset{\mathcal{O}(1/\epsilon)}{=} 0 \ , \\
& \int \frac{d^4k}{(2\pi)^4} \frac{k^\mu k^\nu}{(p-k)^2 k^2} \overset{\mathcal{O}(1/\epsilon)}{=} \frac{i}{(4\pi)^2 \epsilon} \left( \frac{1}{3} p^\mu p^\nu -\frac{1}{12} p^2 g^{\mu\nu} \right)  \ , \\
& \int \frac{d^4k}{(2\pi)^4} \frac{k^2k^\mu}{(p-k)^2 k^2} \overset{\mathcal{O}(1/\epsilon)}{=} 0 \ , \\
& \int \frac{d^4k}{(2\pi)^4} \frac{k^4}{(p-k)^2 k^2} \overset{\mathcal{O}(1/\epsilon)}{=} 0 \ .
\end{aligned}
\end{equation}
Next, we consider $ \varphi^+ \varphi^- \rightarrow \varphi^+ \varphi^- $ scattering.
The NLO operators $ \langle L_\mu L^\mu \rangle^2 $ and $ \langle L_\mu L_\nu \rangle^2 $ give local contributions to this process since they have the expanded forms

\begin{equation}
\langle L_\mu L^\mu \rangle^2 = \frac{16}{v^4} \, \partial_\mu \varphi^+ \partial^\mu \varphi^- \, \partial_\nu \varphi^+ \partial^\nu \varphi^- + ... \ ,
\end{equation}
\begin{equation}
\langle L_\mu L_\nu \rangle^2 = \frac{16}{v^4} \, \partial_\mu \varphi^+ \partial_\nu \varphi^- \, \partial^\mu \varphi^+ \partial^\nu \varphi^- + ... \ .
\end{equation}
The contribution of these operators to the scattering process is thus

\begin{equation}
\begin{aligned}
& \frac{16i}{v^4} \int d^4x \langle 0 | T \lbrace a_{+3} a_{-4} \, (\partial_\mu \varphi^+ \partial^\mu \varphi^- \, \partial_\nu \varphi^+ \partial^\nu \varphi^-) \, a^\dagger_{+1} a^\dagger_{-2} \rbrace | 0 \rangle \doteq (2\pi)^4 \delta^4(\sum_i p_i) i \mathcal{M} \\
& \Rightarrow i \mathcal{M} = i \frac{16}{v^4} \left(  2( p_1 \cdot p_2) (p_3 \cdot p_4) + 2 (p_1 \cdot p_3) (p_2 \cdot p_4)\right)  = \frac{8i}{v^4} (s^2+t^2) \
\end{aligned}
\end{equation}
and 

\begin{equation}
\begin{aligned}
& \frac{16i}{v^4} \int d^4x \langle 0 | T \lbrace a_{+3} a_{-4} \, ( \partial_\mu \varphi^+ \partial_\nu \varphi^- \, \partial^\mu \varphi^+ \partial^\nu \varphi^-) \,  a^\dagger_{+1} a^\dagger_{-2} \rbrace | 0 \rangle \doteq (2\pi)^4 \delta^4(\sum_i p_i) i \mathcal{M} \\
& \Rightarrow i \mathcal{M} = i \frac{16}{v^4} \left( 4( p_1 \cdot p_4) (p_2 \cdot p_3) \right) = \frac{16i}{v^4} u^2 \ .
\end{aligned}
\end{equation}
We concentrate on the situation of two singlet fields as in the $ SO(6)/SO(5) $ CHM, where the Goldstone kinetic term has the generic form

\begin{equation}
\mathcal{L}_{\text{G,kin}} = \frac{v^2}{4} \langle D_\mu U^\dagger D^\mu U \rangle \left( F_{10} \frac{h}{v} + F_{01} \frac{S}{v} + F_{11} \frac{hS}{v^2} + F_{20} \frac{h^2}{v^2}+ F_{02} \frac{S^2}{v^2} + ...\right) .
\end{equation}
In particular, we calculate diagrams with singlet loops as shown in Figure \ref{fig:gg_singlet}.

\if false
\begin{figure}[ht]
	\centering
	\begin{tikzpicture}
	\begin{feynman}
	\vertex (a);
	\vertex (b);
	\vertex [above left=of a] (p1) {\( \varphi^+_{p_1} \)};
	\vertex [below left=of a] (p2) {\( \varphi^-_{p_2} \)};
	\vertex [right =1.5cm of a] (b) ;
	\vertex [above right=of b] (f1) {\( \varphi^+_{p_3} \)};
	\vertex [below right=of b] (f2) {\( \varphi^-_{p_4} \)};

	\diagram*{
		(a) -- (p1);
		(a) -- (p2);
		(b) --  (f1);
		(b) --  (f2);
		(a) -- [scalar, edge label = \( h \), half right] (b) ;
		(a) -- [scalar,edge label = \( h \), half left] (b) ;
	};
	
	\end{feynman}
	\end{tikzpicture}
	\begin{tikzpicture}
	\begin{feynman}
	\vertex (a);
	\vertex (b);
	\vertex [above left=of a] (p1) {\( \varphi^+_{p_1} \)};
	\vertex [below left=of a] (p2) {\( \varphi^-_{p_2} \)};
	\vertex [right =1.5cm of a] (b) ;
	\vertex [above right=of b] (f1) {\( \varphi^+_{p_3} \)};
	\vertex [below right=of b] (f2) {\( \varphi^-_{p_4} \)};

	\diagram*{
		(a) -- (p1);
		(a) -- (p2);
		(b) --  (f1);
		(b) --  (f2);
		(a) -- [ghost, edge label = \( S \), half right] (b) ;
		(a) -- [ghost, edge label = \( S \), half left] (b) ;
	};
	
	\end{feynman}
	\end{tikzpicture}
	%\\
	\begin{tikzpicture}
	\begin{feynman}
	\vertex (a);
	\vertex (b);
	\vertex [above left=of a] (p1) {\( \varphi^+_{p_1} \)};
	\vertex [below left=of a] (p2) {\( \varphi^-_{p_2} \)};
	\vertex [right =1.5cm of a] (b) ;
	\vertex [above right=of b] (f1) {\( \varphi^+_{p_3} \)};
	\vertex [below right=of b] (f2) {\( \varphi^-_{p_4} \)};

	\diagram*{
		(a) -- (p1);
		(a) -- (p2);
		(b) --  (f1);
		(b) --  (f2);
		(a) -- [scalar, edge label = \( h \), half right] (b) ;
		(a) -- [ghost, edge label = \( S \), half left] (b) ;
	};
	
	\end{feynman}
	\end{tikzpicture}
	\\
	(+ t-channels)
	\caption{Goldstone-Goldstone-Scattering via Higgs loops.}
	\label{fig:gg_singlet}
\end{figure}
\fi
%\if false
 \begin{figure}[ht]
	\centering
	\begin{subfigure}{.3\textwidth}
		\centering
		\includegraphics[width=0.8\linewidth]{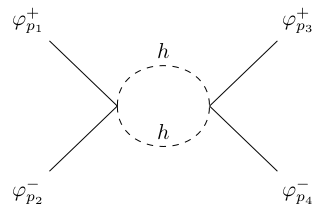}
	\end{subfigure}
	\begin{subfigure}{.3\textwidth}
		\centering
		\includegraphics[width=0.8\linewidth]{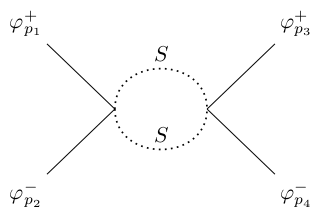}
	\end{subfigure}
	\begin{subfigure}{.3\textwidth}
		\centering
		\includegraphics[width=0.8\linewidth]{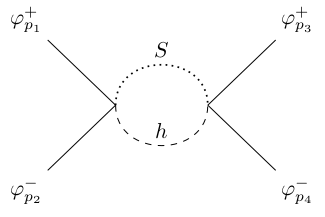}
	\end{subfigure}
	\caption{Goldstone-Goldstone-Scattering via Higgs loops. The t-channel is implicit.}
	\label{fig:gg_singlet}
\end{figure}
%\fi
The signature of the singlet loop diagrams are the respective couplings since the vertices contain two Goldstones and two singlets.
The coupling parameters are, in the expansion above, $ F_{20}, \, F_{02} $, and $ F_{11} $.
There is only one operator in our final result (\ref{finalN}) providing these interactions.
In detail, we have the divergent contribution

\begin{equation}
\begin{aligned}
& 2 \tilde{\mathcal{B}}^{ij}\tilde{\mathcal{B}}^{ij} \langle L_\mu L^\mu \rangle^2 \supset \frac{2}{64} v^4 F_U^{(ij)}F_U^{(ij)} \langle L_\mu L^\mu \rangle^2 = \frac{1}{32} \left( (2F_{20})^2 + (2F_{02})^2 + 2F_{11}^2  \right) \langle L_\mu L^\mu \rangle^2 \\
&= \frac{1}{8} (F_{20}^2 +F_{02}^2 + \frac{1}{2} F_{11}^2 ) \langle L_\mu L^\mu \rangle^2 {\Rightarrow} \ i \mathcal{M} = \frac{i}{v^4}(F_{20}^2 + F_{02}^2 +\frac{1}{2} F_{11}^2 ) (s^2+t^2) \ ,
\end{aligned}
\end{equation}
which we should find by the computation of the above diagrams.
Indeed, for an $ h $-loop we find

\begin{equation}
\begin{aligned}
i \mathcal{M} &= \left( \frac{i2F_{20}}{v^2}\right)^2 \frac{s^2}{4} \frac{1}{2} \int \frac{d^4q}{(2\pi)^4} \frac{i}{(p_1+p_2+q)^2-m^2} \frac{i}{q^2-m^2} + (t-\text{channel}) \\
& \overset{\mathcal{O}(1/\epsilon)}{\longrightarrow} \frac{F_{20}^2}{v^4} \frac{s^2+t^2}{2} \frac{i}{(4\pi)^2\epsilon} = \frac{1}{32\pi^2 \epsilon} \frac{iF_{20}^2}{v^4} (s^2+t^2) \ .
\end{aligned}
\end{equation}
This result can also be compared to the corresponding divergence of the EWCLh \eqref{eqn:finalewclh}.
The result is trivially amended for an $ S $-loop.
The loop consisting of both an $ h $- and an $ S $-propagator misses the symmetry factor 2.
Our diagrammatic and functional computations agree.

\subsection{Running of NLO Operator Coefficients}

To cancel the divergences, counterterms have to be introduced:

\begin{equation}
\mathcal{L}_{\text{counter}} = \sum_k c_k O_k \ ,
\end{equation}
where the coefficients $ c_k $ fulfill

\begin{equation}
c_k = \mu^{2\epsilon} \left(  c_k^r(\mu) - \frac{1}{32\pi^2 \epsilon} \Gamma_k \right) \ .
\end{equation}
The $ \Gamma_k $ are the coefficients of the divergences as they are given in our formulas and the $ c_k^r $ are the renormalized, finite coefficients.

In the previous section we reproduced diagrammatically some coefficients of NLO operators appearing in the divergence.
A divergence with a momentum scaling $ \sim p^4 $ cannot be absorbed in the renormalization of the LO Lagrangian, since it corresponds to next-to-leading order in chiral dimensions.
We only get rid of them by the renormalization of the NLO Lagrangian.
This is a peculiarity of the energy expansion of the effective theory, where loops are suppressed through the derivative couplings.
The divergences produced by loops are kind of spurious as they are generated out of ignorance \cite{donoghue}.
The correct high energy behavior is unknown to the effective theory.
We can then see that the running of the coefficients of NLO operators, whose actual values are determined by the underlying theory, is fully determined by the LO Lagrangian through the $ \Gamma_k $:

\begin{equation}
\begin{aligned}
& 0 = \mu \frac{d}{d\mu} c_k = 2\epsilon \mu^{2\epsilon} \left( c_k^r(\mu) - \frac{1}{32\pi^2\epsilon} \Gamma_k \right) +\mu^{2\epsilon} \, \mu \frac{d}{d\mu} c_k^r(\mu) \\
& \overset{\epsilon \rightarrow 0}{\longrightarrow} \mu \frac{d}{d \mu} c_k^r(\mu)= \frac{1}{16\pi^2} \Gamma_k \ .
\end{aligned}
\end{equation}
Note the loop suppression factor $ 1/16\pi $ on the right.
The running of the coefficients has non-negligible impact in chiral perturbation theory, where the dimensional analysis is built on \textit{naive dimensional analysis} \cite{nda}.
The influence of the loops may be as big or even bigger than that of NLO operators at tree level.
For the scaling of an amplitude we find

\begin{equation}
\mathcal{M} \sim \frac{p^2}{v^2} + \left( c_k^r(\mu) + \frac{1}{16\pi^2} \Gamma_k \, \ln \frac{p}{\mu} + ... \right) \frac{p^4}{v^4} + \mathcal{O}(p^6) \ .
\end{equation}
In the bracket we have the contribution from the corresponding NLO operators and that from leading order loops which both scale as $ \sim p^4 $ by naive dimensional analysis \cite{manohar}.
The running has to be taken into account when comparing to data for the tests of a model.

\section{Conclusions}
The renormalization of the scalar sector of the EWCLNh developed in this work together with the full renormalization of the EWCLh in \cite{completerenorm} constitute a framework for a complete one-loop renormalization of any pure scalar extension to the electroweak chiral theory, thereby encompassing composite Higgs models and many extensions such as scalar dark matter candidates.
The applicability is demonstrated at the hand of the next-to-minimal composite Higgs model with SO(6)/SO(5) coset structure in a follow-up work \cite{LindnerMuzakka2022b}.

\section*{Acknowledgments}
We want to thank Gerhard Buchalla for the supervision of our Master's theses that this work is based on.
We are also grateful for the discussions with him during the production process of this paper.

\section*{Supplementary Material}
The Mathematica notebook at \href{https://notebookarchive.org/2022-04-1fbhb4o}{https://notebookarchive.org/2022-04-1fbhb4o} is a collection of some supplementary and extra calculations for the joint project of the formula developed here and the application on a specific composite Higgs model investigated in \cite{LindnerMuzakka2022b}.
The calculations in the present work have all been carried out by hand.

%\appendix

\ifx false

\subsection{Quadratic Expansion}
We proceed analogously to the scenario of two singlets, expanding the scalar fields  up to quadratic order.
The scalar fields $ h_{i} $ is expanded as $ h_i \rightarrow h_{i} + \tilde{h}_i $ into classical background and quantum fluctuations.
Hence, the one-loop effective Lagrangian takes the form

\begin{equation*}
\begin{aligned}
\mathcal{L}_{\text{eff}}^{\text{1loop}}=& - \frac{1}{2} \tilde{\varphi}^a \Delta^{ab} \, \tilde{\varphi}^b - \frac{1}{2} \tilde{\varphi}^a \Delta^{ai} \, \tilde{h}^i- \frac{1}{2} \tilde{h}^i \Delta^{ia} \, \tilde{\varphi}^a - \frac{1}{2} \tilde{h}^i \Delta^{ij} \, \tilde{h}^j \ .
\end{aligned}
\end{equation*}
Derivatives with respect to the singlet field $ h_i $ are denoted by an over-script (i), e.g. $F^{(i)}_U=\frac{\partial F_U}{\partial h^i}$ etc.
The extension of the quadratic expansion from the preceding scenarios to the new one is quite straight-forward.

\begin{align*}
\Delta^{ab} &= F_U \Bigg[  \delta^{ab}\left( \partial^2 + \frac{1}{F_U}(\partial^\mu F_U)\partial_\mu\right) 
+ \left( g\epsilon^{cab}W^{\mu c} +g'B^\mu \epsilon^{3ab} \right) \partial_\mu \\
& +gg'B^\mu\left(W_\mu^a \delta^{b3}-W^3_\mu \delta^{ab}\right) \Bigg] +\left[\frac{-4}{v^2}\bar{\psi}_L T^a T^b M \psi_R+\textrm{h.c.}\right]\ \\
&\overset{PI}{\longrightarrow} F_U \delta^{ab} \partial^2 
+ F_U\left( g\epsilon^{cab}W^{\mu c} +g'B^\mu \epsilon^{3ab} \right) \partial_\mu \\
&+F_U gg'B^\mu\left(\frac{1}{2}W_\mu^a \delta^{b3}+\frac{1}{2}W_\mu^b \delta^{a3}-W^3_\mu \delta^{ab}\right)\\
&+\delta^{ab}\left[ - \frac{1}{2}(\partial^2 F_U)+\frac{-1}{v^2}\bar{\psi}_L M \psi_R+\textrm{h.c.}\right] \ ,\\
\Delta^{ai} &= \left[  \frac{4i}{v}\bar{\psi}_L T^aM^{(i)}\psi_R+ \textrm{h.c.} \right]  -  F_U^{(i)} vgg'B^\mu\left(W^1_\mu\delta^{a2}-W^2_\mu \delta^{a1}\right) \ \\
&\overset{\text{sym} \, PI}{\longrightarrow} \left[  \frac{2i}{v}\bar{\psi}_L T^aM^{(i)}\psi_R+ \textrm{h.c.} \right]  -  \frac{F_U^{(i)}}{2} vgg'B^\mu\left(W^1_\mu\delta^{a2}-W^2_\mu \delta^{a1}\right) \\
&+ \frac{1}{4} \partial_\mu \left[F^{(i)}_U v\left( gW^{\mu a}-g'B^\mu\delta^{a3} \right)\right]+   \frac{F^{(i)}_U}{2} v\left( gW^{\mu a}-g'B^\mu\delta^{a3} \right)\partial_\mu  \ , \\
\Delta^{ia} &=  - F^{(i)}_U v\left( gW^{\mu a}-g'B^\mu\delta^{a3} \right)\partial_\mu \\
&\overset{\text{sym} \, PI}{\longrightarrow} \left[  \frac{2i}{v}\bar{\psi}_L T^aM^{(i)}\psi_R+ \textrm{h.c.} \right]  -  \frac{F_U^{(i)}}{2} vgg'B^\mu\left(W^1_\mu\delta^{a2}-W^2_\mu \delta^{a1}\right) \\
&+ \frac{1}{4} \partial_\mu \left[F^{(i)}_U v\left( gW^{\mu a}-g'B^\mu\delta^{a3} \right)\right] - \frac{F^{(i)}_U}{2} v\left( gW^{\mu a}-g'B^\mu\delta^{a3} \right)\partial_\mu\, . \\
\end{align*}
The last one is a bit more intriguing.

\begin{align*}
\Delta^{ij} &=(\delta_{ij}+F_{ij}(h))\partial^2+\left(F^{(j)}_{ik}-F^{(i)}_{jk}\right)(\partial_\mu h_k)\,\partial^\mu +\left[ \bar{\psi}_L M^{(ij)} \psi_R+\textrm{h.c.}\,\right] \\
& + V^{(ij)}-\partial_\mu\left[(F^{(k)}_{ij}-F^{(i)}_{jk}-F^{(j)}_{ik})\partial^\mu h_k\right]-\frac{1}{2}F^{(ij)}_{kl}(\partial_\mu h_k)(\partial^\mu h_l)   \\
&  -\left(  \frac{1}{8}g^2W_\mu^a W^{\mu a}+\frac{1}{8}g'^2B_\mu B^\mu -\frac{gg'}{4}B^\mu W_\mu^3\right)v^2 F_U^{(ij)} \\
& \doteq A_{ij}\partial^2+2B^\mu_{ij}\partial_\mu +C_{ij}\, .
\end{align*}
Now, defining a new index $\alpha=\{a,i\}$, the above quadratic fluctuation can be recast into the form

\begin{equation*}
\Delta^{\alpha\beta}=\mathbf{A}^{\alpha\beta}\partial^2+2\mathbf{B}^{\alpha\beta}\partial_\mu+\mathbf{C}^{\alpha\beta}\, .
\end{equation*}
We use again
$
C^a= gW^{\mu a}-g'B^\mu\delta^{a3} \ 
$
and define

\begin{eqnarray}
\mathbf{A} &= \left(\begin{tabular}{c c c| c}
$F_U$ & 0 & 0 & 0\\
0 & $ F_U$ & 0 & 0  \\
0 &0  & $F_U$ & 0\\
\hline
0&0& 0& $A_{ij}$\\
\end{tabular}
\right) \ , \nonumber \\
\mathbf{B}_\mu &= \left(\begin{tabular}{c c c| c}
. & . & . & .\\
. & $ \frac{ F_U}{2}\epsilon^{abc}\left( gW^c_\mu +g'B^c_\mu\right)   $ & . & $   \frac{vF^{(i)}_U}{4} C^a_\mu $ \\
. &. & . & .\\
\hline
.&$-\frac{vF^{(i)}_U}{4}C^a_\mu$ & .& $B_\mu^{ij}$\\
\end{tabular} \
\right) \ , 
\label{A}
\end{eqnarray}

\begin{align*}
\mathbf{C}^{ab} &=F_U gg'B^\mu\left(W_\mu^a \delta^{b3}-W^3_\mu \delta^{ab}\right) +\delta^{ab}\left[ - \frac{1}{2}(\partial^2 F_U)+\frac{-1}{v^2}\bar{\psi}_L M \psi_R+\textrm{h.c.}\right] \\
& = \mathbf{C}^{ba} \ , \\
\mathbf{C}^{ai} &=\left[  \frac{2i}{v}\bar{\psi}_L T^aM^{(i)}\psi_R+ \textrm{h.c.} \right]  -  \frac{F_U^{(i)}}{2} vgg'B^\mu\left(W^1_\mu\delta^{a2}-W^2_\mu \delta^{a1}\right) \\
& + \frac{1}{4} \partial_\mu \left[F^{(i)}_U v\left( gW^{\mu a}-g'B^\mu\delta^{a3} \right)\right] =\mathbf{C}^{ia}  \ , \\
\mathbf{C}^{ij} &= \left[ \bar{\psi}_L M^{(ij)} \psi_R+\textrm{h.c.}\,\right]+V^{(ij)}-\partial_\mu\left[(F^{(k)}_{ij}-F^{(i)}_{jk}-F^{(j)}_{ik})\partial^\mu h_k\right] \\
& -\frac{1}{2}F^{(ij)}_{kl}(\partial_\mu h_k)(\partial^\mu h_l) \\
&-\left(  \frac{1}{8}g^2W_\mu^a W^{\mu a}+\frac{1}{8}g'^2B_\mu B^\mu -\frac{gg'}{4}B^\mu W_\mu^3\right)v^2 F_U^{(ij)} = \mathbf{C}^{ji} \ .
\end{align*}
We recall from Section \ref{generaltwoscalars} the formula for $ X_\mu $ and $ Y $ (\ref{eqn:XandY}) and obtain the final result after lengthy calculations.

\subsection{Extraction of $ X $ and $ Y $}
Recall from Section \ref{generaltwoscalars} the formula for $ X_\mu $ and $ Y $ (\ref{eqn:XandY}).
We deduce

\begin{equation*}
\begin{aligned}
X_\mu &= \mathbf{A}^{-1/2} \mathbf{B}^\mu \mathbf{A}^{-1/2} +\frac{1}{2}\mathbf{A}^{1/2} (\partial^\mu \mathbf{A}^{-1/2})- \frac{1}{2}(\partial^\mu \mathbf{A}^{-1/2}) \mathbf{A}^{1/2} \ , \\
Y &=\mathbf{A}^{-1/2} \mathbf{C}\mathbf{A}^{-1/2} +\mathbf{A}^{-1/2} \mathbf{B}^\mu (\partial_\mu \mathbf{A}^{-1/2}) -   (\partial_\mu \mathbf{A}^{-1/2}) \mathbf{B}^\mu  \mathbf{A}^{-1/2} \\
& -\frac{1}{2}  \left(\partial_\mu \mathbf{A}^{-1/2} \,\partial^\mu \mathbf{A}^{1/2}+  \partial_\mu \mathbf{A}^{1/2} \,\partial^\mu \mathbf{A}^{-1/2}\right)-X_\mu X^\mu\\
&\doteq Z-X_\mu X^\mu \ .
\end{aligned}
\end{equation*}
From (\ref{A}), we obtain 

\begin{eqnarray*}
\mathbf{A}^{1/2} = \left(\begin{tabular}{c c c| c}
	$\sqrt{F_U}$ & 0 & 0 & 0\\
	0 & $ \sqrt{F_U}$ & 0 & 0  \\
	0 &0  & $\sqrt{F_U}$ & 0\\
	\hline
	0&0& 0& $A^{1/2}$\\
\end{tabular} 
\right) \ , \\
\mathbf{A}^{-1/2} = \left(\begin{tabular}{c c c| c}
	$\frac{1}{\sqrt{F_U}}$ & 0 & 0 & 0\\
	0 & $\frac{1}{\sqrt{F_U}}$ & 0 & 0  \\
	0 &0  & $\frac{1}{\sqrt{F_U}}$ & 0\\
	\hline
	0&0& 0& $A^{-1/2}$\\
\end{tabular}
\right)  \ .
\end{eqnarray*}
By direct computation, we find 

\begin{align*}
X_\mu^{ab} &= \frac{ 1}{2}\epsilon^{abc}\left( gW^c_\mu +g'B^c_\mu\right) = \frac{i}{2}f^c_{ab} \left( gW^c_\mu +g'B^c_\mu\right)  \ ,\\
X_\mu^{ai} &= \frac{vF^{(j)}}{4\sqrt{F_U}}C^a_\mu A^{-1/2}_{ji}\doteq \frac{1}{2}\kappa^i C^a_\mu \ ,\\
X_\mu^{ij} &= A^{-1/2}_{ik}B^\mu_{kl} A^{-1/2}_{lj}+\frac{1}{2}A^{1/2}_{ik} \partial_\mu A^{-1/2}_{kj}-\frac{1}{2}\partial_\mu A^{-1/2}_{ik}A^{1/2}_{kj}\doteq \frac{1}{2}\mathcal{X}_\mu^{ij} \ .
\end{align*}	
Where we have the generalized $ \kappa_i $,

\begin{equation*}
\kappa_i= \frac{vF^{(j)}_U}{2\sqrt{F_U}} A^{-1/2}_{ji} \ .
\end{equation*}
We compute 

\begin{align*}
X_{\mu\nu}^{ab} &= if^c_{ab} \left[ \frac{1}{2}\partial_\mu (gW+g'B)^c_\nu -\frac{1}{2}\partial_\nu (gW+g'B)^c_\mu\right]  \\
& -\frac{1}{4} (gW+g'B)^d_\mu (gW+g'B)^e_\nu \left[f^d, f^e\right]_{ab}-\frac{1}{4}\kappa^2\left(C^a_\mu C^b_\nu-C^a_\nu C^b_\mu\right)\\
&= if^c_{ab} \Bigg[ \frac{1}{2}\partial_\mu (gW+g'B)^c_\nu -\frac{1}{2}\partial_\nu (gW+g'B)^c_\mu \\
& -\frac{1}{4}\epsilon^{cde} (gW+g'B)^d_\mu (gW+g'B)^e_\nu -\frac{1}{2} \epsilon^{cde} g^2W^d_\mu W^e_\nu +\frac{1}{2} \epsilon^{cde} g^2W^d_\mu W^e_\nu \Bigg]\\
&-\frac{1}{4}\kappa^2 \left(C^a_\mu C^b_\nu-C^a_\nu C^b_\mu\right)\\
&= \frac{1}{2}if^c_{ab} \left(gW^c_{\mu\nu}+g'B^c_{\mu\nu}\right)-\frac{1}{4}\left(\kappa^2-1\right)\left(C^a_\mu C^b_\nu-C^a_\nu C^b_\mu\right) \ , \\
X_{\mu\nu}^{ai} &= \frac{1}{2}\partial_\mu (\kappa_i C^a_\nu)- \frac{1}{2}\partial_\nu (\kappa_i C^a_\mu) +\frac{\kappa_i }{4}if^c_{ad}\left[(gW+g'B)^c_\mu C^d_\nu-(gW+g'B)^c_\nu C^d_\mu\right] \\
& -\frac{1}{4} \kappa_j\left(C^a_\mu\mathcal{X}^{ji}_{\nu}-C^a_\nu\mathcal{X}^{ji}_{\mu}\right)\\
&= \frac{1}{2} (\partial_\mu \kappa_i)C^a_\nu-\frac{1}{2} (\partial_\nu \kappa_i)C^a_\mu+\frac{\kappa_i}{2}\left(gW^a_{\mu\nu}-g'B^a_{\mu\nu}\right) \\
&+\underbrace{\frac{\kappa_i}{2}\epsilon^{acd} g^2W^c_\mu W^d_\nu+\frac{\kappa_i }{4}if^c_{ad}\left[(gW+g'B)^c_\mu C^d_\nu-(gW+g'B)^c_\nu C^d_\mu\right]}_{=0}\\
&-\frac{1}{4} \kappa_j\left(C^a_\mu\mathcal{X}^{ji}_{\nu}-C^a_\nu\mathcal{X}^{ji}_{\mu}\right)\\
& = \frac{1}{2} (\partial_\mu \kappa_i)C^a_\nu-\frac{1}{2} (\partial_\nu \kappa_i)C^a_\mu+\frac{\kappa_i}{2}\left(gW^a_{\mu\nu}-g'B^a_{\mu\nu}\right) \\
& -\frac{1}{4} \kappa_j\left(C^a_\mu\mathcal{X}^{ji}_{\nu}-C^a_\nu\mathcal{X}^{ji}_{\mu}\right) \ , \\
X_{\mu\nu}^{ij} &= \frac{1}{2}\partial_\mu \mathcal{X}^{ij}_{\nu}- \frac{1}{2}\partial_\nu \mathcal{X}^{ij}_{\mu}+\frac{1}{4}\left[\mathcal{X}^{ik}_\mu  \mathcal{X}^{kj}_\nu-\mathcal{X}^{ik}_\nu  \mathcal{X}^{kj}_\mu   \right] \ . 
\end{align*}
With

\begin{equation*}
\kappa^2=\sum_i \kappa_i^2 \ .
\end{equation*}
Next we compute the divergence $\text{tr} X_{\mu\nu} X^{\mu\nu}$ using \eqref{eqn:trX}

\begin{equation*}
\text{tr} X_{\mu\nu}^{\alpha \beta} X^{\mu\nu \alpha \beta} = X^{ab} X^{ba}-2X^{ai}X^{ai}-X^{ij}X^{ij} \ ,
\end{equation*}
We compute in parts,

\begin{align*}
&X^{ab} X^{ba} \\
&=-\frac{1}{4}(gW^c_{\mu\nu}+g'B^c_{\mu\nu})(gW^{\mu\nu d}+g'B^{\mu\nu d})\left\langle f^c f^d\right\rangle \\
& +\frac{\kappa^2-1}{2}(gW^c_{\mu\nu}+g'B^c_{\mu\nu})\epsilon^{cab} C^{\mu a} C^{\nu b} -\frac{(\kappa^2-1)^2}{8} C^{\mu a} C^{\nu b} \left(C^a_\mu C^b_\nu -C^a_\nu C^b_\mu\right)\\
&\overset{IST}{\longrightarrow} - g^2 \left \langle W_{\mu\nu} W^{\mu\nu}\right\rangle -\frac{1}{2}g^2 B_{\mu\nu} B^{\mu\nu} -2gg'\left\langle W_{\mu\nu} \tau_L\right\rangle B^{\mu\nu} \\
&-i\left(\kappa^2-1\right)\Big( g\left \langle W_{\mu\nu} \left[L^\mu, L^\nu \right]\right\rangle+g'B_{\mu\nu}\left\langle \tau_L \left[L^\mu, L^\nu \right]\right\rangle\Big)\\
& -\frac{(\kappa^2-1)^2}{2}\Big(\left\langle L_\mu L^\mu\right\rangle^2-\left\langle L_\mu L_\nu\right\rangle^2\Big) \ ,
\\[0.8cm ]
& -2X^{ai} X^{ai} \\
&= -2 \Bigg[ \frac{1}{2}\left((\partial_\mu \kappa_i) C^a_\nu-(\partial_\nu \kappa_i) C^a_\mu\right)(\partial^\mu \kappa_i)C^{\nu a}+\frac{\kappa^2}{4}(gW^a_{\mu\nu}-g'B^a_{\mu\nu})^2 \\
&+\frac{\kappa_j \kappa_k}{16} (C^a_\mu \mathcal{X}^{ji}_{\nu}-C^a_\nu \mathcal{X}^{ji}_{ \mu})(C^{\mu a} \mathcal{X}^{\nu ki}-C^{\nu a} \mathcal{X}^{\mu ki}) +\kappa_i(\partial^\mu \kappa_i) C^{\nu a} (gW^a_{\mu\nu}-g'B^a_{\mu\nu}) \\
& -\frac{\kappa_j}{2}(\partial^\mu \kappa_i) C^{\nu a} (C^a_\mu \mathcal{X}^{ji}_{\nu}-C^a_\nu \mathcal{X}^{ji}_{\mu})-\frac{\kappa_i\kappa_j}{2}(gW^a_{\mu\nu}-g'B^a_{\mu\nu})C^{\mu a} \mathcal{X}^{\nu ji}\Bigg]\\
&\overset{IST}{\longrightarrow} -2(\partial_\mu\kappa_i)^2\left\langle L_\mu L^\mu\right\rangle +2 (\partial_\mu \kappa_i)(\partial_\nu \kappa_i)\left\langle L^\mu L^\nu\right\rangle -\kappa^2g^2 \left\langle W_{\mu\nu} W^{\mu\nu}\right\rangle \\
& -\frac{\kappa^2}{2}g'^2B_{\mu\nu}B^{\mu\nu}+2\kappa^2 gg' \left\langle W_{\mu\nu}\tau_L\right\rangle B^{\mu\nu}\\
&-\frac{\kappa_j\kappa_k}{2}\Big(\left\langle L_\mu L^\mu\right\rangle \mathcal{X}^{ji}_{\mu}\mathcal{X}^{\mu ki}-\left\langle L^\mu L^\nu\right\rangle \mathcal{X}^{ji}_{\mu}\mathcal{X}^{ki}_\nu\Big)  \\
&-4\kappa_i(\partial_\mu \kappa_i)\Big( g\left\langle W^{\mu\nu}L_\nu\right\rangle -g'B^{\mu\nu}\left\langle \tau_L L_\nu\right\rangle\Big)\\
&-2\kappa_j \mathcal{X}^{ji}_\mu (\partial^\mu \kappa_i)\left\langle L_\nu L^\nu\right\rangle+\kappa_j\left(\mathcal{X}^{ji}_\mu (\partial_\nu\kappa_i)+\mathcal{X}^{ji}_\nu (\partial_\mu\kappa_i)\right) \left\langle L^\mu L^\nu \right\rangle \\
& +2\kappa_i\kappa_j  \mathcal{X}^{ji}_{\nu}\Big(g\left\langle L_\mu W^{\mu\nu} \right\rangle-\left\langle L_\mu \tau_L \right\rangle g'B^{\mu\nu}\Big) \ , 
\\[0.8cm]
& -X^{ij} X^{ij} \\
& =-\Bigg[ \frac{1}{2}\partial_\mu \mathcal{X}^{ij}_{\nu}- \frac{1}{2}\partial_\nu \mathcal{X}^{ij}_{\mu}+\frac{1}{4}\left[\mathcal{X}^{ik}_\mu  \mathcal{X}^{kj}_\nu-\mathcal{X}^{ik}_\nu  \mathcal{X}^{kj}_\mu   \right]\Bigg]^2 \ .
\end{align*}
Therefore, we obtain

\begin{align*}
& \frac{1}{12}\text{tr}\, X_{\mu\nu} X^{\mu\nu} \\
& =  -\frac{1+\kappa^2}{12}g^2\left\langle W_{\mu\nu} W^{\mu\nu}\right\rangle -\frac{1+\kappa^2}{24}g'^2 B_{\mu\nu} B^{\mu\nu} +\frac{ \kappa^2-1}{6}gg'\left\langle W_{\mu\nu} \tau_L\right\rangle B^{\mu\nu}\\
&-i\frac{\kappa^2-1 }{12}\Big(g\left\langle W_{\mu\nu} \left[L^\mu, L^\nu \right]\right\rangle+g'B_{\mu\nu}\left\langle \tau_L \left[L^\mu, L^\nu \right]\right\rangle\Big) \\
& -\frac{(\kappa^2-1)^2}{24}\Big(\left\langle L_\mu L^\mu\right\rangle^2-\left\langle L_\mu L_\nu\right\rangle^2\Big)\\
&+\frac{1}{12}\left(-2(\partial_\mu\kappa_i)^2 -\frac{\kappa_i\kappa_j}{2} \mathcal{X}^{ji}_{\nu}\mathcal{X}^{ji}_{\nu}-2\kappa_j \mathcal{X}^{ji}_{\mu} (\partial_\mu \kappa_i)  \right) \left\langle L_\mu L^\mu\right\rangle \\
& +\frac{1}{12}\Big(2 (\partial_\mu \kappa_i)(\partial_\nu \kappa_i)  +\frac{\kappa_j\kappa_k}{2}\mathcal{X}^{ji}_{\mu}\mathcal{X}^{ki}_\nu +\kappa_j\left(\mathcal{X}^{ji}_\mu (\partial_\nu\kappa_i)+\mathcal{X}^{ji}_\nu (\partial_\mu\kappa_i)\right)    \Big)\left\langle L^\mu L^\nu\right\rangle \\
& -\frac{1}{3}\kappa_i(\partial_\mu \kappa_i) \Big( g\left\langle W^{\mu\nu}L_\nu\right\rangle -g'B^{\mu\nu}\left\langle \tau_L L_\nu\right\rangle\Big) \\
& - \frac{1}{12}\Bigg[ \frac{1}{2}\partial_\mu \mathcal{X}^{ij}_{\nu}- \frac{1}{2}\partial_\nu \mathcal{X}^{ij}_{\mu}+\frac{1}{4}\left[\mathcal{X}^{ik}_\mu  \mathcal{X}^{kj}_\nu-\mathcal{X}^{ik}_\nu  \mathcal{X}^{kj}_\mu   \right]\Bigg]^2 \ .
\end{align*}

\noindent Next we compute $Z$.

\begin{align*}
Z^{ab} &= gg'B^\mu\left(\frac{1}{2}W_\mu^a \delta^{b3}+\frac{1}{2}W_\mu^b \delta^{a3}-W^3_\mu \delta^{ab}\right)\\
& +\delta^{ab}\left[\frac{(\partial_\mu F_U)^2}{4F_U^2} - \frac{1}{2}\frac{\partial^2 F_U}{F_U}+\frac{-1}{v^2F_U}\bar{\psi}_L M \psi_R+\textrm{h.c.}\right] \ , \\
Z^{ai} &= \frac{1}{\sqrt{F_U}}\Bigg[\left[  \frac{2i}{v}\bar{\psi}_L T^aM^{(j)}\psi_R+ \textrm{h.c.} \right]  -  \frac{F_U^{(j)}}{2} vgg'B^\mu\left(W^1_\mu\delta^{a2}-W^2_\mu \delta^{a1}\right)\\
& + \frac{1}{4} \partial_\mu \left[F^{(j)}_U v\left( gW^{\mu a}-g'B^\mu\delta^{a3} \right)\right]\Bigg]A^{-1/2}_{ji}\\
&+\frac{vF^{(j)}_U}{4\sqrt{F_U}} C^a_\mu  \partial^\mu A^{-1/2}_{ji}-\left(\partial^\mu\frac{1}{\sqrt{F_U}}\right) \frac{vF^{(j)}_U}{4} C^a_\mu  A^{-1/2}_{ji}\\
&= \frac{1}{\sqrt{F_U}}\Bigg[\left[  \frac{2i}{v}\bar{\psi}_L T^aM^{(j)}\psi_R+ \textrm{h.c.} \right]  \\
&-  \frac{F_U^{(j)}}{2} vgg'B^\mu\left(W^1_\mu\delta^{a2}-W^2_\mu \delta^{a1}\right)\Bigg]A^{-1/2}_{ji} \\
& +\partial^\mu\left(\frac{vF^{(j)}_U}{4\sqrt{F_U}} C^a_\mu   A^{-1/2}_{ji}\right) +\frac{vF^{(j)}_U}{2\sqrt{F_U}} C^a_\mu  A^{-1/2}_{ji}\frac{\partial^\mu F_U}{2F_U} \ , \\
Z^{ij} &=A^{-1/2}_{im}\Bigg[ \left[ \bar{\psi}_L M^{(mn)} \psi_R+\textrm{h.c.}\,\right]+V^{(mn)}-\partial_\mu\left[(F^{(k)}_{mn}-F^{(m)}_{nk}-F^{(n)}_{mk})\partial^\mu h_k\right] \\
& -\frac{1}{2}F^{(mn)}_{kl}(\partial_\mu h_k)(\partial^\mu h_l) \Bigg] A^{-1/2}_{nj}\\
&-A^{-1/2}_{im}\left(  \frac{1}{8}g^2W_\mu^a W^{\mu a}+\frac{1}{8}g'^2B_\mu B^\mu -\frac{gg'}{4}B^\mu W_\mu^3\right)v^2 F_U^{(mn)} A^{-1/2}_{nj}\\
&+A^{-1/2}_{ik} B^\mu_{kl} \partial_\mu A^{-1/2}_{kj}-(\partial_\mu A^{-1/2}_{ik})B^\mu_{kl} A^{-1/2}_{kj} \\
& -\frac{1}{2}\left[\partial_\mu A^{-1/2}_{ik}\,\partial^\mu A^{1/2}_{kj}+\partial_\mu A^{1/2}_{ik}\,\partial^\mu A^{-1/2}_{kj}\right] \ . 
\end{align*}
We compute $X_\mu X^\mu$.

\begin{align*}
(X^\mu X_\mu)^{ab} &= -\frac{1}{4} (gW+g'B)^{\mu c} (gW+g'B)^d_\mu \left(f^c f^d\right)_{ab}-\frac{1}{4}\kappa^2C^{\mu a} C^{b}_\mu \ , \\
(X^\mu X_\mu)^{ai} &= -\frac{\kappa_i}{4}\epsilon^{acd} (gW+g'B)^{\mu c} C^d_\mu-\frac{1}{4}\kappa_j C^{\mu a} \mathcal{X}^{ji}_\mu \ , \\
(X^\mu X_\mu)^{ij} &= -\frac{1}{4} \kappa_i\kappa_j C^{\mu a} C^a_\mu -\frac{1}{4}\mathcal{X}^{\mu ik} \mathcal{X}^{jk}_\mu \ .
\end{align*}
This gives us $Y=Z-X_\mu X^\mu$.

\begin{align*}
Y^{ab}
&= gg'B^\mu\left(\frac{1}{2}W_\mu^a \delta^{b3}+\frac{1}{2}W_\mu^b \delta^{a3}-W^3_\mu \delta^{ab}\right) \\
& +\delta^{ab}\left(\frac{(\partial_\mu F_U)^2}{4F_U^2} - \frac{1}{2}\frac{\partial^2 F_U}{F_U}+\frac{-1}{v^2F_U}[\bar{\psi}_L M \psi_R+\textrm{h.c.}] \right)\\
& +\frac{1}{4} (gW+g'B)^{\mu c} (gW+g'B)^d_\mu \left(f^c f^d\right)_{ab}+\frac{1}{4}\kappa^2C^{\mu a} C^b_\mu \ , \\
Y^{ai} 
&= \frac{1}{\sqrt{F_U}}\Bigg[\left[  \frac{2i}{v}\bar{\psi}_L T^aM^{(j)}\psi_R+ \textrm{h.c.} \right]  \\
& -  \frac{F_U^{(j)}}{2} vgg'B^\mu\left(W^1_\mu\delta^{a2}-W^2_\mu \delta^{a1}\right)\Bigg]A^{-1/2}_{ji} +\partial^\mu\left(\frac{vF^{(j)}_U}{4\sqrt{F_U}} C^a_\mu   A^{-1/2}_{ji}\right) \\
&+\frac{vF^{(j)}_U}{2\sqrt{F_U}} C^a_\mu  A^{-1/2}_{ji}\frac{\partial^\mu F_U}{2F_U} +\frac{\kappa_i}{4}\epsilon^{acd} (gW+g'B)^{\mu c} C^d_\mu+\frac{1}{4}\kappa_j C^{\mu a} \mathcal{X}^{ji}_\mu \ , \\
Y^{ij} &=A^{-1/2}_{im}\Bigg[\left[ \bar{\psi}_L M^{(mn)} \psi_R+\textrm{h.c.}\,\right]+ V^{(mn)} \\
&-\partial_\mu\left[(F^{(k)}_{mn}-F^{(m)}_{nk}-F^{(n)}_{mk})\partial^\mu h_k\right]  -\frac{1}{2}F^{(mn)}_{kl}(\partial_\mu h_k)(\partial^\mu h_l) \Bigg] A^{-1/2}_{nj}\\
&-A^{-1/2}_{im}\left(  \frac{1}{8}g^2W_\mu^a W^{\mu a}+\frac{1}{8}g'^2B_\mu B^\mu -\frac{gg'}{4}B^\mu W_\mu^3\right)v^2 F_U^{(mn)} A^{-1/2}_{nj}\\
&+A^{-1/2}_{ik} B^\mu_{kl} \partial_\mu A^{-1/2}_{kj}-(\partial_\mu A^{-1/2}_{ik})B^\mu_{kl} A^{-1/2}_{kj} \\
& -\frac{1}{2}\left[\partial_\mu A^{-1/2}_{ik}\,\partial^\mu A^{1/2}_{kj}+\partial_\mu A^{1/2}_{ik}\,\partial^\mu A^{-1/2}_{kj}\right]+\frac{1}{4} \kappa_i\kappa_j C^{\mu a} C^a_\mu +\frac{1}{4}\mathcal{X}^{\mu ik} \mathcal{X}^{jk}_\mu \ .
\end{align*}
We can simplify them using our generalized shorthands

\begin{equation*}
\kappa_i= \frac{vF^{(j)}_U}{2\sqrt{F_U}} A^{-1/2}_{ji} \, , \ \mathcal{B}_U = \frac{(\partial_\mu F_U)^2}{4F_U^2} - \frac{1}{2}\frac{\partial^2 F_U}{F_U} \ .
\end{equation*}
Else, we use our old known manipulations to obtain

\begin{align*}
Y^{ab} &=  \delta^{ab}\left(\mathcal{B}_U -\frac{1}{v^2F_U}[\bar{\psi}_L M \psi_R+\textrm{h.c.}] +\frac{1}{4} C^c C^c \right) + \frac{1}{4} (\kappa^2-1) C^aC^b \ ,\\
Y^{ai} &=  \frac{A^{-1/2}_{ji}}{\sqrt{F_U}} \left[  \frac{2i}{v}\bar{\psi}_L T^aM^{(j)}\psi_R+ \textrm{h.c.} \right]  + \kappa_i  \, gg'B^\mu W_\mu^b \epsilon^{ab3} \\
&+\frac{1}{2} \partial^\mu\left( \kappa_i C^a_\mu \right) +\kappa_i C^a_\mu \frac{\partial^\mu F_U}{2F_U} +\frac{\kappa_i}{4}\epsilon^{acd} (gW+g'B)^{\mu c} C^d_\mu+\frac{1}{4}\kappa_j C^{\mu a} \mathcal{X}^{ji}_\mu \ , \\
Y^{ij} &=A^{-1/2}_{im}\Bigg[\left[ \bar{\psi}_L M^{(mn)} \psi_R+\textrm{h.c.}\,\right]+ V^{(mn)} \\
& -\partial_\mu\left[(F^{(k)}_{mn}-F^{(m)}_{nk}-F^{(n)}_{mk})\partial^\mu h_k\right] -\frac{1}{2}F^{(mn)}_{kl}(\partial_\mu h_k)(\partial^\mu h_l) \Bigg] A^{-1/2}_{nj}\\
&- \frac{1}{8}v^2A^{-1/2}_{im}F_U^{(mn)}A^{-1/2}_{nj} C^aC^a +\frac{1}{4} \kappa_i\kappa_j C^a C^a\\
&+A^{-1/2}_{ik} B^\mu_{kl} \partial_\mu A^{-1/2}_{kj}-(\partial_\mu A^{-1/2}_{ik})B^\mu_{kl} A^{-1/2}_{kj} \\
& -\frac{1}{2}\left[\partial_\mu A^{-1/2}_{ik}\,\partial^\mu A^{1/2}_{kj}+\partial_\mu A^{1/2}_{ik}\,\partial^\mu A^{-1/2}_{kj}\right] +\frac{1}{4}\mathcal{X}^{ik}_\mu \mathcal{X}^{jk}_\mu \ .
\end{align*}
In analogy to before, we can define

\begin{equation*}
\mathcal{B_H}^{ij} \doteq \frac{1}{4} \kappa^i\kappa^j - \frac{1}{8}v^2A^{-1/2}_{im}F_U^{(mn)}A^{-1/2}_{nj} \ .
\end{equation*}
As compactor for $ Y^{ij} $ we can introduce another new shorthand,

\begin{equation*}
Y^{ij} = A^{-1/2}_{im}\left[ \bar{\psi}_L M^{(mn)} \psi_R+\textrm{h.c.}\,\right] A^{-1/2}_{nj} + \mathcal{K}^{ij} + \mathcal{B_H}^{ij} C^a C^a \ , \\
\end{equation*}
where we have defined

\begin{align*}
\mathcal{K}^{ij} & = A^{-1/2}_{im}\Bigg[ V^{(mn)}-\partial_\mu\left[(F^{(k)}_{mn}-F^{(m)}_{nk}-F^{(n)}_{mk})\partial^\mu h_k\right] \\
& -\frac{1}{2}F^{(mn)}_{kl}(\partial_\mu h_k)(\partial^\mu h_l) \Bigg] A^{-1/2}_{nj} +A^{-1/2}_{ik} B^\mu_{kl} \partial_\mu A^{-1/2}_{lj}-(\partial_\mu A^{-1/2}_{ik})B^\mu_{kl} A^{-1/2}_{lj} \\
& -\frac{1}{2}\left[\partial_\mu A^{-1/2}_{ik}\,\partial^\mu A^{1/2}_{kj}+\partial_\mu A^{1/2}_{ik}\,\partial^\mu A^{-1/2}_{kj}\right] +\frac{1}{4}\mathcal{X}^{\mu ik} \mathcal{X}^{jk}_\mu \ .
\end{align*}
We remember,

\begin{equation*}
\frac{\kappa}{4}\epsilon^{acd} (gW+g'B)^{\mu c} C^d_\mu = \frac{\kappa}{4}g^2 \underbrace{\epsilon^{acd} W_\mu^c W^{\mu d}}_{=0} - \frac{\kappa}{2}gg' B^\mu \epsilon^{ab3} W_\mu^b  \ .
\end{equation*}
$ Y^{ai} $ becomes

\begin{equation*}
\begin{aligned}
Y^{ai} &= \frac{2A^{-1/2}_{ji}}{v\sqrt{F_U}} \left[  i\bar{\psi}_L T^aM^{(j)}\psi_R+ \textrm{h.c.} \right]  + \frac{1}{2}\kappa_i  gg'B^\mu W_\mu^b \epsilon^{ab3} \\
&+\frac{1}{2} \partial^\mu\left( \kappa_i C^a_\mu \right) +\kappa_i C^a_\mu \frac{\partial^\mu F_U}{2F_U} +\frac{1}{4}\kappa_j C^{\mu a} \mathcal{X}^{ji}_\mu \ ,
\end{aligned}
\end{equation*}
We can rewrite it by the use of the covariant derivative, which acts on $ C $ as \eqref{eqn:covariantDC}

\begin{equation*}
D_\mu C^a= \partial_\mu C^a +gg'B_\mu W^b \epsilon^{ab3} \ ,
\end{equation*}
and on $ F_U $, a function of singlets, as usual derivative.
Then,

\begin{align*}
Y^{ai} &= \frac{2A^{-1/2}_{ji}}{v\sqrt{F_U}}  \left[  i\bar{\psi}_L T^aM^{(j)}\psi_R+ \textrm{h.c.} \right]  +\frac{1}{4}\kappa_j C^a_\mu \mathcal{X}^{\mu ji} +\frac{1}{2} (\partial^\mu  \kappa_i ) C^a_\mu \\
&+ \frac{1}{2} \kappa_i D_\mu C^{\mu a} +\kappa_i  \frac{1}{2F_U} C^{\mu a} D_\mu F_U  \\
& = \frac{2A^{-1/2}_{ji}}{v\sqrt{F_U}}  \left[  i\bar{\psi}_L T^aM^{(j)}\psi_R+ \textrm{h.c.} \right]  +\frac{1}{4}\kappa_j C^a_\mu \mathcal{X}^{\mu ji} \\
& +\frac{1}{2} (\partial_\mu  \kappa_i ) C^{\mu a} + \frac{\kappa_i}{2F_U}  D_\mu (C^{\mu a} F_U)  \ .
\end{align*}
We can simplify $ Y^{ai} $ further via the equations of motion of the Goldstone fields (\ref{eqn:eomGold1}),

\begin{equation*}
D_\mu (C^a F_U) + \frac{4}{v^2} \left[ i\bar{\psi}_L T^aM\psi_R+ \textrm{h.c.} \right] = 0 \ .
\end{equation*}
We end up with

\begin{align*}
Y^{ai} & = \frac{2A^{-1/2}_{ji}}{v\sqrt{F_U}}  \left[  i\bar{\psi}_L T^aM^{(j)}\psi_R+ \textrm{h.c.} \right]  +\frac{1}{4}\kappa_j C^a_\mu \mathcal{X}^{\mu ji} \\
& +\frac{1}{2} (\partial^\mu  \kappa_i ) C^a_\mu - \frac{\kappa_i}{F_U} 2\left[  i\bar{\psi}_L T^aM \psi_R+ \textrm{h.c.} \right]   \\
&= \frac{2A^{-1/2}_{ji}}{v } \left[  i\bar{\psi}_L T^a \left( \frac{M}{\sqrt{F_U}} \right)^{(j)} \psi_R+ \textrm{h.c.} \right]  +\frac{1}{4}\kappa_j C^a_\mu \mathcal{X}^{\mu ji} +\frac{1}{2} (\partial^\mu  \kappa_i ) C^a_\mu \ .
\end{align*}
We compute tr $ Y^2= Y^{ab}Y^{ab} + 2 Y^{ai}Y^{ai} + Y^{ij}Y^{ij} $ in parts.

\begin{align*}
&Y^{ab}Y^{ab} \\
&= 3\left( \mathcal{B}_U - \frac{1}{v^2F_U} [\bar{\psi}_L M \psi_R +\textrm{h.c.} ]+ \frac{1}{4}C^cC^c \right)^2 \\
& + \frac{1}{16} \left(\kappa^2 -1 \right)^2 C_\mu^a C^{\nu a} C_\mu^b C^{\nu b} \\
&+\frac{1}{2} \left( \mathcal{B}_U - \frac{1}{v^2F_U} [\bar{\psi}_L M \psi_R +\textrm{h.c.} ]+ \frac{1}{4}C^cC^c \right) \left(\kappa^2-1\right) C^a C^{ a} \\
& = 3 \mathcal{B}_U^2 + \frac{3}{v^4F_U^2} [\bar{\psi}_L M \psi_R +\textrm{h.c.} ]^2 +\frac{3}{16}(C^c C^c)^2 \\
& -6 \frac{\mathcal{B}_U}{v^2F_U} [\bar{\psi}_L M \psi_R +\textrm{h.c.} ] + \frac{3\mathcal{B}_U}{2} C^cC^c \\
&-  \frac{3}{2v^2F_U} [\bar{\psi}_L M \psi_R +\textrm{h.c.} ] \ C^cC^c +\frac{1}{16} \left(\kappa^2-1 \right)^2 C_\mu^a C^{\nu a} C_\mu^b C^{\nu b} \\
&+ \frac{1}{2}\mathcal{B}_U \left(\kappa^2-1 \right) C^aC^a - \frac{1}{2v^2F_U} [\bar{\psi}_L M \psi_R +\textrm{h.c.} ]\left( \kappa^2-1 \right) C^aC^a \\
& +\frac{1}{8} \left( \kappa^2-1\right) (C^aC^a)^2\\
&\overset{IST}{\longrightarrow} \ 3 \mathcal{B}_U^2 + \frac{3}{v^4F_U^2} [\bar{\psi}_L U M \psi_R +\textrm{h.c.} ]^2 +\frac{3}{4} \langle L_\mu L^\mu \rangle^2 \\
& -6 \frac{\mathcal{B}_U}{v^2F_U} [\bar{\psi}_LU M \psi_R +\textrm{h.c.} ] + 3\mathcal{B}_U \langle L_\mu L^\mu \rangle \\
&-  \frac{3}{v^2F_U} [\bar{\psi}_L U M \psi_R +\textrm{h.c.} ] \langle L_\mu L^\mu \rangle +\frac{1}{4}\left(\kappa^2-1 \right)^2 \langle L_\mu L_\nu \rangle^2 \\
&+\mathcal{B}_U \left( \kappa^2-1 \right) \langle L_\mu L^\mu \rangle- \frac{1}{v^2F_U} [\bar{\psi}_L M \psi_R +\textrm{h.c.} ]\left(\kappa^2-1 \right) \langle L_\mu L^\mu \rangle \\
&+ \frac{1}{2}\left(\kappa^2-1\right) \langle L_\mu L^\mu \rangle^2 \ ,
\\[0.8cm]
&Y^{ai}Y^{ai}\\
&= \frac{4}{v^2} \left( A^{-1/2}_{ji} \left[  i\bar{\psi}_L T^a \left( \frac{M}{\sqrt{F_U}} \right)^{(j)}  \psi_R+ \textrm{h.c.} \right]\right) ^2 \\
&+ \frac{1}{16}\kappa_n \kappa_m (C^a_\mu \mathcal{X}_{ni}^\mu )(C^a_\mu \mathcal{X}_{mi}^\mu )  + \frac{1}{4} \partial^\mu \kappa_i \partial^\nu \kappa_i C_\mu^aC_\nu^a \\
&+  \frac{2A^{-1/2}_{ji}}{v}\left[  i\bar{\psi}_L T^a \left( \frac{M}{\sqrt{F_U}} \right)^{(j)}  \psi_R+ \textrm{h.c.} \right](\partial_\mu \kappa_i) C^{\mu a} \\
& + \frac{\kappa_n A^{-1/2}_{ji} }{v}\left[  i\bar{\psi}_L T^a \left( \frac{M}{\sqrt{F_U}} \right)^{(j)}  \psi_R+ \textrm{h.c.} \right]\mathcal{X}_{ni}^\mu C_\mu^a \\
& + \frac{1}{4} \kappa_j (\partial_\mu \kappa_i)C^{\mu a} (\mathcal{X}_{ji}^\nu C_\nu^a) \\
& \overset{IST}{\longrightarrow} \  \frac{4}{v^2}\left( A^{-1/2}_{ji}\left[  i\bar{\psi}_L U \left( \frac{M}{\sqrt{F_U}} \right)^{(j)}  \psi_R+ \textrm{h.c.} \right]\right) ^2 \\
&+ \frac{1}{8}\kappa_n \kappa_m \mathcal{X}_{ni}^\mu \mathcal{X}_{mi}^\nu \langle L_\mu L_\nu\rangle + \frac{1}{2} \partial^\mu \kappa_i \partial^\nu \kappa_i \langle L_\mu L_\nu \rangle \\
&+ \frac{2A^{-1/2}_{ji}}{v}\left[  i\bar{\psi}_L L_\mu U \left( \frac{M}{\sqrt{F_U}} \right)^{(j)}  \psi_R+ \textrm{h.c.} \right](\partial^\mu \kappa_i) \\
& + \frac{\kappa_n A^{-1/2}_{ji}}{v}\left[  i\bar{\psi}_L L_\mu U \left( \frac{M}{\sqrt{F_U}} \right)^{(j)}  \psi_R+ \textrm{h.c.} \right] \mathcal{X}_{ni}^\mu \\
& + \frac{1}{4} \kappa_j \left( (\partial^\mu \kappa_i)\mathcal{X}_{ji}^\nu + (\partial^\nu \kappa_i) \mathcal{X}^\mu_{ji}\right)  \langle L_\mu L_\nu \rangle \ ,
\\[0.8cm ]
&Y^{ij}Y^{ij} \\
& =  \, \overset{IST}{\longrightarrow}  \left[ \bar{\psi}_L UM^{(mn)} \psi_R+\textrm{h.c.}\,\right]A^{-1}_{mp}A^{-1}_{nq}\left[ \bar{\psi}_L U M^{(pq)} \psi_R+\textrm{h.c.}\,\right] \\
&+\mathcal{K}^{ij}\mathcal{K}^{ij} + 4 \mathcal{B_H}^{ij}\mathcal{B_H}^{ij} \langle L_\mu L^\mu \rangle^2 \\
& +4 \mathcal{K}^{ij}\mathcal{B_H}^{ij}\langle L_\mu L^\mu \rangle + 2A^{-1/2}_{im}\left[ \bar{\psi}_L U M^{(mn)} \psi_R+\textrm{h.c.}\,\right]A^{-1/2}_{jn} \mathcal{K}^{ij} \\
& + 4A^{-1/2}_{im}\left[ \bar{\psi}_LU  M^{(mn)} \psi_R+\textrm{h.c.}\,\right]A^{-1/2}_{jn}\mathcal{B_H}^{ij} \langle L_\mu L^\mu \rangle \ .
\end{align*}
Therefore, we obtain

\begin{align*}
& \frac{1}{2}\text{tr}\, Y^2 \\
& =  \frac{3}{2} \mathcal{B}_U^2 + \frac{3}{2v^4F_U^2} [\bar{\psi}_L U M \psi_R +\textrm{h.c.} ]^2 +\frac{3}{8} \langle L_\mu L^\mu \rangle^2 -3 \frac{\mathcal{B}_U}{v^2F_U} [\bar{\psi}_LU M \psi_R +\textrm{h.c.} ] \\
&+ \frac{3}{2}\mathcal{B}_U \langle L_\mu L^\mu \rangle -  \frac{3}{2v^2F_U} [\bar{\psi}_L U M \psi_R +\textrm{h.c.} ] \langle L_\mu L^\mu \rangle +\frac{1}{8}\left(\kappa^2-1 \right)^2 \langle L_\mu L_\nu \rangle^2 \\
&+\frac{1}{2}\mathcal{B}_U \left( \kappa^2-1 \right) \langle L_\mu L^\mu \rangle- \frac{1}{2v^2F_U} [\bar{\psi}_L M \psi_R +\textrm{h.c.} ]\left(\kappa^2-1 \right) \langle L_\mu L^\mu \rangle \\
&+ \frac{1}{4}\left(\kappa^2-1\right) \langle L_\mu L^\mu \rangle^2 +\frac{4}{v^2}\left( A^{-1/2}_{ji}\left[  i\bar{\psi}_L U \left( \frac{M}{\sqrt{F_U}} \right)^{(j)}  \psi_R+ \textrm{h.c.} \right]\right) ^2 \\
& + \frac{1}{8}\kappa_n \kappa_m \mathcal{X}_{ni}^\mu \mathcal{X}_{mi}^\nu \langle L_\mu L_\nu\rangle + \frac{1}{2} \partial^\mu \kappa_i \partial^\nu \kappa_i \langle L_\mu L_\nu \rangle \\
&+ \frac{2A^{-1/2}_{ji}}{v}\left[  i\bar{\psi}_L L_\mu U \left( \frac{M}{\sqrt{F_U}} \right)^{(j)}  \psi_R+ \textrm{h.c.} \right](\partial^\mu \kappa_i) \\
& + \frac{\kappa_n A^{-1/2}_{ji}}{v}\left[  i\bar{\psi}_L L_\mu U \left( \frac{M}{\sqrt{F_U}} \right)^{(j)}  \psi_R+ \textrm{h.c.} \right] \mathcal{X}_{ni}^\mu \\
&+ \frac{1}{4} \kappa_j \left( (\partial^\mu \kappa_i)\mathcal{X}_{ji}^\nu + (\partial^\nu \kappa_i) \mathcal{X}^\mu_{ji}\right)  \langle L_\mu L_\nu \rangle \\
& + \frac{1}{2}\left[ \bar{\psi}_L UM^{(mn)} \psi_R+\textrm{h.c.}\,\right]A^{-1}_{mp}A^{-1}_{nq}\left[ \bar{\psi}_L U M^{(pq)} \psi_R+\textrm{h.c.}\,\right]+\frac{1}{2}\mathcal{K}^{ij}\mathcal{K}^{ij} \\
&+ 2 \mathcal{B_H}^{ij}\mathcal{B_H}^{ij} \langle L_\mu L^\mu \rangle^2 +2 \mathcal{K}^{ij}\mathcal{B_H}^{ij}\langle L_\mu L^\mu \rangle \\
& + A^{-1/2}_{im}\left[ \bar{\psi}_L U M^{(mn)} \psi_R+\textrm{h.c.}\,\right]A^{-1/2}_{jn} \mathcal{K}^{ij} \\
& + 2A^{-1/2}_{im}\left[ \bar{\psi}_LU  M^{(mn)} \psi_R+\textrm{h.c.}\,\right]A^{-1/2}_{jn}\mathcal{B_H}^{ij} \langle L_\mu L^\mu \rangle \ .
\end{align*}

\subsection{Final Divergence}	

For the divergence of the theory with respect to scalar loops, we combine our results in the master formula by t'Hooft \eqref{eqn:tHooft},

\begin{equation*}
\mathcal{L}_{\text{eff,div}}^{\text{1loop}} = \dfrac{1}{32\pi^2 \epsilon} \text{tr} \left[ \frac{1}{12} X_{\mu\nu}X^{\mu\nu} + \frac{1}{2} Y^2 \right] \ .
\end{equation*}
This gives us

\begin{align*}
& \mathcal{L}_{\text{eff,div}}^{\text{1loop}} = \frac{1}{32\pi^2 \epsilon} \ \times \\
& \Bigg[  -\frac{1+\kappa^2}{12}g^2\left\langle W_{\mu\nu} W^{\mu\nu}\right\rangle -\frac{1+\kappa^2}{24}g'^2 B_{\mu\nu} B^{\mu\nu} +\frac{ \kappa^2-1}{6}gg'\left\langle W_{\mu\nu} \tau_L\right\rangle B^{\mu\nu}\\
&-i\frac{\kappa^2-1 }{12}\Big(g\left\langle W_{\mu\nu} \left[L^\mu, L^\nu \right]\right\rangle+g'B_{\mu\nu}\left\langle \tau_L \left[L^\mu, L^\nu \right]\right\rangle\Big)  \\
& -\frac{1}{3}\kappa_i(\partial_\mu \kappa_i) \Big( g\left\langle W^{\mu\nu}L_\nu\right\rangle -g'B^{\mu\nu}\left\langle \tau_L L_\nu\right\rangle\Big)+ \frac{1}{6} (\kappa^2-1)^2 \langle L_\mu L_\nu \rangle^2 \\
& +\left[\frac{3}{8}+ \frac{1}{4} (\kappa^2-1) - \frac{1}{24} (\kappa^2-1)^2 + 2 \mathcal{B_H}^{ij}\mathcal{B_H}^{ij} \right] \langle L_\mu L^\mu \rangle^2 \\
& + \Bigg[\frac{1}{12}\left(-2(\partial_\nu\kappa_i)^2 -\frac{\kappa_i\kappa_j}{2} \mathcal{X}^{ji}_{\nu}\mathcal{X}^{\nu ji}-2\kappa_j \mathcal{X}^{ji}_{\nu} (\partial^\nu \kappa_i)  \right) \\
& -  \frac{3}{2v^2F_U} [\bar{\psi}_L U M \psi_R +\textrm{h.c.} ] +\frac{1}{2}\mathcal{B}_U \left( \kappa^2+2 \right) \\
&- \frac{1}{2v^2F_U} [\bar{\psi}_L M \psi_R +\textrm{h.c.} ]\left(\kappa^2-1 \right) \\ 
& + 2A^{-1/2}_{im}\left[ \bar{\psi}_LU  M^{(mn)} \psi_R+\textrm{h.c.}\,\right]A^{-1/2}_{jn}\mathcal{B_H}^{ij} +2 \mathcal{K}^{ij}\mathcal{B_H}^{ij} \Bigg]\langle L_\mu L^\mu \rangle \\
& \Bigg[ \frac{1}{12}\Big(2 (\partial^\mu \kappa_i)(\partial^\nu \kappa_i) +\frac{\kappa_j\kappa_k}{2}\mathcal{X}^{\mu ji}\mathcal{X}^{\nu ki} +\kappa_j\left(\mathcal{X}^{\mu ji} (\partial^\nu\kappa_i)+\mathcal{X}^{\nu ji} (\partial^\mu\kappa_i)\right)    \Big) \\
& + \frac{1}{8}\kappa_n \kappa_m \mathcal{X}_{ni}^\mu \mathcal{X}_{mi}^\nu  + \frac{1}{2} \partial^\mu \kappa_i \partial^\nu \kappa_i  \\
&+ \frac{1}{4} \kappa_j \left( (\partial^\mu \kappa_i)\mathcal{X}_{ji}^\nu + (\partial^\nu \kappa_i) \mathcal{X}^\mu_{ji}\right)  \Bigg]\langle L_\mu L_\nu\rangle\quad \\
&+ \frac{3}{2v^4F_U^2} [\bar{\psi}_L U M \psi_R +\textrm{h.c.} ]^2 -3 \frac{\mathcal{B}_U}{v^2F_U} [\bar{\psi}_LU M \psi_R +\textrm{h.c.} ]  \\
& +\frac{4}{v^2}\left(A^{-1/2}_{ji} \left[  i\bar{\psi}_L U \left( \frac{M}{\sqrt{F_U}} \right)^{(j)}  \psi_R+ \textrm{h.c.} \right]\right) ^2 \\
&+ \frac{2 A^{-1/2}_{ji}}{v}\left[  i\bar{\psi}_L L_\mu U \left( \frac{M}{\sqrt{F_U}} \right)^{(j)}  \psi_R+ \textrm{h.c.} \right](\partial^\mu \kappa_i) \\
& + \frac{\kappa_n A^{-1/2}_{ji}}{v}\left[  i\bar{\psi}_L L_\mu U \left( \frac{M}{\sqrt{F_U}} \right)^{(j)}  \psi_R+ \textrm{h.c.} \right] \mathcal{X}_{ni}^\mu\\
& + \frac{1}{2}\left[ \bar{\psi}_L UM^{(mn)} \psi_R+\textrm{h.c.}\,\right]A^{-1}_{mp}A^{-1}_{nq}\left[ \bar{\psi}_L U M^{(pq)} \psi_R+\textrm{h.c.}\,\right] \\
& + A^{-1/2}_{im}\left[ \bar{\psi}_L U M^{(mn)} \psi_R+\textrm{h.c.}\,\right]A^{-1/2}_{jn} \mathcal{K}^{ij} \\
& +\frac{3}{2} \mathcal{B}_U^2 +\frac{1}{2}\mathcal{K}^{ij}\mathcal{K}^{ij}- \frac{1}{12}\left(  \frac{1}{2}\partial_\mu \mathcal{X}^{ij}_{\nu}- \frac{1}{2}\partial_\nu \mathcal{X}^{ij}_{\mu}+\frac{1}{4}\left[\mathcal{X}^{ik}_\mu  \mathcal{X}^{kj}_\nu-\mathcal{X}^{ik}_\nu  \mathcal{X}^{kj}_\mu   \right]\right) ^2  \Bigg] \ .
\end{align*}	
We give our collection of shorthands:

\begin{align*}
\kappa_i &= \frac{vF^{(j)}_U}{2\sqrt{F_U}} A^{-1/2}_{ji}\,,\quad\quad \kappa^2=\sum_i \kappa_i^2 \ , \\
\mathcal{X}_\mu^{ij}& = A^{-1/2}_{ik}\left(F^{(l)}_{km}-F^{(k)}_{lm}\right)(\partial_\mu h_m) A^{-1/2}_{lj}+A^{1/2}_{ik} \partial_\mu A^{-1/2}_{kj}-\partial_\mu A^{-1/2}_{ik}A^{1/2}_{kj} \, , \\
\mathcal{B}_U &= \frac{(\partial_\mu F_U)^2}{4F_U^2} - \frac{1}{2}\frac{\partial^2 F_U}{F_U}\,,\quad\quad \mathcal{B_H}^{ij} = \frac{1}{4} \kappa^i\kappa^j - \frac{1}{8}v^2A^{-1/2}_{im}F_U^{(mn)}A^{-1/2}_{nj} \ , \\
\mathcal{K}^{ij} & = A^{-1/2}_{im}\Bigg[ V^{(mn)}-\partial_\mu\left[(F^{(k)}_{mn}-F^{(m)}_{nk}-F^{(n)}_{mk})\partial^\mu h_k\right] \\
& -\frac{1}{2}F^{(mn)}_{kl}(\partial_\mu h_k)(\partial^\mu h_l) \Bigg] A^{-1/2}_{nj}\\
&+\frac{1}{2}\Big[A^{-1/2}_{ik}\left(F^{(l)}_{km}-F^{(k)}_{lm}\right)(\partial^\mu h_m) \partial_\mu A^{-1/2}_{lj} \\
&-(\partial_\mu A^{-1/2}_{ik})\left(F^{(l)}_{km}-F^{(k)}_{lm}\right)(\partial^\mu h_m) A^{-1/2}_{lj}   \Big]\\
&-\frac{1}{2}\left[\partial_\mu A^{-1/2}_{ik}\,\partial^\mu A^{1/2}_{kj}+\partial_\mu A^{1/2}_{ik}\,\partial^\mu A^{-1/2}_{kj}\right] +\frac{1}{4}\mathcal{X}^{ik}_\mu \mathcal{X}^{\mu jk} \ . 
\end{align*}
We can rewrite our result again by the use of the equations of motion of the singlets.
These are given by 

\begin{equation}
\begin{aligned}
\partial^2 h_i =& (\partial h_m)(\partial h_n) A^{-1}_{ij} \left(F^{(j)}_{mn}-\frac{1}{2}F^{(n)}_{jm}-\frac{1}{2}F^{(m)}_{jn}\right)+\frac{v^2}{4}\langle L_\mu L^\mu \rangle A^{-1}_{ij} F^{(j)}_U \\
& -A^{-1}_{ij} V^{(j)}-A^{-1}_{ij} \left[ \bar{\psi}_L U M^{(j)}\psi_R +\textrm{h.c.}\right]        \\
\overset{S.T.}{\longrightarrow}& (\partial h_m)(\partial h_n) A^{-1}_{ij} \left(F^{(j)}_{mn}-\frac{1}{2}F^{(n)}_{jm}-\frac{1}{2}F^{(m)}_{jn}\right)+\frac{v^2}{8} A^{-1}_{ij} F^{(j)}_U C^a C^a \\& -A^{-1}_{ij} V^{(j)}
-A^{-1}_{ij} \left[ \bar{\psi}_L M^{(j)}\psi_R +\textrm{h.c.}\right]    \ .
\end{aligned}
\end{equation}
Therefore, we can express $B_U$ as 

\begin{align*}
\mathcal{B}_U & =\frac{(\partial F_U)^2}{4F_U^2} - \frac{1}{2}\frac{\partial^2 F_U}{F_U}=\left(\frac{F^{(i)}F^{(j)}}{4F_U^2} -\frac{F_U^{(ij)}}{2F_U} \right) (\partial h_i)(\partial h_j)-\frac{F_U^{(i)}}{2F_U}\partial^2 h_i\\
&\overset{e.o.m}{\longrightarrow} \left(\frac{F^{(i)}F^{(j)}}{4F_U^2} -\frac{F_U^{(ij)}}{2F_U} \right) (\partial h_i)(\partial h_j) \\
& -(\partial h_m)(\partial h_n)\frac{F_U^{(i)}}{2F_U} A^{-1}_{ij}\left(F^{(j)}_{mn}-\frac{1}{2}F^{(n)}_{jm}-\frac{1}{2}F^{(m)}_{jn}\right) \\
& -\frac{v^2}{16 F_U} F_U^{(i)}A^{-1}_{ij} F^{(j)}_U C^a C^a +\frac{1}{2 F_U} F^{(i)}_U A^{-1}_{ij} \left(V^{(j)}+ \left[ \bar{\psi}_L M^{(j)}\psi_R +\textrm{h.c.} \right] \right)\\
& \doteq (\partial h_i)(\partial h_j) D^{ij}-\frac{v^2}{16 F_U} F_U^{(i)}A^{-1}_{ij} F^{(j)}_U C^a C^a \\
&+\frac{1}{2 F_U} F^{(i)}_U A^{-1}_{ij} \left(V^{(j)}+\left[ \bar{\psi}_L M^{(j)}\psi_L +\textrm{h.c.} \right] \right)
\end{align*}
Then, $Y^{ab}$ becomes 

\begin{equation*}
\begin{aligned}
Y^{ab} &=  \delta^{ab}\left(\mathcal{B}_U -\frac{1}{v^2F_U}[\bar{\psi}_L M \psi_R+\textrm{h.c.}] +\frac{1}{4} C^c C^c \right) + \frac{1}{4} (\kappa^2-1) C^aC^b  \\
&\overset{e.o.m}{\longrightarrow} \delta^{ab}\Bigg[ (\partial h_i)(\partial h_j) D^{ij}+\mathcal{Y} -\frac{1}{4}\left(\kappa^2-1\right) C^c C^c \Bigg] + \frac{1}{4} (\kappa^2-1) C^aC^b \ ,\\
\end{aligned}
\end{equation*}
where 

\begin{align*}
D^{ij} &= \left(\frac{F^{(i)}F^{(j)}}{4F_U^2} -\frac{F_U^{(ij)}}{2F_U}\right) -\frac{F_U^{(m)}}{2F_U} A^{-1}_{mn}\left(F^{(n)}_{ij}-\frac{1}{2}F^{(j)}_{ni}-\frac{1}{2}F^{(i)}_{nj}\right)  \\
& \doteq \mathcal{B}^{ij}-\frac{F_U^{(m)}}{2F_U} A^{-1}_{mn}\left(F^{(n)}_{ij}-\frac{1}{2}F^{(j)}_{ni}-\frac{1}{2}F^{(i)}_{nj}\right) \ , \\
\mathcal{Y} &=\frac{1}{2 F_U} F^{(i)}_U A^{-1}_{ij} \left(V^{(j)}+\left[ \bar{\psi}_L M^{(j)}\psi_R +\textrm{h.c.} \right] \right)  -\frac{1}{v^2F_U}[\bar{\psi}_L M \psi_R+\textrm{h.c.}] \ .
\end{align*}
Now, we compute 

\begin{align*}
& \frac{1}{2}Y^{ab}Y^{ab} \\
&= \frac{1}{2}\Bigg\{3\left[ (\partial h_i)(\partial h_j) D^{ij}+\mathcal{Y} -\frac{1}{4}\left(\kappa^2-1\right) C^c C^c \right]^2 \\
& + \frac{1}{2} (\kappa^2-1) \left[ (\partial h_i)(\partial h_j) D^{ij}+\mathcal{Y} -\frac{1}{4}\left(\kappa^2-1\right) C^c C^c \right] C^aC^a \\
&+\frac{1}{16} (\kappa^2-1)^2 C^a_\mu C^{\mu b}  C^a_\nu C^{\nu b} 
\Bigg\}\\
&\overset{IST}{\longrightarrow} \frac{1}{2}\Bigg\{\frac{1}{4}\left(\kappa^2-1\right)^2\left[ \langle L_\mu L^\mu\rangle^2+ \langle L_\mu L_\nu\rangle^2 \right] \\
& -2\left(\kappa^2-1\right)\left[ (\partial h_i)(\partial h_j) D^{ij}+\tilde{\mathcal{Y}} \right]\langle L_\mu L^\mu \rangle + 3\left[(\partial h_i)(\partial h_j) D^{ij}\right]^2\\
&+3\mathcal{Y}^2+6\mathcal{Y}(\partial h_i)(\partial h_j) D^{ij}
\Bigg\}\\
&= \frac{1}{8}\left(\kappa^2-1\right)^2\left[ \langle L_\mu L^\mu\rangle^2+ \langle L_\mu L_\nu\rangle^2 \right]-\left(\kappa^2-1\right)\left[ (\partial h_i)(\partial h_j) D^{ij}+\tilde{\mathcal{Y}} \right]\langle L_\mu L^\mu \rangle \\ & + \frac{3}{2}\left[(\partial h_i)(\partial h_j) D^{ij}\right]^2
+\frac{3}{2} \tilde{\mathcal{Y}}^2+3 \tilde{\mathcal{Y}}(\partial h_i)(\partial h_j) D^{ij} \ .
\end{align*}
We defined the $ \mathcal{Y} $ after inverting the Stückelberg transformation to be

\begin{equation*}
\tilde{\mathcal{Y}} =\frac{1}{2 F_U} F^{(i)}_U A^{-1}_{ij} \left(V^{(j)}+\left[ \bar{\psi}_L U M^{(j)}\psi_R +\textrm{h.c.} \right] \right)  -\frac{1}{v^2F_U}[\bar{\psi}_L U M \psi_R+\textrm{h.c.}] \ .
\end{equation*}
We also have another source for $\partial^2 h_i$ terms.
From 

\begin{align*}
& \mathcal{K}^{ij} \\
& = A^{-1/2}_{im}\Bigg[ V^{(mn)}-\partial_\mu\left[(F^{(k)}_{mn}-F^{(m)}_{nk}-F^{(n)}_{mk})\partial^\mu h_k\right] \\
& -\frac{1}{2}F^{(mn)}_{kl}(\partial_\mu h_k)(\partial^\mu h_l) \Bigg] A^{-1/2}_{nj}\\
&+\frac{1}{2}\Big[A^{-1/2}_{ik}\left(F^{(l)}_{km}-F^{(k)}_{lm}\right)(\partial^\mu h_m) \partial_\mu A^{-1/2}_{lj} \\
& -(\partial_\mu A^{-1/2}_{ik})\left(F^{(l)}_{km}-F^{(k)}_{lm}\right)(\partial^\mu h_m) A^{-1/2}_{lj}   \Big]\\
&-\frac{1}{2}\left[\partial_\mu A^{-1/2}_{ik}\,\partial^\mu A^{1/2}_{kj}+\partial_\mu A^{1/2}_{ik}\,\partial^\mu A^{-1/2}_{kj}\right] +\frac{1}{4}\mathcal{X}^{ik}_\mu \mathcal{X}^{jk}_\mu\\
& =  \Bigg\{A^{-1/2}_{im}\Bigg[ V^{(mn)}-\left(F^{(kl)}_{mn}-F^{(ml)}_{nk}-F^{(nl)}_{mk}  -\frac{1}{2}F^{(mn)}_{kl} 			\right)(\partial_\mu h_k)(\partial^\mu h_l) \Bigg] A^{-1/2}_{nj}\\
&+\frac{1}{2}\Big[A^{-1/2}_{ik}\left(F^{(l)}_{km}-F^{(k)}_{lm}\right)(\partial_\mu h_m) \partial_\mu A^{-1/2}_{lj} \\
& -(\partial_\mu A^{-1/2}_{ik})\left(F^{(l)}_{km}-F^{(k)}_{lm}\right)(\partial_\mu h_m) A^{-1/2}_{lj}   \Big]\\
&-\frac{1}{2}\left[\partial_\mu A^{-1/2}_{ik}\,\partial^\mu A^{1/2}_{kj}+\partial_\mu A^{1/2}_{ik}\,\partial^\mu A^{-1/2}_{kj}\right] +\frac{1}{4}\mathcal{X}^{ik}_\mu \mathcal{X}^{\mu jk} \Bigg\} \\
& -A^{-1/2}_{im}(F^{(k)}_{mn}-F^{(m)}_{nk}-F^{(n)}_{mk})A^{-1/2}_{nj} \partial^2 h_k\\
&\doteq \mathcal{Q}^{ij}-\mathcal{T}^{ijk} \partial^2 h_k\\
&\overset{e.o.m}{\longrightarrow} \mathcal{Q}^{ij}-\mathcal{T}^{ijk}A^{-1}_{kr}\left(F^{(r)}_{pq}-F^{(p)}_{rq}- F^{(q)}_{rp} \right)(\partial h_p)(\partial h_q)+   \mathcal{T}^{ijk} A^{-1}_{kp}V^{(p)} \\
& - \frac{v^2}{8}\mathcal{T}^{ijk} A^{-1}_{kp} F^{(p)}_{U} C^a C^a+ \mathcal{T}^{ijk} A^{-1}_{kp} \left[ \bar{\psi}_L M^{(p)}\psi_R +\textrm{h.c.}\right] \\
& \doteq \mathcal{S}^{ij}- \frac{v^2}{8}\mathcal{T}^{ijk} A^{-1}_{kp} F^{(p)}_{U} C^a C^a+ \mathcal{T}^{ijk} A^{-1}_{kp} \left[ \bar{\psi}_L M^{(p)}\psi_R +\textrm{h.c.} \right]  \ ,
\end{align*}
where

\begin{align*}
& \mathcal{S}^{ij} \\
&=A^{-1/2}_{im}\Bigg[ V^{(mn)}-\left(F^{(kl)}_{mn}-F^{(ml)}_{nk}-F^{(nl)}_{mk}  -\frac{1}{2}f^{(mn)}_{kl} 			\right)(\partial_\mu h_k)(\partial^\mu h_l) \Bigg] A^{-1/2}_{nj}\\
&+\frac{1}{2}\Big[A^{-1/2}_{ik}\left(F^{(l)}_{km}-F^{(k)}_{lm}\right)(\partial^\mu h_m) \partial_\mu A^{-1/2}_{lj} \\
& -(\partial_\mu A^{-1/2}_{ik})\left(F^{(l)}_{km}-F^{(k)}_{lm}\right)(\partial^\mu h_m) A^{-1/2}_{lj}   \Big]\\
&-\frac{1}{2}\left[\partial_\mu A^{-1/2}_{ik}\,\partial^\mu A^{1/2}_{kj}+\partial_\mu A^{1/2}_{ik}\,\partial^\mu A^{-1/2}_{kj}\right] +\frac{1}{4}\mathcal{X}^{ik}_\mu \mathcal{X}^{\mu jk} \\
& -\mathcal{T}^{ijk}A^{-1}_{kr}\left(F^{(r)}_{pq}-F^{(p)}_{rq}- F^{(q)}_{rp} \right)(\partial h_p)(\partial h_q)+   \mathcal{T}^{ijk} A^{-1}_{kp}V^{(p)} \ , \\
&\mathcal{T}^{ijk} = A^{-1/2}_{im}(F^{(k)}_{mn}-F^{(m)}_{nk}-F^{(n)}_{mk})A^{-1/2}_{nj} \ .
\end{align*}
We therefore have

\begin{align*}
& Y^{ij} \\
&= A^{-1/2}_{im}\left[ \bar{\psi}_L M^{(mn)} \psi_R+\textrm{h.c.}\,\right] A^{-1/2}_{nj} + \mathcal{K}^{ij} + \mathcal{B_H}^{ij} C^a C^a \ , \\
&\overset{e.o.m}{\longrightarrow} A^{-1/2}_{im}\left[ \bar{\psi}_L M^{(mn)} \psi_R+\textrm{h.c.}\,\right] A^{-1/2}_{nj} + \mathcal{S}^{ij}- \frac{v^2}{8}\mathcal{T}^{ijk} A^{-1}_{kp} F^{(p)}_{U} C^a C^a \\
& + \mathcal{T}^{ijk} A^{-1}_{kp} \left[ \bar{\psi}_L M^{(p)}\psi_R+\textrm{h.c.}\right]  + \mathcal{B_H}^{ij} C^a C^a \\
& = \mathcal{S}^{ij}+\left(\mathcal{B_H}^{ij}  - \frac{v^2}{8}\mathcal{T}^{ijk} A^{-1}_{kp} F^{(p)}_{U} \right) C^a C^a \\
& +A^{-1/2}_{im}\left[ \bar{\psi}_L M^{(mn)} \psi_R+\textrm{h.c.}\,\right] A^{-1/2}_{nj}	+ \mathcal{T}^{ijk} A^{-1}_{kp} \left[  \bar{\psi} M^{(p)}\psi_R +\textrm{h.c.} \right] \\
&\doteq \mathcal{S}^{ij}+\tilde{\mathcal{B}}^{ij}  C^a C^a +\mathcal{P}^{ij} \ ,
\end{align*}
with

\begin{align*}
&\tilde{\mathcal{B}}^{ij}=\mathcal{B_H}^{ij}  - \frac{v^2}{8}\mathcal{T}^{ijk} A^{-1}_{kp} F^{(p)}_{U} \ , \\
& \mathcal{P}^{ij}=A^{-1/2}_{im}\left[ \bar{\psi}_L M^{(mn)} \psi_R+\textrm{h.c.}\,\right] A^{-1/2}_{nj}	+ \mathcal{T}^{ijk} A^{-1}_{kp} \left[ \bar{\psi}_L M^{(p)}\psi_R +\textrm{h.c.} \right]  \ .
\end{align*}
Thus,

\begin{align*}
& \frac{1}{2} Y^{ij}Y^{ij} \\
&=  \frac{1}{2} \mathcal{S}^{ij}\mathcal{S}^{ij}+\frac{1}{2} \tilde{\mathcal{B}}^{ij} \tilde{\mathcal{B}}^{ij}  C^a_\mu C^{\mu a} C^b_\nu C^{\nu b} +\frac{1}{2} \mathcal{P}^{ij}\mathcal{P}^{ij}  +\mathcal{S}^{ij}\tilde{\mathcal{B}}^{ij}  C^a C^a +\mathcal{S}^{ij}\mathcal{P}^{ij} \\
&+\tilde{\mathcal{B}}^{ij}  C^a C^a \mathcal{P}^{ij}\\
& \overset{IST}{\longrightarrow} \frac{1}{2} \mathcal{S}^{ij}\mathcal{S}^{ij}+2\tilde{\mathcal{B}}^{ij} \tilde{\mathcal{B}}^{ij} \langle L_\mu L^\mu\rangle^2+\frac{1}{2} \tilde{\mathcal{P}}^{ij}\tilde{\mathcal{P}}^{ij} +2\left[\mathcal{S}^{ij}\tilde{\mathcal{B}}^{ij}+\tilde{\mathcal{P}  }^{ij} \tilde{\mathcal{B}}^{ij}\right] \langle L_\mu L^\mu\rangle \\
& + \mathcal{S}^{ij}\tilde{\mathcal{P}}^{ij} \ .
\end{align*}
Therefore,

\begin{align*}
& \frac{1}{2}\text{tr}\, Y^2 \\
& =\Big[ \frac{1}{8}\left(\kappa^2-1\right)^2  +2\tilde{\mathcal{B}}^{ij} \tilde{\mathcal{B}}^{ij}  \Big]\langle L_\mu L^\mu\rangle^2+\frac{1}{8}\left(\kappa^2-1\right)^2 \langle L_\mu L_\nu\rangle^2 \\
&+\Big[-\left(\kappa^2-1\right)\left[ (\partial h_i)(\partial h_j) D^{ij}+ \tilde{\mathcal{Y}} \right]  +2\left[\mathcal{S}^{ij}\tilde{\mathcal{B}}^{ij}+\tilde{\mathcal{P}  }^{ij} \tilde{\mathcal{B}}^{ij}\right]\Big]\langle L_\mu L^\mu \rangle\\
&+\Big[\frac{1}{8}\kappa_n \kappa_m \mathcal{X}_{ni}^\mu \mathcal{X}_{mi}^\nu  + \frac{1}{2} \partial^\mu \kappa_i \partial^\nu \kappa_i + \frac{1}{4} \kappa_j \left( (\partial^\mu \kappa_i)\mathcal{X}_{ji}^\nu + (\partial^\nu \kappa_i) \mathcal{X}^\mu_{ji}\right)  \Big]\langle L_\mu L_\nu \rangle  \\
& +\frac{3}{2}\tilde{\mathcal{Y}}^2+3 \tilde{\mathcal{Y}}(\partial h_i)(\partial h_j) D^{ij} +\frac{4}{v^2}\left( A^{-1/2}_{ji}\left[  i\bar{\psi}_L U \left( \frac{M}{\sqrt{F_U}} \right)^{(j)}  \psi_R+ \textrm{h.c.} \right]\right) ^2 \\
&+ \frac{2A^{-1/2}_{ji}}{v}\left[  i\bar{\psi}_L L_\mu U \left( \frac{M}{\sqrt{F_U}} \right)^{(j)}  \psi_R+ \textrm{h.c.} \right](\partial^\mu \kappa_i) \\
& + \frac{\kappa_n A^{-1/2}_{ji}}{v}\left[  i\bar{\psi}_L L_\mu U \left( \frac{M}{\sqrt{F_U}} \right)^{(j)}  \psi_R+ \textrm{h.c.} \right] \mathcal{X}_{ni}^\mu +\frac{1}{2} \tilde{\mathcal{P}}^{ij}\tilde{\mathcal{P}}^{ij}+ \mathcal{S}^{ij}\tilde{\mathcal{P}}^{ij} \\
&+ \frac{1}{2} \mathcal{S}^{ij}\mathcal{S}^{ij}+ \frac{3}{2}\left[(\partial h_i)(\partial h_j) D^{ij}\right]^2 \ .
\end{align*}		
Finally,

\begin{align*}
&\mathcal{L}_{\text{eff,div}}^{\text{1loop}} = \frac{1}{32\pi^2 \epsilon}\ \times \\
&\Bigg[
-\frac{1+\kappa^2}{12}g^2\left\langle W_{\mu\nu} W^{\mu\nu}\right\rangle -\frac{1+\kappa^2}{24}g'^2 B_{\mu\nu} B^{\mu\nu} +\frac{ \kappa^2-1}{6}gg'\left\langle W_{\mu\nu} \tau_L\right\rangle B^{\mu\nu} \\
&-i\frac{\kappa^2-1 }{12}\Big(g\left\langle W_{\mu\nu} \left[L^\mu, L^\nu \right]\right\rangle+g'B_{\mu\nu}\left\langle \tau_L \left[L^\mu, L^\nu \right]\right\rangle\Big) \\
&-\frac{1}{3}\kappa_i \kappa_i^{(n)} (\partial_\mu h_n) \Big( g\left\langle W^{\mu\nu}L_\nu\right\rangle -g'B^{\mu\nu}\left\langle \tau_L L_\nu\right\rangle\Big) \\
&+\Bigg[ \frac{1}{12}\left(\kappa^2-1\right)^2  +2\tilde{\mathcal{B}}^{ij} \tilde{\mathcal{B}}^{ij}  \Bigg]\langle L_\mu L^\mu\rangle^2+\frac{1}{6}\left(\kappa^2-1\right)^2 \langle L_\mu L_\nu\rangle^2 \\
&+\Bigg[\frac{1}{12}\left(-2(\kappa_i^{(n)} \partial_\mu h_n)^2 -\frac{\kappa_i\kappa_j}{2} \mathcal{X}^{ji}_{\nu}\mathcal{X}^{\nu ji}-2\kappa_j \mathcal{X}^{ji}_{\mu} \kappa_i^{(n)} (\partial^\mu h_n)  \right) \\
&	-\left(\kappa^2-1\right)\Big[ (\partial h_i)(\partial h_j) D^{ij}+\tilde{\mathcal{Y}} \Big] +2\left[\mathcal{S}^{ij}\tilde{\mathcal{B}}^{ij}+\tilde{\mathcal{P}  }^{ij} \tilde{\mathcal{B}}^{ij}\right]\Bigg]\langle L_\mu L^\mu \rangle\\
&+\Bigg[ \frac{2}{3} \kappa_i^{(n)} \kappa_i^{(m)} (\partial_\mu h_n)(\partial_\nu h_m) +\frac{\kappa_j\kappa_k}{6}\mathcal{X}^{ji}_{\mu}\mathcal{X}^{ki}_\nu \\
& +\frac{1}{3}\kappa_j\left(\mathcal{X}^{ji}_\mu \kappa_i^{(n)} (\partial_\nu h_n)+\mathcal{X}^{ji}_\nu \kappa_i^{(m)} (\partial_\mu h_m)\right)   \Bigg]\langle L^\mu L^\nu \rangle  \\
& +\frac{3}{2}\mathcal{\tilde{Y}}^2 +3\mathcal{\tilde{Y}}(\partial h_i)(\partial h_j) D^{ij} \\
&+\frac{4}{v^2}\left( A^{-1/2}_{ji}\left[  i\bar{\psi}_L U \left( \frac{M}{\sqrt{F_U}} \right)^{(j)}  \psi_R+ \textrm{h.c.} \right]\right) ^2 \\
&+ \frac{2A^{-1/2}_{ji}}{v}\left[  i\bar{\psi}_L L_\mu U \left( \frac{M}{\sqrt{F_U}} \right)^{(j)}  \psi_R+ \textrm{h.c.} \right]\kappa_i^{(n)} (\partial^\mu h_n) \\
& + \frac{\kappa_n A^{-1/2}_{ji}}{v}\left[  i\bar{\psi}_L L_\mu U \left( \frac{M}{\sqrt{F_U}} \right)^{(j)}  \psi_R+ \textrm{h.c.} \right] \mathcal{X}_{ni}^\mu +\frac{1}{2} \tilde{\mathcal{P}}^{ij}\tilde{\mathcal{P}}^{ij}+ \mathcal{S}^{ij}\tilde{\mathcal{P}}^{ij} \\
&+ \frac{1}{2} \mathcal{S}^{ij}\mathcal{S}^{ij}+ \frac{3}{2}\left[(\partial h_i)(\partial h_j) D^{ij}\right]^2  \\
&- \frac{1}{12}\Bigg[ \frac{1}{2}\partial_\mu \mathcal{X}^{ij}_{\nu}- \frac{1}{2}\partial_\nu \mathcal{X}^{ij}_{\mu}+\frac{1}{4}\left[\mathcal{X}^{ik}_\mu  \mathcal{X}^{kj}_\nu-\mathcal{X}^{ik}_\nu  \mathcal{X}^{kj}_\mu   \right]\Bigg]^2 \ 
\Bigg] \ .
\end{align*}
The shorthands used, read	

\begin{align*}
\kappa_i &= \frac{vF^{(j)}_U}{2\sqrt{F_U}} A^{-1/2}_{ji}\,,\quad\quad \kappa^2=\sum_i \kappa_i^2 \ , \\
\mathcal{X}_\mu^{ij}& = A^{-1/2}_{ik}\left(F^{(l)}_{km}-F^{(k)}_{lm}\right)(\partial_\mu h_m) A^{-1/2}_{lj} \\
& +A^{1/2}_{ik}  A^{-1/2(n)}_{kj} (\partial_\mu h_n)- A^{-1/2(n)}_{ik}A^{1/2}_{kj}(\partial_\mu h_n) \ , \\
\mathcal{S}^{ij} &=A^{-1/2}_{im}\Bigg[ V^{(mn)}-\left(F^{(kl)}_{mn}-F^{(ml)}_{nk}-F^{(nl)}_{mk}  -\frac{1}{2}F^{(mn)}_{kl} 			\right)(\partial_\mu h_k)(\partial^\mu h_l) \Bigg] A^{-1/2}_{nj}\\
&+\frac{1}{2}\Big[A^{-1/2}_{ik}\left(F^{(l)}_{km}-F^{(k)}_{lm}\right)(\partial_\mu h_m) A^{-1/2(n)}_{lj}(\partial^\mu h_n) \\
&- A^{-1/2(n)}_{ik}(\partial_\mu h_n)\left(F^{(l)}_{km}-F^{(k)}_{lm}\right)(\partial^\mu h_m) A^{-1/2}_{lj}   \Big]\\
&-\frac{1}{2}\left[ A^{-1/2(n)}_{ik}(\partial_\mu h_n)\, A^{1/2(m)}_{kj}(\partial^\mu h_m)+ A^{1/2(n)}_{ik}(\partial_\mu h_n)\, A^{-1/2(m)}_{kj}(\partial^\mu h_m)\right] \\
&-  A^{-1/2}_{im}(F^{(k)}_{mn}-F^{(m)}_{nk}-F^{(n)}_{mk})A^{-1/2}_{nj} A^{-1}_{kr}\left(F^{(r)}_{pq}-F^{(p)}_{rq}- F^{(q)}_{rp} \right)(\partial h_p)(\partial h_q)\\
&+\frac{1}{4}\mathcal{X}^{ik}_\mu \mathcal{X}^{jk}_\mu +    A^{-1/2}_{im}(F^{(k)}_{mn}-F^{(m)}_{nk}-F^{(n)}_{mk})A^{-1/2}_{nj}  A^{-1}_{kp}V^{(p)} \ ,\\
\tilde{\mathcal{B}}^{ij} &=\frac{1}{4} \kappa^i\kappa^j - \frac{1}{8}v^2A^{-1/2}_{im}F_U^{(mn)}A^{-1/2}_{nj}  \\
& - \frac{v^2}{8} A^{-1/2}_{im}(F^{(k)}_{mn}-F^{(m)}_{nk}-F^{(n)}_{mk})A^{-1/2}_{nj}  A^{-1}_{kp} F^{(p)}_{U}\,, \\
\tilde{\mathcal{Y}} &=\frac{1}{2 F_U} F^{(i)}_U A^{-1}_{ij} \left(V^{(j)}+\left[ \bar{\psi}_L U M^{(j)}\psi_R +\textrm{h.c.} \right] \right)  -\frac{1}{v^2F_U}[\bar{\psi}_L U M \psi_R+\textrm{h.c.}] \, , \\ 
\tilde{\mathcal{P}}^{ij} &=A^{-1/2}_{im}\left[ \bar{\psi}_L U M^{(mn)} \psi_R+\textrm{h.c.}\,\right] A^{-1/2}_{nj}	\\
& + A^{-1/2}_{im}(F^{(k)}_{mn}-F^{(m)}_{nk}-F^{(n)}_{mk})A^{-1/2}_{nj}  A^{-1}_{kp} \left[ \bar{\psi}_L U M^{(p)}\psi_R +\textrm{h.c.} \right]  \ , \\
D^{ij} &= \left(\frac{F^{(i)}_UF^{(j)}_U}{4F_U^2} -\frac{F_U^{(ij)}}{2F_U}\right) -\frac{F_U^{(m)}}{2F_U} A^{-1}_{mn}\left(F^{(n)}_{ij}-\frac{1}{2}F^{(j)}_{ni}-\frac{1}{2}F^{(i)}_{nj}\right) \ .
\end{align*}
We give them in a more compact form in Section \ref{NSinglets}.

\fi

\bibliography{Bibliography}

\end{document}